\documentclass[letterpaper,11pt]{article}
\pdfoutput=1 % if your are submitting a pdflatex (i.e. if you have
  \input xy
  \xyoption{all}

\usepackage{jheppub} % for details on the use of the package, please
\usepackage{graphicx}
\usepackage{epstopdf}
\usepackage{amsmath, amssymb}

\usepackage{caption}
\usepackage{subcaption}
\usepackage{mathtools}

\usepackage{tikz}

\newcommand{\be}{\begin{eqnarray}}
\newcommand{\ee}{\end{eqnarray}}
\newcommand{\nn}{\nonumber}
\newcommand{\bn}{\begin{enumerate}}
\newcommand{\en}{\end{enumerate}}

\def\IC{\mathbb{C}}

\def\IP{\mathbb{P}}
\def\IR{\mathbb{R}}
\def\IZ{\mathbb{Z}}
\def\Z{\mathbb{Z}}

\def\CA{{\cal A}}

\def\CC{{\cal C}}

\def\CF{{\cal F}}
\def\CG{{\cal G}}

\def\CI{{\cal I}}

\def\CM{{\cal M}}
\def\CN{{\cal N}}
\def\CO{{\cal O}}

\def\CQ{{\cal Q}}
\def\CR{{\cal R}}

\def\CT{{\cal T}}

\def\CV{{\cal V}}

\def\CZ{{\cal Z}}

\def\a{\alpha}
\def\b{\beta}
\def\g{\gamma}

\def\z{\zeta}
\def\k{\kappa}

\def\G{\Gamma}

\def\S{\Sigma}

\def\half{\frac{1}{2}}

\def\p{\partial}

\newcommand{\suup}{1}
\newcommand{\sudown}{2}

\def\Tr{{\rm Tr}}
\def\tr{{\rm Tr}}
\def\det{{\rm det}}

\usetikzlibrary{arrows}
\usetikzlibrary{positioning}
\usetikzlibrary{decorations.pathmorphing}
\usetikzlibrary{decorations.markings}
\def\arrowhead{angle 90}
\tikzset{>=\arrowhead}
\pgfarrowsdeclaredouble{<<}{>>}{\arrowhead}{\arrowhead}
\pgfarrowsdeclaretriple{<<<}{>>>}{\arrowhead}{\arrowhead}
\pgfarrowsdeclaredouble{<<<<}{>>>>}{<<}{>>}
\tikzstyle{W}=[draw, circle, minimum size=1em, scale=1, inner sep=2pt]
\tikzstyle{B}=[draw,circle,fill=black,scale=1]
\tikzstyle{D}= [circle, minimum size=1em]
\tikzstyle{H}=[draw,circle,fill=gray,scale=1]
\tikzstyle{every picture}=[scale=1,baseline=(current bounding box.south)]

\title{Superconformal Index, BPS Monodromy \\ and Chiral Algebras}

\author[a]{Sergio Cecotti,}
\author[b]{Jaewon Song,}
\author[c]{Cumrun Vafa,}
\author[c, d, e]{and Wenbin Yan}

\affiliation[a]{Scuola Internazionale Superiore di Studi Avanzati, via Bonomea 265, 34100 Trieste, ITALY}
\affiliation[b]{Department of Physics, University of California, San Diego, La Jolla, CA 92093, USA}
\affiliation[c]{Jefferson Physical Laboratory, Harvard University, Cambridge, MA 02138, USA}
\affiliation[d]{Center for Mathematical Sciences and Applications, Harvard University,\\ Cambridge, MA 02138, USA}
\affiliation[e]{Walter Burke Institute for Theoretical Physics, California Institute of Technology,\\ Pasadena, CA 91125, USA}

\emailAdd{cecotti@sissa.it}
\emailAdd{jsong@physics.ucsd.edu}
\emailAdd{vafa@physics.harvard.edu}
\emailAdd{wbyan@cmsa.fas.harvard.edu}

\abstract
{
We show that specializations of the 4d $\CN=2$ superconformal index labeled by an integer $N$
is given by $\tr\,{\cal M}^N$ where ${\cal M}$ is the Kontsevich-Soibelman monodromy operator for BPS states on the Coulomb branch.
We provide evidence that the states enumerated by these limits of the index lead to a family of 2d chiral algebras $\mathcal{A}_{N}$.
This generalizes the recent results for the $N=-1$ case which corresponds to the Schur limit of the superconformal index.
We show that this specialization of the index leads to the same integrand as that of the elliptic genus of compactification of the superconformal theory on $S^2\times T^2$ where we turn on $\half N$ units
of $U(1)_r$ flux on $S^2$.
}

\begin{document}
\maketitle
%%%%%%%%%%%%%%%%%%%%%%%%%%%%%%%%%%%%%%%%%%%%%%%%%%%
\section{Introduction}

Four dimensional superconformal field theories (SCFT's) with ${\cal N}=2$ supersymmetry have common features with the simpler 2d SCFT's
with ${\cal N}=(2,2)$ supersymmetry:  Upon deformations away from the conformal point, both typically lead to Bogomol'nyi-Prasad-Sommerfield (BPS)
states whose mass is given by absolute value of a central charge in the SUSY algebra which is a complex number.
Both cases undergo wall-crossing where the number of BPS states change \cite{Cecotti:1992rm, Seiberg:1994rs, Seiberg:1994aj}.  Despite the
wall-crossing, one can form an invariant from monodromy operators, ordered by the phases of the central charge,
whose invariance under wall crossing characterizes the nature of jumps in the spectrum of BPS states \cite{Cecotti:1992rm, Cecotti:2010qn, Kontsevich:2008fj}.
The monodromy operator is invariant up to conjugation and so its eigenvalues, captured by the trace of its integer
powers, lead to invariants of the SCFT.  It was shown in \cite{Cecotti:1992rm,Cecotti:2010qn} that the eigenvalues of the monodromy matrix capture
 $\exp(2\pi i Q)$ where $Q$ is the $U(1)_R$ charge of the ground states of the theory.  This result motivated
the parallel question in the 4d case \cite{Cecotti:2010fi} where the trace of the powers of the monodromy operator $\CM(q)$ were computed.
It was found that if the $U(1)_r$ charges of the conformal theory are integer multiples of $1/k$, then the monodromy operator
to the power $k$ acts as the identity\footnote{ More precisely it commutes with the line operators.}, very much as in the 2d case, and an explanation for this
phenomenon was provided.  More surprising was that the trace of powers of the monodromy operator $\CM(q)$ (including
fractional powers of $\CM(q)$ when the theory had additional discrete R-symmetries)
in many cases were related to characters of rational 2d conformal theories including minimal models and coset models.
Furthermore it was found that insertion of line operators in the monodromy trace acts by changing the characters of the 2d
conformal theory.  In addition, it was suggested in \cite{Iqbal:2012xm} that an integer sequence of specializations
of the superconformal index should lead to the trace of various powers of the monodromy.
In later work \cite{Beem:2013sza} it was found that, by considering operators contributing to the Schur limit of the superconformal index,
one obtains chiral algebras in 2d, with a very specific central charge:  $c^{2d}=-12 c^{4d}$, where $c^{2d}$
is the 2d central charge of the Virasoro algebra and $c^{4d}$ is the $c$-function of the 4d SCFT.
Motivated by this, in a recent work \cite{Cordova:2015nma} it was shown that the Schur index specialization is equal to the trace of the inverse monodromy operator $\tr\,{\cal M}(q)^{-1}$.

One main motivation for this paper is to generalize this for arbitrary powers of $N$ by revisiting the specialization of
the superconformal index suggested in \cite{Iqbal:2012xm}.  We argue that a modification of that proposal identifies the trace of the $N$-th power of the monodromy operator
$\tr\,{\cal M}(q)^N$ as the superconformal index $\CI(q,p,t)$ with $t=qp^{N+1}$ and $p \rightarrow \exp(2\pi i)$ extending
the $N=-1$ case.  The $N=-1$ is the special case where the index is automatically $p$-independent.
For other values of $N$ one needs to insert some operators for the superconformal
index to lead to a non-vanishing finite answer.  The reason for this is that typically there are some elements
of chiral algebra which correspond to either non-compact bosons, or fermionic zero modes which, if not absorbed,
would make the index diverge or vanish.
We provide evidence for the identification of this limit of the superconformal index with $\mathrm{Tr}\,\CM(q)^N$ by providing general arguments and also by showing explicitly that it works for Lagrangian theories. Moreover we present evidence for the existence of a 2d chiral algebra $\CA_N$ and construct this chiral algebra in the limit of the extreme weak coupling.
 For non-Lagrangian theories we broaden the scope of examples studied in \cite{Cecotti:2010fi} and identify
the corresponding chiral algebras for a number of Argyres-Douglas type theories.
There are two notions of central charge one can associate to these chiral algebras:  One is the Casimir
of the vacuum character, which leads to
\be
c^{2d}=12N c^{4d},
\ee
 generalizing the Schur index case with $N=-1$.  The other one is the growth in the number of
chiral algebra elements of given level which leads (when $c^{4d}-a^{4d}>0$ which is the more typical case)
\begin{align}
 c_{\textrm{eff}}^{2d} &=
 \begin{cases}
 -48N(c^{4d}-a^{4d}) \qquad & {\rm for}\ N<0 \\
 12N c^{4d} \qquad &{\rm for}\ N>0.
 \end{cases}
\end{align}
The equality $c^{2d}\equiv c^{2d}_\text{eff}$ for $N>0$ is consistent with $\mathrm{Tr}\,\CM(q)^N$  being a character of a unitary 2d CFT in this case, as found in various examples \cite{Cecotti:2010fi}.

We also consider partial topological twisting of 4d ${\cal N}=2$ superconformal theory
on $S^2$ with $\half N$ units of $U(1)_r$ flux.  This leads to $(0,2)$ supersymmetric theories in 2d.  Studying
the $(0,2)$ elliptic genus of these theories leads essentially to the same expression for the integrand as in the above specialization of the index.
However, while one uses the Jeffrey--Kirwan residue prescription to compute the elliptic genus, in the index case the integral is
over the unit circle.  In addition one finds that (for $N>0$) the central charge of the 2d theory is $12N c^{4d}$. For $N<0$ one obtains instead the $c^{2d}_\text{eff}$ as the central charge of the resulting
2d theory.

It turns out that there are two competing versions of the Kontsevich-Soibelman operator:  One involves
the compact version of the quantum dilogarithm, which is the main focus of the present paper, and the other one
uses the non-compact version of it.  The trace of the $N$-th powers of the non-compact version of the
monodromy is related to compactifying the 4d theory on $S^1\times S^3$ where as we go around the circle $S^1$ we twist
by $\exp(2\pi i N (r-R))$, and computes its partition function on the squashed $S^3$.  This
is the connection proposed in \cite{Cecotti:2010fi} for the non-compact version.

The organization of this paper is as follows:  In section \ref{sec:review2d} we review the relation between the elliptic genus of $(2,2)$
theories and the BPS monodromy.  In section \ref{sec:BPSmonodromyAndIndex} we discuss the 4d case and outline the argument for the connection
between the BPS monodromy and specializations of the superconformal index as well as its compactification on $S^1$
twisted by $U(1)_r$ charge.   In section \ref{sec:LagTheo} we discuss the 4d $\CN=2$ models with a Lagrangian description.  In section \ref{sec:T2S2} we study
the compactification of the theory on $S^2\times T^2$ with $U(1)_r$ flux through $S^2$ for the Lagrangian models.
In section \ref{sec:BPSandSIndex} we discuss how the traces of the monodromy operators are formulated and computed.
In section \ref{s:monodromiesAD} we give a number of examples for Argyres-Douglas theories. In section \ref{sec:conclusion} we present some
concluding thoughts.  Some technical discussions are postponed to the appendices.

\section{Review 2d case}
\label{sec:review2d}
Consider a $(2,2)$  superconformal theory in 2 dimensions.  Let $J_0,L_0$ denote the left-moving
$U(1)_L$ R-charge and left-moving Hamiltonian $L_0=H-P$ in the NS sector.  Let $H_L=L_0-{1\over 2} J_0$
be the Hamiltonian in the R-sector,  and similar quantities for the right-movers.
 The superconformal index, which can be viewed as the Witten index, in this case is known as the elliptic genus and is given by
 \be
 Z_{ell}(z,q)=\Tr(-1)^F z^{J_0} q^{H_L} {\overline {q}}^{H_R} \ .
 \ee
 The elliptic genus is independent of ${\overline q}$ since $H_R=\{Q_R,Q_R^\dagger\}$ and $Q_R$ commutes
 with $J_0,H_L$.  Let us consider the specialization of this elliptic genus to
 \be
 z={\rm exp}(2\pi i N) \ ,
 \ee
 where $N$ is an integer.  Then $Z_{ell}$ will become independent of $q$ because in this limit
also  $Q_L$ commutes with $z^{J_0}=\exp(2\pi i N J_0)$ (as it carries $J_0$ charge 1).  So in this limit
 the partition function becomes just a number.  Moreover since it is independent of $q$ and ${\overline q}$
 we can compute it in the limit $q,{\overline q}\rightarrow 0$ which implies that it is the index restricted to Ramond ground states:
 \be\label{specialization}
 I_N=Z_{ell}\big({\rm exp}(2\pi i N),q\big)=\Tr_{\text{ground}\atop \text{states}\hfill}(-1)^F {\rm exp}(2\pi iN{J_0})
 \ee
 From this definition it is not \emph{a priori} clear why $I_N$ has to be an integer, but it is.  To see this \cite{Vafa:1989xc}
 note that, by a modular transformation of the torus, this computation is the same as counting the Ramond ground states
 where the space is twisted by ${\rm exp}(2\pi i  N J_0)$.  Note that if the spectrum of $J_0$
is rational of the form $r/k$ then
\be
I_N=I_{N+k}\,,
\ee
so the specialized indices $I_N$ compute only a finite number of independent invariants.

So far we have been studying the conformal point.  Now suppose that the SCFT admits a deformation
with $m$ supersymmetric vacua
 having a mass--gap.
This massive deformation of $(2,2)$ theory will have some BPS solitons $\alpha_{ij}$ with a complex valued central charge $Z_\alpha$
connecting the $i$-th vacuum to the $j$-th one.
 Associate an $m\times m$ upper triangular matrix
 $M_{\alpha_{ij}}$ with 1 along the diagonal and $ij$ entry $n_{ij}$,  which is a signed version of the number of solitons from $i$-th vacuum to the $j$-th
 (see \cite{Cecotti:1992rm} for details).
 Order the BPS solitons according to the phase $\arg Z_\alpha$ and consider the ordered product
 \be
 M=\prod_{ij} T(M_{\alpha_{ij}}) \ ,
 \ee
where $T$ denotes the phase ordering.   When one changes the parameters of the massive theory the number
of BPS states change in such a way that $M$ simply gets conjugated.  In particular the traces of all its powers
are invariant.  Moreover, as shown in \cite{Cecotti:1992rm}
\be
 \Tr\, M^N=I_N=\Tr_{\text{ground} \atop\text{states}\hfill}(-1)^F {\rm exp}(2\pi iN{J_0}) \ .
\ee
In other words, the specialization of the superconformal index \eqref{specialization} has an extension away
from the superconformal point, which is captured by the BPS spectrum of the massive theory.
The generalization of this idea to the 4d ${\cal N}=2$ theory, was the motivation of \cite{Cecotti:2010fi} which we next turn to.

\section{BPS monodromy and 4d index}
\label{sec:BPSmonodromyAndIndex}
We now move on to the 4d ${\cal N}=2$ supersymmetric theories.  Here again we wish to connect the superconformal index computation
to some computation when we move away from the conformal point, i.e.\! as we move on to the Coulomb branch.  Just as in 2d there are
BPS states, which undergo wall crossing, and we wish to connect some invariant data on both sides.

Let us first start with the superconformal side.  The superconformal index is defined as \cite{Kinney:2005ej, Romelsberger:2005eg}(see appendix \ref{app:index} for more detail)
\begin{equation}
\label{eq:def:indi}
\CI(p,q,t)=\Tr(-1)^Fp^{J_{34}+r} q^{J_{12}+r}t^{R-r}e^{-\beta H} \ ,
\end{equation}
where $J_{12}, J_{34}$ are the Cartan generators of the $SO(4)\subset SO(4,2)$ conformal group,
$R$ is the Cartan generator of the $SU(2)_R$ symmetry of ${\cal N}=2$ theories and $r$ is the $U(1)_r$ symmetry
of the conformal theory.  The index can be viewed as the partition function of the conformal theory on $S^1\times S^3$ where
as we go around $S^1$ we rotate the $S^3$ and we have turned on specific fugacities for the $R$, $r$ symmetries.
The Hamiltonian $H$ is a $Q$ commutator, where $Q$ is a supercharge commuting with all  operators inserted in the trace
\eqref{eq:def:indi}, and so the partition function does not depend on $\beta$ and for this reason we sometimes omit writing the $e^{-\beta H}$ insertion.

It is important to understand better this geometry.  For this it is more convenient to view $S^1\times S^3$ as a complex
manifold, known as the Hopf surface, with complex moduli parameterized by $(p,q)$ which we now describe.

\subsection{The Hopf surface}
Consider  the space
\be
W=(\IC \times \IC -\{0,0\})/{\IZ} \ ,
\ee
where the generator of $\IZ$ acts as
\be
(z_1,z_2)\rightarrow (qz_1,pz_2) \, ,
\ee
with $0<|q|,|p|<1$.  $W$ is a complex manifold which is topologically $S^3\times S^1$.  To see this, view $S^3$ as the loci
in the complex 2-plane where $1=|w_1|^2+|w_2|^2$.     Now consider the map
\be
f:\ S^3\times \IR \rightarrow \IC \times \IC -\{0,0\} \ ,
\ee
defined by $\{(w_1,w_2),t\}\mapsto (z_1,z_2)=(q^t w_1,p^t w_2)$.  This map is a bijection.  To see this note that given any $(z_1,z_2)$
there is a unique $\{(w_1,w_2),r\}$ which maps to it, i.e.\! the unique\footnote{ To see that such $t$
is unique, note that $|z_1/q^t|^2+|z_2/p^t|$ is a monotonic function of $t$ which varies in the range $(0,\infty)$ for $t\in\IR$.}  $t$ such that $|z_1/q^t|^2+|z_2/p^t|^2=1$.  To get $W$, note
that modding out by $\IZ$ simply identifies  $t\sim t+1$, and thus $W$ has the topology of $S^3\times \IR/\IZ\equiv S^3\times S^1$.

The Hopf surface $W$ contains two natural tori:  Consider $z_1=0$ (which corresponds also to $w_1=0$).  Over this point the manifold is given by $z_2\sim pz_2$ which defines a torus with
complex structure given by $p$.  Similarly over the point $z_2=0$ (which corresponds to $w_2=0$) we get $ z_1\sim q z_1$ which defines a torus with complex structure $q$.

Note that if you delete the circle given by $w_2=0$ (i.e.\! the circle $|w_1|^2=1$) from $S^3$, you get a non-compact space which has the topology of
\be
{1\over 2} S^3 \times S^1 ,
\ee
where ${1\over 2}S^3$ has the topology of  a solid torus: ${\IC}\times S^1$.

\subsection{A specialization of the 4d index}
We now consider the specialization of the 4d index by setting
$t=qp^{N+1}$.  This leads to

\begin{equation}
\label{eq:def:speci}
\CI'(p,q)=\Tr(-1)^F p^{N(R-r)} p^{J_{34}+R}q^{J_{12}+R} \ .
\end{equation}
There are two special cases of this specialization:  If $N=0$, this partition function is the same as the partition function on the Hopf surface
of the ${\cal N}=2$ theory with Witten's topological twist \cite{Witten:1988ze}, because in the topologically
twisted theory the $SO(4)$ generators are $J'_{12}=J_{12}+R$ and $J'_{34}=J_{34}+R$.    In this case the partition function should not depend
on the metric nor the complex structure of the manifold, and so the partition function  \eqref{eq:def:speci} with $N=0$  does not depend on $p,q$.
 For $N=-1$ we get the Schur limit of the index studied in \cite{Gadde:2011ik, Gadde:2011uv}
which is expected to give a reduction of the index to the partition function of a 2d chiral algebra \cite{Beem:2013sza}.  In that limit the index will not
depend on $p$ but will still depend on $q$ and is characterized by a chiral algebra.  For other values of $N$ we
have an object which a priori depends on both $p,q$.  It can be interpreted as the topologically twisted
theory on $S^1\times S^3$ where in addition as we go around the $S^1$ we mod out by the action of $p^{N(R-r)}$. In other
words, in addition to the usual twist we have introduced a chemical potential for the $(R-r)$ charge.
This is very similar to the 2d case where we inserted $z^{J_0}$.  Here $R-r$ plays the role of $J_0$.  Just as in the 2d case we need a further
specialization to connect to BPS spectra:
Thus we further specialize $p\rightarrow e^{2\pi i}$, i.e.\!
we define the index
\begin{equation}
\label{eq:def:specfin}
\CI_N(q)=\Tr(-1)^F e^{2\pi i N(R-r)} e^{2\pi i({J_{34}+R})} q^{J_{12}+R} \ .
\end{equation}
Note that this specialization can only be realized as a limiting instance of the Hopf surface.
  In particular the torus over $w_1=0$ degenerates.   Over $w_1=0$ there are two circles
which form a torus:  One is the circle $w_2\to w_2\,p^{t+\phi}$.  The other is $w_2\rightarrow w_2 \,e^{i\theta}$.
In the limit $p \rightarrow e^{2\pi i}$ the size of the circle corresponding to phase of $w_2$ is much bigger than that of the circle $t\to t+\phi$.
If we wish to keep the size of $t$-circle finite, we effectively make $w_2$ infinitely large, and so the corresponding circle cannot shrink.  This effectively
deletes the point $w_2=0$ where the corresponding circle would have shrunk from the geometry.  Thus, we get an ${1\over 2} S^3 \times S^1$ geometry
in this limiting case.  View the ${1\over 2} S^3$ as $\big(w_1, {w_2\over |w_2|}\big)={ \IC}\times S^1$.  As we go around $S^1$
we mod out this geometry according to
\be
t\rightarrow t+1: \qquad \left(w_1,{w_2\over |w_2|}\right)\rightarrow  \left(q w_1,   {w_2\over |w_2|}\right) \, .
\ee
In other words we have a Melvin cigar geometry.  This is the geometric part of the interpretation of the
$\CI_N(q)$.  This is also accompanied with the Witten twist due to the $R$ action.  But now the action is not
purely topological because, as we go around the circle, we have  in addition the action of the operator $e^{2\pi i N (r-R)}$ .
In other words we have a geometry of the form
\be
 W'= S^1_{q,N(R-r)}\times S^1\times {\IC} \ ,
\ee
where the notation $S^1_{q,N(R-r)}$ means that as we go around that $S^1$ we both rotate ${\IC}$ by $q$ and has a fugacity action by $e^{2\pi iN(R-r)}$.

\subsection{BPS mondromy and the specialized index}
If we consider a deformation of a ${\cal N}=2$ SCFT in $d=4$ to its Coulomb branch, we find a tower
of BPS states.  This tower of BPS states can change as we move in the Coulomb branch as we cross
walls of marginal stability \cite{Seiberg:1994rs, Seiberg:1994aj},  as was shown in \cite{Kontsevich:2008fj} (and its refinement in \cite{Dimofte:2009bv, Dimofte:2009tm}). If one
consider an ordered product of BPS states, ordered by their central charge phases, where each BPS particle of helicity $s$ and
charge $\gamma$ contributes the factor
\be\label{factorZgamma}
\CZ_\gamma= \prod_{n=0}^\infty(1 -q^{n+{1\over 2}+s} {X}_\gamma)^{(-1)^{2s}}
\ee
to the monodromy operator
\be
M(q)=T\big(\prod_{\gamma} \CZ_\gamma\big).
\ee
Here ${X}_\gamma$ form a quantum torus algebra $\mathbb{T}$:
\be
{X}_\gamma {X}_{\gamma'}=q^{\langle \gamma,\gamma'\rangle} {X}_{\gamma'} {X}_\gamma\qquad \gamma,\gamma^\prime\in\Gamma
\ee
and $\langle \gamma,\gamma'\rangle\in \IZ$ is the electro--magnetic symplectic pairing on the lattice  $\Gamma$ of conserved charges.
$M(q)$ is a wall-crossing invariant up to conjugation.
There is another version of $M(q)$, closely related to the above operator, which is also invariant \cite{Cecotti:2010fi}:  One replaces the compact quantum dilogarithm \eqref{factorZgamma} with its non-compact version
\be
\CZ^\text{nc}_\gamma=\CZ_\gamma (q,X_\gamma)/\CZ_\gamma(\hat q, \hat X_\gamma)
\ee
where now ${X}_\gamma$ and $\hat X_\gamma$ form a dual pair of quantum torus algebras
\be
{X}_\gamma {X}_{\gamma'}=q^{\langle \gamma,\gamma'\rangle} {X}_{\gamma'} {X}_\gamma\qquad  X_\gamma \hat X_{\gamma^\prime}=\hat X_{\gamma^\prime} X_\gamma,\qquad  {\hat X}_\gamma {\hat X}_{\gamma'}=\hat q^{-\langle \gamma,\gamma'\rangle} {\hat X}_{\gamma'} {\hat X}_\gamma,
\ee
related as follows
\begin{equation}
\begin{aligned}
q&=e^{2\pi i\tau}, &X_\gamma&=e^{ix_\gamma},\\
\hat q&=e^{-2\pi i/\tau},
&\hat X_\gamma&=e^{ix_\gamma/\tau}
\end{aligned}\qquad \text{with}\quad\big[x_\gamma,x_{\gamma^\prime}\big]=-2\pi i\tau\,\langle\gamma,\gamma^\prime\rangle.
\end{equation}

 As one
can see this is very similar to the 2d story.  The main difference is that now the monodromy operator  depends on a parameter
$q$.  In the above definition of $M(q)$ it would be natural to include also the massless BPS state, namely the photons of the Coulomb branch.
They would correspond to $\gamma =0$ and $s=\pm 1/2$ leading to $\eta^{-2{r}}(q)$ additional factor to $M(q)$.
  More precisely
we need to delete a zero mode (corresponding to $n=0,s=-1/2$) to get that.  This will be important to keep in mind. We shall write $\CM(q)$ for the monodromy $M(q)$ dressed by  the massless photon factor.

An explanation of this result was provided in \cite{Cecotti:2010fi} as follows:  Consider an ${\cal N}=2$ theory in 4 dimensions.
Compactify it on the Melvin cigar which is the geometry $S^1_q\times {\IC}$ where as you go around $S^1$
you rotate $\IC$ by multiplying it by $q$.  One obtains a theory in 1d with 4 supercharges.  In terms of supersymmetric quantum mechanics (SQM)
data, this theory has infinitely many vacua and the analog of the finite dimensional matrix contribution to the monodromy
operator \cite{Cecotti:1992rm}, gets promoted to the operator $\CZ_\gamma$ given in \eqref{factorZgamma}.
One can view the parameter $q$ as a regularization of the computation\footnote{ For a more detailed discussion of the role of $q$ in this context, see \cite{Cecotti:2014wea}.}.
  Thus consider the following geometry:
\be
S^1_{N(R-r)} \times S^1_q\times {\IC} .
\ee
This gets naturally interpreted as follows \cite{Cecotti:2010fi}:  We compactify the theory from
4d to 3d on a circle where we turn on the global symmetry $g=\exp(2\pi iN (r-R))$ as
we go around the circle.  Then we obtain a 3d theory.  If we consider the partition function of this 3d theory on $\half S^3$ this would give us the trace of the monodormy operator to the $N$-th power.  To make the theory compact, and completing the
${ 1\over 2} S^3$ to $S^3$, we proceed in two different ways.  One way is simply to consider the
squashed partition function on $S^3$ \cite{Closset:2012ru} where  the R-twisting needed to define it is
inherited from the 4d $SU(2)_R$ symmetry (rather than the one natural from 3d SCFT perspective).  This gets identified with the trace of the non-compact version of the monodromy operator $M(q)^\text{nc}$ made of non-compact quantum dilogs.  This was already
suggested in \cite{Cecotti:2010fi} and will not be the main focus of this paper.\footnote{ This is
nevertheless an interesting construction because the $(M(q)^\text{nc})^k$ is strictly the identity operator if the $r$ charges are multiples of $1/k$ and so by studying this operator
we can deduce what fractions appear in the R-charges.  Therefore this will only lead
to inequivalent result for $N$ mod $k$.  In particular for all the Lagrangian
theories where $k=1$ there is nothing interesting to compute in this version.  This is similar to the case of 2d for $(2,2)$ supersymmetric gauged linear sigma models where
the R-charges are integer (or half-integer).}

However to connect to the index it is best to modify this construction slightly and compactify to $S^3$ differently:  We first
compactify the theory on ${\IC}$ which is the cigar geometry with the usual topological twisting.
This leads to a 2d theory with $(2,2)$ supersymmetry due to the non-trivial cigar geometry.  This geometry has an extra $U(1)$ symmetry involving rotation of $\IC$.   In this reduced theory we can
consider the 2d BPS monodromy, but keep track also of the extra $U(1)$ symmetry of the BPS solitons by weighting them with $q^{\rm charge}$.
This will lead again to the $\CZ_\gamma$ factors in eqn.\eqref{factorZgamma}.   In other words, we consider instead
\be
 S^1_{q,N(R-r)}\times S^1\times {\IC}  ,
\ee
but this is precisely the geometry $W'$ that our limiting case of the index computes.  We thus conclude
\be
{\rm Tr}\,{\cal M}(q)^N=Z(S^1_{q,N(R-r)}\times S^1\times {\IC})=\CI_N(q)=\CI(p,q,t)\big|_{t\rightarrow qp^{N+1},\;p\rightarrow e^{2\pi i}} .
\ee
This is essentially the connection anticipated in \cite{Iqbal:2012xm} with a minor modification.  The case $N=-1$, the Schur case, is a special
case of this which was already conjectured and checked in some examples in \cite{Cordova:2015nma}.

%%%%%%%%%%%%%%%%%%%%%%%%%%%%%%%%%%%%%%%%%%%%%%%%%%%%%%
%%%%%%%%%%%%%%%%%%%%%%%%%%%%%%%%%%%%%%%%%%%%%%%%%%%%%%
\section{Lagrangian theories}
\label{sec:LagTheo}

The 4d $\CN=2$ superconformal index is a refined Witten index on $S^3\times S^1$ evaluated by a trace formula \cite{Kinney:2005ej, Romelsberger:2005eg},
\begin{equation}
\label{eq:def:indexi}
\CI(p,q,t)=\Tr(-1)^Fp^{j_1-j_2+r}q^{j_1+j_2+r}t^{R-r}e^{-\beta\delta_{1-}} \ ,\hspace{1cm}\delta_{1-}=2\{\CQ_{1-},\CQ_{1-}^\dag\},
\end{equation}
where $\CQ_{1-}$ is one of the eight supercharges in 4d $\CN=2$ superconformal algebra (SCA), $j_1$, $j_2$, $R$, and $r$ label the $SU(2)_1\times SU(2)_2\times SU(2)_R\times U(1)_r$ symmetry of 4d $\CN=2$ SCA.   Note that $J_{12}=j_1+j_2,\ J_{34}=j_1-j_2$.
By definition the index \eqref{eq:def:indexi} gets contribution from the states annihilated by $\CQ_{1-}$, or equivalently, states satisfying
\begin{equation}
\delta_{1-}=\Delta-2j_1-2R-r=0.
\end{equation}
We can further twist the index with fugacities $\mu_i$ dual to the Cartan generators $F_i$ of the flavor symmetry $\CF$,
\begin{equation}
\label{eq:def:indexFlavor}
\CI(p,q,t,\mu_i)=\Tr(-1)^Fp^{j_1-j_2+r}q^{j_1+j_2+r}t^{R-r}\prod_i\mu_i^{F_i}\,e^{-\beta\delta_{1-}}.
\end{equation}

The superconformal index is invariant under exactly marginal deformations. Therefore for Lagrangian theories we can always compute the exact index in the free theory limit. For a Lagrangian theory with gauge group $\CG$ and matter in the representation $\oplus_i (\CR_i\otimes\CR^F_i)$ of gauge group $\CG$ and flavor group $\CF$, the index is given by
\begin{equation}
\label{eq:def:lagrangianIndex}
\CI(p,q,t)=\int[d\CG]\,\CI^{V}\!(p,q,t; \CG)\prod_i\CI^{H}\!(p,q,t; \CR_i\otimes\CR^F_i),
\end{equation}
$[d\CG]$ being the Haar measure of the gauge group $\CG$, $\CI^{H}$ and $\CI^{V}$ the indices of $\CN=2$ hypermultiplets and vector-multiplets respectively. They are given by
\begin{equation}\label{IHIV}
\begin{split}
&\CI^{H}(p, q, t; \CR\otimes\CR^F) =\prod_{w \in \CR}\prod_{v\in\CR^F} \prod_{m, n=0}^{\infty} \frac{(1- z^w\mu^v t^{-\half} p^{m+1} q^{n+1})}{(1- z^w\mu^v t^{\half} p^m q^n)} \ ,\\
&\CI^{V}(p, q, t; \CG) = \prod_{\a \in \Delta(\CG)} \prod_{m=0}^{\infty} (1-z^{\a} p^{m+1})(1-z^{\a} q^{m+1}) \prod_{m, n=0}^{\infty} \frac{1 - t p^m q^n z^\a }{1-t^{-1} p^{m+1} q^{n+1} z^\a }.
\end{split}
\end{equation}
Here we introduced the flavor fugacity $\mu$ and gauge fugacity $z$. In equation \eqref{IHIV} $w\in\CR$ (resp.\! $v\in\CR^F$) are the weights of the gauge (flavor) representation $\CR$ (resp.\! $\CR^F$), and $\Delta(\CG)$ is the set of all roots of the gauge group $\CG$ (the zero root being counted with multiplicity $r=\mathrm{rank}\,\CG$). $z^\a$ is a short-hand notation for $z^\a \equiv \prod_{i=1}^r z_i^{\a_i}$. For example, for $\CG=SU(2)$, $z^{\a}$ takes the values $z^2,\;1,\;z^{-2}$.

The physical meaning of equation \eqref{eq:def:lagrangianIndex} is now clear. $\CI^V$ and $\CI^H$ are partition functions of supersymmetric states of the respective multiplets. The integrand of \eqref{eq:def:lagrangianIndex} is the partition function of all supersymmetric states of the given theory at zero coupling, and $\int[d\CG]$ projects onto the gauge singlet part of the supersymmetric spectrum.

%One can also compute the index with Wilson line operators wrapping the $S^1$ circle. For two Wilson line operators in $R$ and $\bar{R}$ representation of $\CG$ respective inserted on antipodal points of the $S^3$, the index is
%\begin{equation}
%\CI(p,q,t)=\int[d\CG]\chi_R(z)\chi_{\bar{R}}(z)\CI^{V}(p,q,t,\CG)\prod_i\CI^{H}(p,q,t,\CR_i\otimes\CR^F_i),
%\end{equation}
%with $\chi_R$ and $\chi_{\bar{R}}$ being the character of $R$ and $\bar{R}$ representation of $\CG$ respectively.

\subsection{Specialized index for the Lagrangian theory}

As discussed in previous section, to relate the superconformal index to chiral algebra we need to take a specialization of the index by setting $t=qp^{N+1}$. From the equation \eqref{eq:def:indexFlavor}, the specialized index is
\begin{equation}
\CI_N(p,q;\mu_i)=\Tr(-1)^Fp^{j_1-j_2+R+N(R-r)}q^{j_1+j_2+R}\prod_i\mu_i^{F_i}e^{-\beta\delta_{1-}}.
\end{equation}
Since the indices of Lagrangian theories are built upon indices of free hypermultiplets and vector multiplets, it is enough to discuss these two cases.

When $N=-1$ the specialized index is the Schur index \cite{Gadde:2011ik,Gadde:2011uv} and the dependence on $p$ drops off automatically.

\subsubsection{Insertion of Wilson loop line operators}
Insertion of half-BPS operators in Schur index was discussed by \cite{Dimofte:2011py, Gang:2012yr}. The line operators discussed there preserve the supercharge $\CQ_{1-}+\tilde{\CQ}_{1\dot{-}}$ and commute with $J_{12}+R$. These line operators cannot be inserted in the index for arbitrary values of $p,q$.\footnote{ We would like to thank
Kimyeong Lee for a discussion on this point.}  However something remarkable happens precisely for the specializations we are interested in.
Even though $\CQ_{1-}$ has $J_{34}+R+N(R-r)$ charge $0$,
the corresponding charge of $\tilde{\CQ}_{1\dot{-}}$ is
$N+1$, and the operator
$J_{34}+R+N(R-r)$
 will not commute with the supercharge
 $\CQ_{1-}+\tilde{\CQ}_{1\dot{-}}$ used to define the index in the presence
of line operators (except for $N=-1$).
Nevertheless $\CQ_{1-}+\tilde{\CQ}_{1\dot{-}}$ still commutes with the discrete symmetry
 $$e^{2\pi i (J_{34}+R+N(R-r))}$$ which
is the specialization of interest to us, $p\rightarrow \exp (2\pi i)$.   Thus our index specializations could have been pointed out \emph{a priori}
as the class of superconformal indices consistent with the insertion of supersymmetric line operators.

For Lagrangian theories, the specialized index with insertion of Wilson line operators  in representation $R$ and $\bar{R}$ wrapping $S^1$ and placed at antipodal points of $S^3$ is
\begin{equation}
\label{eq:def:lagrangianSpecializedIndex}
\CI_N(q)=\int[d\CG]\;\chi_R(q)\,\chi_{\bar{R}}(q)\,\CI_N^{V}(q; \CG)\prod_i\CI_N^{H}(q; \CR_i\otimes\CR^F_i),
\end{equation}
with $\CI_N^H$ and $\CI_N^V$ the specialized indices for hyper and vector multiplets which will be discussed in detail below, and $\chi_R(q)$ ($\chi_{\bar{R}}(q)$) the character of the representation $R$ ($\bar{R}$) which contributed from Wilson line operators. The other line operators which play an important role later are
\begin{equation}\label{det1pU}
\det(1-p\, U)_{\mathrm{Adj}}^{\pm}
\end{equation}
with $U$ the holonomy of gauge group along $S^1$. For gauge group $SU(k)$ the operator
\eqref{det1pU} can be understood as a combination of Wilson line operators in fundamental and anti-fundamental representations.
One can also similarly consider insertion of 't Hooft and dyonic line operators, which we will not spell out here as we will not need them for the present paper.

\subsubsection{Free hypermultiplets}\label{ss:Lagrangianfreehypers}

The 4d index of a free (full) hypermultiplet is given by
\begin{equation}
\CI^{H}=
\prod_{i,j=0}^{\infty}
\frac{(1- t^{-\half} q^{i+1}p^{j+1}u)}{(1-t^{\half} q^{i}p^{j}u)}
\frac{(1- t^{-\half} q^{i+1}p^{j+1}u^{-1})}{(1-t^{\half} q^{i}p^{j}u^{-1})},
\end{equation}
with $u$ the fugacity of $U(1)$ flavor symmetry of a free hypermultiplet. After setting $t=qp^{N+1}$ we have
\begin{align} \label{eq:specialIndexHyp}
\begin{split}
\CI^{H}_{N}= \begin{dcases}
\prod_{i=0}^{|N|-1}\prod_{m=0}^{\infty}
\frac{1}{(1-q^{m+\half}p^{\frac{N+1}{2}+i}u)}\frac{1}{(1-q^{m+\half}p^{\frac{N+1}{2}+i}u^{-1})} &\mathrm{for }~~N<0,\\
\prod_{i=0}^{N-1}\prod_{m=0}^{\infty}
\left(1-q^{m+\half}p^{\half-\frac{N}{2}+i}u\right)\left(1-q^{m+\half}p^{\half-\frac{N}{2}+i}u^{-1} \right) & \mathrm{for }~~ N>0 .
\end{dcases}
\end{split}
\end{align}
We can write this in terms of a theta function as
\be \label{cIhn}
\CI^H_N =(q)_\infty^{-N}\!\!\!\! \prod_{-\frac{|N|-1}{2} \le j \le \frac{|N|-1}{2}} \theta (q^{\half} p^j u; q)^{\textrm{sign}(N)} \xrightarrow{\;p \to 1\;} (q)^{-N}_\infty\;\theta(q^\half u; q)^N \ ,
\ee
where
\be\label{whattheta}
\theta(z; q) = \prod_{i \ge 0} (1-q^i)(1-zq^i)(1-z^{-1}q^{i+1})\equiv (q;q)_\infty\, (z;q)_\infty\, (q/z;q)_\infty,
\ee
and we used the standard short--hand notations
\be
(a)_\infty\equiv (a;q)_\infty:=\prod_{k=0}^\infty (1-a q^k).
\ee
In particular, \eqref{cIhn} yields the Schur index for $N=-1$. We see that the specialized index $\CI_N^H$ looks like $N$ copies of the same object (a negative number of copies meaning, as always, $|N|$ copies
with the opposite Fermi/Bose statistics).

To understand the specialized index from the viewpoint of operator counting, we can check in table \ref{tabel:letters} of appendix B the indices of single letters contributing to the index. When $N<0$, we find that letters contributing to the specialized index of hypers are
\begin{equation}
\begin{split}
\partial^i_z\partial^j_wq, \quad \partial^i_z\partial^j_w\tilde{q}, \qquad
\mathrm{with }~~  i\in \IZ_{\ge 0}, \ j=0,1,\cdots,|N|-1,
\end{split}
\end{equation}
where we use the notation $\partial_z \equiv \partial_{+\dot{+}}$ and $\partial_w \equiv \partial_{+\dot{-}}$. The derivatives $\p_z$ and $\p_w$ contribute to the index by $q$ and $p$ respectively. When $N>0$, the letters contributing to the specialized index are
\begin{equation}
\partial^i_z\partial^j_w\psi_+, \quad \partial^i_z\partial^j_w\tilde{\psi}_+, \qquad \mathrm{with}\,\,\ i \in \IZ_{\ge 0}, \  j=0,1,\cdots,N-1.
\end{equation}
In other words, the specialized index counts the operator $q$, $\tilde{q}$ or $\psi_+$, $\tilde{\psi}_+$ with arbitrary number of derivatives $\partial_{+\dot{+}}$ and up to $|N|-1$ derivatives $\partial_{+\dot{-}}$.

It is clear from the equation \eqref{eq:specialIndexHyp} that the specialized index of a free hypermultiplet $\CI_N^H$ is the same as the partition function of $N$ 2d complex bosons with spin $\half$ for $N<0$. When $N>0$, the specialized index describes $N$ 2d complex fermions with the correct spin $1/2$. In both cases, $p$ becomes a fugacity for the $U(1)_p$ charge corresponding to the angular momentum in the $w$-direction.
This shows that for the specialization of the index at $t=qp^{N+1}$ there corresponds
a 2d chiral algebra $\CA_N$, at least for the free theory. This generalizes the 2d chiral algebra associated to the 4d $\CN=2$ theories discussed in Beem \emph{et al.}\! \cite{Beem:2013sza}. In our language, their case corresponds to picking $N=-1$ among all possible choices of $N$. The relation between the specialized index $\CI_N$ and the 2d chiral algebra $\CA_N$ will be discussed further in section \ref{sec:chiralAlgebra}.

It is not difficult to write down the specialized index for hypermultiplets transforming in the representation $\CR$,
\begin{align}
\begin{split}
\CI^{H}_{N}(\CR)=
\begin{dcases}
\prod_{w\in\CR}
\prod_{i=0}^{|N|-1}\prod_{m=0}^{\infty}
\frac{1}{(1-q^{m+\half}p^{\frac{N+1}{2}+i}z^{w})}\frac{1}{(1-q^{m+\half}p^{\frac{N+1}{2}+i}z^{-w})} & \mathrm{for}~~N<0,\\
\prod_{w\in\CR}\prod_{i=0}^{N-1}\prod_{m=0}^{\infty}
\left(1-q^{m+\half}p^{\half-\frac{N}{2}+i}z^{w}\right)\left(1-q^{m+\half}p^{\half-\frac{N}{2}+i}z^{-w} \right) & \mathrm{for}~~ N>0 .
\end{dcases}
\end{split}
\end{align}
This gives the partition function of $|N|$ copies of $2d$ complex spin-$\half$ bosons/fermions in representation $\CR$.

\subsubsection{Free vector multiplets}
Now, let us consider the case of free vector multiplets. As we will see later, we encounter subtle new features involving the nature of zero modes.
The index of a free vector multiplet is given by
\begin{equation}
\label{eq:indexU(1)Vector}
\CI^{V} = \prod_{i=0}^{\infty} (1- p^{i+1})(1-q^{i+1}) \prod_{i, j=0}^{\infty} \frac{1 - t p^i q^j }{1-t^{-1} p^{i+1} q^{j+1} }.
\end{equation}
After specialization to $t=qp^{N+1}$, it becomes
\be
 \CI^V_N = \prod_{i=0}^{\infty} (1- p^{i+1})(1-q^{i+1}) \prod_{i, j=0}^{\infty} \frac{1 -  p^{i+N+1} q^{j+1} }{1- p^{i-N} q^{j} } .
\ee
For $N \ge 0$ the denominator vanishes and the specialized index diverges. Let us first consider the simpler case $N<0$.

\paragraph{Negative $N$.}
When $N<0$, the specialized index is
\begin{equation}
\begin{split}
 \CI^V_{N}
 =&\prod_{i=1}^{|N|-1}(1-p^{i})\prod_{i=0}^{|N|-1}\prod_{j=0}^{\infty}(1-p^i q^{j+1})(1-p^{-i}q^{j+1}) .
 \end{split}
\end{equation}
It reduces to the Schur index at $N=-1$. The 4d letters contributing to the specialized index of vector multiplets are
\begin{equation}
\begin{aligned}
&\partial^j_{z}\partial^{|N|-1}_w\lambda_{1+}, &&\partial^j_{z}\bar{\lambda}_{1\dot{-}}, && \mathrm{with}\,\,\ j \in \IZ_{\ge 0},\\
&\partial^j_z\partial^{i-1}_w\lambda_{1+}, &&\partial^j_z\partial^{i-1}_w\bar{\lambda}_{1\dot{+}}, &&\mathrm{with}\,\,\ j \in \IZ_{\ge 0},\,\,\,\,\,i=1,2,\cdots,|N|-1 \,.
\end{aligned}
\end{equation}
Through the state-operator correspondence, the specialized index is a partition function over the corresponding states on $S^3$. The prefactor $\prod_{i=1}^{|N|-1}(1-p^i)$ represents the contribution of the zero modes $\partial^i_w \bar{\lambda}_{1\dot{+}}$ ($i=0,\,2,\,\cdots,\,|N|-2$). As in the hypermultiplet case, arbitrary number of derivatives $\partial_z$ contribute to the index while $\partial_w$ contribute to the specialized index up to some finite number. The specialized index can also be interpreted as the partition function of $|N|$ copies of $bc$-system $b^{(i)}$ and $c^{(i)}$ ($i=0,\,1,\,\cdots,\,|N|-1$) with $U(1)_p$ charges $\pm i$.

For a vector multiplet of the non-Abelian gauge group $\CG$ of rank $r$ at zero coupling constant, the specialized index at $N<0$ is
\begin{equation}
\begin{split}
\label{eq:specialIndexVecNeg}
 \CI^V_n(\CG)=&\prod_{\alpha\in\Delta(\CG)}\prod_{i=1}^{|N|-1}(1-z^\a p^{i})\prod_{i=0}^{|N|-1}\prod_{j=0}^{\infty}(1-z^\a p^i q^{j+1})(1-z^\a p^{-i}q^{j+1}) ,
\end{split}
\end{equation}
where, as before, $\Delta(\CG)$ is the set of all roots of $\CG$ the zero root counted with multiplicity $r$.

The index has a zero at $p\rightarrow e^{2\pi i}$ of order $r(|N|-1)$, because in this limit the $r$ Cartan elements of $\partial^{i}_w\bar{\lambda}_{1\dot{+}}$ behave as $r(|N|-1)$ fermionic zero modes. As we normally do in 2d, to avoid the problem, we trace over the states with all fermionic zero modes excited. In other words, to get a non-zero answer,
we have to slightly modify the index computation by inserting a suitable operator.  Of course we need to make sure the operator
we insert is gauge-invariant.  One choice we can make is the operator
\be{\cal O}(z)= \det_{\mathrm{Adj}} \!\left( \prod_{i=1}^{|N|-1}\partial^{i-1}_w\bar{\lambda}_{1\dot{+}}(z,0)\right)\ee
%,
and consider trace over the vacuum with the above operator inserted,
\begin{equation}
\CO(0)
  |0\rangle,
\end{equation}
defined by taking the trace over the corresponding space of states.
This removes the whole prefactor $\prod_{\alpha\in\Delta(\CG)}\prod_{i=1}^{|N|-1}(1-z^\a p^{i})$ in the equation \eqref{eq:specialIndexVecNeg}, and we get
\begin{equation}
\begin{split}
 \CI^{V,\mathrm{finite}}_N(\CG)=&\prod_{\alpha\in\Delta(\CG)}\prod_{i=0}^{|N|-1}\prod_{j=0}^{\infty}(1-z^\a p^i q^{j+1})(1-z^\a p^{-i}q^{j+1}).
\end{split}
\end{equation}
This is the partition function of $|N|-1$ copies of the $bc$--system transforming in the adjoint of the gauge group $\CG$.

Instead of inserting a point-like operator we can also use line operators to absorb the fermionic zero modes:  Notice that the insertion of ${\cal O}$ is equivalent to computing the specialized index with an insertion of the form
\be\label{de1pu}\prod_{i=1}^{|N|-1}\frac{1}{\det_{\mathrm{Adj}} (1-p^{i-1}U)}.\ee
We can expand this insertion in characters of $\CG$ as $\sum_R a_R (p) \Tr_R U$, so that \eqref{de1pu} can be thought of as a generating function for certain line operators. The line operator index has been studied in \cite{Dimofte:2011py,Gang:2012yr}. In order to preserve supersymmetry, the line operators should wrap the time circle $S^1$ and be placed at the north and south pole of the $S^3$. Therefore, all the line operator indices should involve insertions of the form $\tr_{R\otimes \bar{R}} U$ in the integral. In our case, the representations which appear are the tensor powers of the adjoint representation. For example, for $SU(k)$ gauge group we can simply use the fact that $k \otimes \bar{k} = \textrm{adj} \oplus 1$ to construct
line operators which effectively insert the operator \eqref{de1pu} in the index computation.

\paragraph{Positive $N$.}
When $N\geq0$, there is a divergence coming from the term $1-t^{-1}p^{N+1}q$ in the denominator of the equation \eqref{eq:indexU(1)Vector}. This term comes from the contribution of $\partial^N_w \bar{\phi}$. Formally, the specialized index of a $U(1)$ vector multiplet can be written as,
\begin{equation}
\CI^{V}_N=\prod_{i=0}^{N}\frac{1}{1-p^{-i}}\prod_{i=1}^{N}\prod_{j=0}^{\infty}\frac{1}{(1-p^{-i}q^{j+1})(1-p^{i}q^{j+1})}.
\end{equation}
The 4d letters contributing to the index are as follows:
\begin{align}
\begin{split}
\partial^{i}_w \bar{\phi},\qquad & \mathrm{with}\,\,\ i=0,1,2,\cdots,N,\\
\partial^{j+1}_z\partial^{i}_w \bar{\phi},\,\,\,\,\,\partial^j_z\partial^{i}_wF_{++}, \quad & \mathrm{with}\,\,\ j=0,1,\cdots,\infty,\,\,\,\,\,i=0,1,\cdots,N-1.
\end{split}
\end{align}
The prefactor $\prod_{i=0}^{N}\frac{1}{1-p^{-i}}$ originates from the contribution of $N+1$ zero modes $\partial^{i}_w \bar{\phi}$. Now we get the partition function of $N$ complex scalars with $U(1)_p$ charges $\pm1,\,\pm2,\,\cdots,\,\pm N$ together with $N+1$ bosonic zero modes.

For the vector multiplet in non-Abelian gauge group $\CG$ with zero coupling constant, formally the specialized index can be written as
\begin{equation}
\begin{split}
 \CI^V_N(\CG)
 =&\prod_{\alpha\in\Delta(\CG)}\prod_{i=0}^{N}\frac{1}{1-z^\a p^{-i}}\prod_{i=1}^{N}\prod_{j=0}^{\infty}\frac{1}{(1-z^\a p^{-i}q^{j+1})(1-z^\a p^{i}q^{j+1})} \quad \mathrm{for}~N\geq0.
\end{split}
\end{equation}
Again there are $r$ copies of the divergent term $\prod_{i=0}^N \frac{1}{1-p^{-i}}$ from the Cartan part of $\CG$ in the prefactor $\prod_{\alpha\in\Delta(\CG)}\prod_{i=0}^{N}\frac{1}{1-z^\a p^{-i}}$.

Under the limit $p\rightarrow e^{2\pi i}$  we get extra poles because $\partial^i_w \bar{\phi}$ contains bosonic zero modes for each $i$. We remove them by taking the trace over the states with delta function inserted:
\begin{equation}
\prod_{a\in \text{Adj}}\left( \prod_{i=0}^{N}\delta(\partial^{i}_w \bar\phi^a(0))\right)|0\rangle .
\end{equation}
We get
\begin{equation}
\begin{split}
 \CI^{V,\mathrm{finite}}_N(\CG)
 =&\prod_{\alpha\in\Delta(\CG)}\prod_{i=1}^{N}\prod_{j=0}^{\infty}\frac{1}{(1-z^\a p^{-i}q^{j+1})(1-z^\a p^{i}q^{j+1})}~~\qquad \mathrm{for}\,\,N\geq0.
\end{split}
\end{equation}
This can also be realized by insertion of line operators.  It is equivalent to the specialized index with the line operator
\begin{equation}
  \prod_{i=0}^{|N|-1}\det(1-p^{i-1}U)_{\mathrm{Adj}}
\end{equation}
inserted.

%%%%%%%%%%%%%%%%%%%%%%%%%%%%%%%%%%%%

\subsection{Chiral Algebras}
\label{sec:chiralAlgebra}

In this section, we provide further evidence for the existence of sectors in the 4d $\CN=2$ SCFT described by a chiral algebra.  In particular we obtain a chiral algebra $\CA_N$ labelled by an integer $N$ for the theory of free hypermultiplets and free vector multiplets.  We also describe the chiral algebra
for interacting theory in the limit $g\rightarrow 0$, by taking into account gauge invariance.   We show that the partition function of the chiral algebra is given by the specialized superconformal index $\CI_N (q)$ (with operator/line operator insertions).

\subsubsection{Free theories}

We call the coordinates of the plane in the $12$ direction $z$ and $\bar{z}$, and the coordinates of the plane in $34$ direction $w$ and $\bar{w}$. In the discussion below, we are going to confine all the operators in the $12$ plane $w=\bar{w}=0$.

\paragraph{Hypermultiplets.}
When $N<0$,
we define the following operators $Q^i(z)$ and $\bar{Q}^i(z)$ on the $z$ plane,
\begin{align}
\begin{split}
\label{eq:def:QQ}
Q^{(i)}(z) &\equiv
\partial^i_w q(z,\bar{z},0,0)
+\bar{z}(z\bar{z})^{|N|-1-i}\partial^{|N|-1-i}_{\bar{w}}\bar{\tilde{q}}(z,\bar{z},0,0),\\
\bar{Q}^{(i)}(z) &\equiv
\partial^{|N|-1-i}_w\tilde{q}(z,\bar{z},0,0)
-\bar{z}(z\bar{z})^i\partial^{i}_{\bar{w}} \bar{q}(z,\bar{z},0,0),
\end{split}
\end{align}
where $i$ runs through $0,1,\cdots,|N|-1$. Note that $\bar{Q}(z)$ is not defined as the complex conjugate of $Q(z)$. The idea behind this construction is that only $q$'s with limited number of derivatives on $w$ contribute to the specialized index and the specialized index is the same as the partition function of $|N|$ complex chiral bosons with wrong spin-$\half$. At $N=-1$ there is no derivatives in $w$ direction and equation \eqref{eq:def:QQ} reduces to chiral fields defined by equation (3.30) in Beem \emph{et al.} \cite{Beem:2013sza} for a free hypermultiplet.

The $Q\bar{Q}$ operator product expansion (OPE) follows from the  correlation function of $q(x)$,
\begin{equation}
q(x)\bar{q}(0)\sim\frac{1}{|x|^2}+\cdots,
\end{equation}
with $|x|^2=z\bar{z}+w\bar{w}$.
From this, we obtain
\begin{equation}
Q^{(i)}(z)\bar{Q}^{(j)}(0)=-\bar{Q}^{(i)}(z)Q^{(j)}(0)=\frac{\delta^{ij}}{z}+\cdots ,
\end{equation}
where the $\bar{z}$-dependence dropped out.
The OPE is chiral in $z$ and the same as $|N|$ copies of 2d symplectic scalars, which gives central charge $c^{2d} = N$. Therefore, the chiral algebra for the free hypermultiplet is given by
\be
 \CA_N (\CT_H) = \left\{ \left\langle Q^{(i)}(z), \bar{Q}^{(i)} (z) \right\rangle \big| i= 0, \cdots, |N|-1 \right\} \quad \textrm{for} \ N<0 .
\ee
We indeed reproduce the partition function from the specialized index $\CI_N^H = (q)_\infty^{|N|}\, \theta(q^{\half} u; q)^{-|N|}$.

When $N>0$ the we define the corresponding chiral fields as
\begin{align}
\begin{split}
\Psi^{(i)}(z) &\equiv
\partial^{i}_w\psi_+(z,\bar{z},0,0)
+z\bar{z}(z\bar{z})^{|N|-1-i}\partial^{|N|-1-i}_{\bar{w}}\bar{\tilde{\psi}}_{\dot{+}}(z,\bar{z},0,0),\\
\bar{\Psi}^{(i)}(z)&\equiv
\partial^{|N|-1-i}_w\tilde{\psi}_+(z,\bar{z},0,0)
-z\bar{z}(z\bar{z})^i\partial^{i}_{\bar{w}}\bar\psi_{\dot{+}}(z,\bar{z},0,0) ,
\end{split}
\end{align}
where $i=0, 1, \cdots, N-1$.
Again, $\bar{\Psi}(z)$ is not the complex conjugate of $\Psi(z)$.
The OPE of the 4d free fermions is given by
\begin{equation}
\psi_+(z,\bar{z},w,\bar{w})\bar{\psi}_{\dot{+}}(0,0,0,0)=\frac{\bar{z}}{(z\bar{z}+w\bar{w})^2}+\cdots.
\end{equation}
Using this result, we obtain the $\Psi\bar{\Psi}$ OPE to be
\begin{equation}
\Psi^{(i)}(z)\bar{\Psi}^{(j)}(0)=\frac{\delta^{ij}}{z}+\cdots,
\end{equation}
where the $\bar{z}$-dependent terms are cancelled out.
This is the OPE of $|N|$ chiral fermions in 2d. So the chiral algebra is given by
\be
 \CA_N (\CT_H) = \left\{ \left\langle \Psi^{(i)}(z), \bar{\Psi}^{(i)} (z) \right\rangle \big| i= 0, \cdots, N-1 \right\} \quad \textrm{for} \ N>0 ,
\ee
with the central charge $c^{2d}=N$. Therefore the partition function of the chiral algebra is indeed given by the specialized index $\CI_N^H = \theta(q^{\half} u; q)^N/(q)_\infty^N$.

\paragraph{Vector multiplets.}

When $N<0$, we define the chiral fields as
\begin{align}
\begin{split}
\partial c^{(0)}(z) & \equiv\partial^{|N|-1}_w\lambda_{1+}(z,\bar{z},0,0)
+\bar{z}\lambda_{2+}(z,\bar{z},0,0),\\
b^{(0)}(z) & \equiv
\lambda_{1\dot{-}}(z,\bar{z},0,0)
+\bar{z}(z\bar{z})^{|N|-1}\partial^{|N|-1}_{\bar{w}}\bar{\lambda}_{2\dot{-}}(z,\bar{z},0,0),
\end{split}
\end{align}
and
\begin{align}
\begin{split}
\partial c^{(i)}(z) &\equiv
\partial^i_w\lambda_{1+}(z,\bar{z},0,0)
+\bar{z}(z\bar{z})^{|N|-1-i}\partial^{|N|-1-i}_{\bar{w}}\lambda_{2+}(z,\bar{z},0,0),\\
b^{(i)}(z)&\equiv
\partial^{|N|-1-i}_w\lambda_{1\dot{+}}(z,\bar{z},0,0)
+\bar{z}(z\bar{z})^i\partial^{i}_{\bar{w}}\bar{\lambda}_{2\dot{+}}(z,\bar{z},0,0),
\end{split}
\end{align}
with $1\leq i\leq |N|-1$.
Their OPE is the same as $|N|$ copies of the $bc$-system,
\begin{equation}
b^{(i)}(z)c^{(j)}(0)=-c^{(i)}(z)b^{(j)}(0)=\frac{\delta^{ij}}{z}+\cdots.
\end{equation}
So the chiral algebra for the vector multiplet can be written as
\be
 \CA_N (\CT_V) = \left\{ \left\langle b^{(i)}(z), c^{(i)} (z) \right\rangle \big| i= 0, \cdots, |N|-1 \right\} \quad \textrm{for} \ N<0 .
\ee
It has the central charge $c^{2d} = 2N=-2|N|$. At $N=-1$ only $b^0(z)$ and $c^0(z)$ remain and there is no derivative with respect to $w$ in the definition. In this case we again obtain the same $bc$-system obtained by Beem \emph{et al.} \cite{Beem:2013sza} for vector multiplet.

When $N>0$ the we define the chiral fields as
\begin{align}
\begin{split}
\partial_z\Phi^{(i)}(z,0) &\equiv
\partial_z\partial^i_w \bar{\phi}(z,\bar{z},0,0)
+\bar{z}^2(z\bar{z})^{|N|-1-i}\partial^{|N|-1-i}_{\bar{w}}F_{++}(z,\bar{z},0,0),\\
\partial_z\bar{\Phi}^{(i)}(z,0) &\equiv
\partial^{|N|-1-i}_wF_{--}(z,\bar{z},0,0)
+\bar{z}^2(z\bar{z})^i\partial_z\partial^{i}_{\bar{w}}\phi(z,\bar{z},0,0),
\end{split}
\end{align}
with $0\leq i\leq N-1$. Their OPE can be computed as,
\begin{equation}
\partial_z\Phi^{(i)}(z)\partial_z\bar{\Phi}^{(j)}(0)=\frac{\delta^{ij}}{z^2}+\cdots ,
\end{equation}
which is independent of $\bar{z}$. This is the same as the OPE of $N$ $2d$ complex scalars.
Therefore the chiral algebra is given by the $N$ complex scalars (in the adjoint of the gauge group)
\be
 \CA_N (\CT_V) = \left\{ \left\langle \Phi^{(i)}(z), \bar{\Phi}^{(i)} (z) \right\rangle \big| i= 0, \cdots, N-1 \right\} \quad \textrm{for} \ N>0 ,
\ee
with the central charge $c^{2d} = 2N$.

We also see the origin of the zeroes and poles in the specialized index of vector multiplets. They are the zero modes of $bc$-system ($N<0$) or complex scalars ($N>0$).

\subsubsection{Interacting theory}

Let us consider a Lagrangian theory $\CT$ with gauge group $\CG$ and hypermultiplets in $\CR\times\CF$ representation of gauge and flavor symmetry. We can construct the 2d chiral algebra $\CA_N$ with the partition function given by the specialized index $\CI_N$ for any integer $N$. The construction is rather simple. We simply prepare the tensor product of the chiral algebras associated to each free matter multiplets and impose the Gauss law constraint. This will certainly reproduce the index $\CI_N$.

For $N<0$ we get $|N|$ $bc$-systems  $b^{a,(i)}(z)$ and $c^{a,(i)}(z)$ with $i=0,1,\,2,\,\cdots,\,|N|-1$ in the adjoint representation of gauge group $\CG$, and $|N|$ chiral symplectic bosons $Q^{(j)}(z)$ and $\bar{Q}^{(j)}(z)$ in representation $\CR\times\CF$. The chiral algebra is built upon the gauge invariant combination of $b^{a,(i)}(z)$, $c^{a,(i)}(z)$, $Q^{(j)}(z)$ and $\bar{Q}^{(j)}(z)$
\be
\label{eq:def:algNegN}
 \CA_N (\CT) = \left\{ \left\langle b^{a, (i)}(z), c^{a, (i)}(z), Q^{(j)}(z), \bar{Q}^{(j)}(z) \right\rangle \big| i, j=0, \cdots, |N|-1 \right\}/ \CG ,
\ee
with $a$ being the index of adjoint representation and $i$, $j$ the label of different copies of the chiral fields.

For $N>0$ we get $N$ chiral scalars $\partial_z\Phi^{a,(i)}(z)$ and $\partial_z\bar{\Phi}^{a,(i)}(z)$ with $i=0,\,1,\,\cdots,\,N-1$ in the adjoint representation of gauge group $\CG$, and $N$ chiral fermions $\Psi^{(i)}(z)$ and $\bar{\Psi}^{(i)}(z)$ in representation $\CR\times\CF$. The final 2d algebra is built upon the gauge invariant combination of  $\partial_z\Phi^{a,(i)}(z)$, $\partial_z\bar{\Phi}^{a,(i)}(z)$, $\Psi^{(i)}(z)$ and $\bar{\Psi}^{(i)}(z)$
\be
 \CA_N (\CT) = \left\{ \left\langle \partial_z\Phi^{a,(i)}(z), \partial_z\bar{\Phi}^{a,(i)}(z), \Psi^{(j)}(z), \bar{\Psi}^{(j)}(z) \right\rangle \big| i, j=0, \cdots, N-1 \right\}/ \CG .
\ee
We constructed superconformal gauge theory out of the free matter content, and the index is obtained by imposing the Gauss law constraint. The chiral algebra at zero gauge coupling is constructed in this way and it certainly exists.

When the gauge coupling $g$ is non-zero, the chiral algebra has to be modified in a suitable way if it exists. In particular
some elements of the algebra at $g=0$ will pair up and disappear at finite $g$, and it is natural to expect that generically there is a 1-1 correspondence
between the elements of the algebra and the character for the algebra at $g>0$.
Note that the partition function of the chiral algebra is robust because it is a limit of the superconformal index (with line operator insertion), which remains the same as we turn on the gauge coupling. Therefore we conjecture that there is still a chiral algebra even at finite coupling.  For $N=-1$ the cohomological arguments of
 \cite{Beem:2013sza} imply that there is an algebra underlying the character at least for that case.  We will find evidence by studying the Argyres-Douglas
 theories that the characters obtained by the monodromy operator are often characters of well known non-trivial 2d chiral algebras.  This gives
 further evidence for our conjecture.

By construction, the central charge of the chiral algebra corresponding to the gauge theory is given (by considering the central charge of the  integrand
contributing to the index) by the formula
\be
 c^{2d} = N(n_h + 2 n_v) \ .
\ee
We see that it can be easily computed by enumerating number of vector multiplets $n_v$ and hypermultiplets $n_h$. For any $\CN=2$ $d=4$ SCFT, one can define more invariant notions of the number of hypermultiplets and vector multiplets by considering conformal anomaly coefficients $a$ and $c$. It is given by
\be
 n_v = 4(2a^{4d}-c^{4d}) , \qquad n_h = 4(5c^{4d}-4a^{4d}) .
\ee
From this relation, we get the $2d$ central charge to be
\be
 c^{2d} = 12Nc^{4d} \ .
\ee
When $N=-1$, this is the same central charge given in \cite{Beem:2013sza}. We conjecture this relation holds even for the non-Lagrangian theories and for all $N$. We provide evidence for this in section \ref{s:monodromiesAD} by computing $\Tr\CM(q)^N$ for Argyres-Douglas theories of $(A_1,A_n)$ type for some values of $N$.

It is also easy to see that the effective growth of the states should be dictated for positive $N$ by $c^{2d}$ but for $N<0$ just by considering the growth of the integrand, it is
\be
c_{eff}^{2d}=-2N(n_h-n_v)=-48N(c^{4d}-a^{4d})\qquad {\rm for}\ N<0.
\ee
Assuming we have a chiral algebra, this implies that the minimum value of $h_\text{min}$ for each $N<0$ is given by
\be
h_\text{min}={1\over 24}(c^{2d}-c^{2d}_{eff})={N\over 2}(5c^{4d}-4a^{4d})<0
\ee
Note in particular for the Schur case of $N=-1$ we are predicting that there should be a representation of the chiral algebra
of the Schur operators whose dimension $h={1\over 2}(4a^{4d}-5c^{4d})$.  For example for $SU(2)$ with $N_f=4$ we expect $h_\text{min}=-1$.
The fact that $h_\text{min}$ should be negative for $N<0$ implies that $(5c^{4d}-4a^{4d})>0$ which is consistent with the unitarity bound \cite{Shapere:2008zf}.
  Moreover if we
have a rational 2d chiral algebra, it is expected that this combination is rational.

\subsubsection{Modular properties for conformal case} \label{subsec:modular}
A further evidence that we have a 2d chiral algebra comes from the fact that the character is a nice modular object.
Let us discuss modular properties of the specialized index $\CI_N$. Before inserting any line operators, the specialized index can be written in terms of $\theta$-functions. We will see in section \ref{sec:T2S2} that it is exactly the same as the integrand for the $S^2 \times T^2$ partition function with $N$-twist (in the zero flux sector).

The specialized index for the hypermultiplet is given by
\begin{equation}
  \CI^{H}_{N}=\prod_{w\in \CR}\theta(\zeta \cdot w;\tau)^{N},
\end{equation}
with $z_i=e^{2\pi i\zeta_i}$, $\zeta\cdot w=\sum_{i=1}^r\zeta_iw^i$ and $q=e^{2\pi i\tau}$. And for the vectors we have
\begin{equation}
  \CI^{V}_{N}=\eta(q)^{-2rN} \prod_{\a\in\Delta^{\pm}(\CG)}\theta(\zeta \cdot \a;\tau)^{-2N},
\end{equation}
where $\Delta^{\pm}(\CG)$ are the set of non-zero roots of the gauge group $\CG$.

The specialized index of a Lagrangian theory with zeroes or divergences removed is then essentially,
\begin{equation}
\CI_n=\eta^{-2rN} \int d\z_1\cdots d\z_r   \left(\prod_{w\in\CR}\frac{\theta^2(\z\cdot w)}{\prod_{\a\in\Delta^{\pm}(\CG)} \theta^2(\z\cdot \a)}\right)^{N},
\end{equation}
which is a Jacobi form.  Note that we are being a bit sloppy in the above formulation because the boson versus fermions have different shifts
in the argument of theta function. Under modular transformation $\z_i$'s transform as $\z_i/\tau$ and the modular weight of $\eta^{2r}$ cancels the weight of $\prod d\z_i$,
leaving us with the modular weight of $\eta^{-2r(N+1)}$ (which picks up a weight because
of operator insertions when $N\not=-1$).  Moreover in the Jacobi form the modular weight of the numerators cancels
that of denominator if and only if
\begin{equation}
\sum_{w\in\CR}(\z\cdot w)^2-\sum_{\a\in\Delta^{\pm}(\CG)}(\z\cdot\a)^2 = 0,
\end{equation}
which is also the condition of vanishing $\beta$-function. Therefore the integrand of the specialized index is modular if and only if the Lagrangian theory is superconformal. As we shall discuss in section \ref{sec:T2S2}, this is exactly the same as the condition that the $N/2$-twisted compactification of the theory on $S^2$ to be free of gauge anomaly.  Note that because we did not
take into account the shifted arguments this is modular only on a subgroup of $SL(2,\Z)$.
Moreover, even though the integrand is modular, the integral may not be modular because we have to change the integration contour by $\z_i \to \z_i/\tau$. The Schur index is given by the vacuum character of the chiral algebra for $N=-1$ \cite{Beem:2013sza}. It is not modular invariant, and will transform to different characters. Notable fact is that the modular transform for a Lagrangian theory is implemented by a change of contour.  The partition function of the chiral algebra $\CA_N$ for $N \neq -1$ is not written in terms of the $\theta$-functions, since we insert appropriate line operators to the integrand to remove zero modes. Therefore we do not expect it to be strictly modular.
This is consistent with its transforming into combination of characters of modules of the same algebra as is expected for chiral algebras.

%%%%%%%%%%%%%%%%%%%%%%%%%%%%%%%%%%%%%%%%%%%%%%%%%%%%%%%%%%%

\subsection{Lagrangian examples}

\subsubsection{Abelian theories with matter}\label{ss:abelianmatter}
Let us start with the simplest example of an interacting theory, even though it is only an effective theory.
The index for the $U(1)$ theory with a hypermultiplet can be evaluated easily by taking the index of a hypermultiplet and then integrating over $U(1)$ gauge group. It is given by
\be
 \CI(p, q, t) = \CI^V \oint \frac{dz}{2\pi i z} \prod_{i, j \ge 0} \left(\frac{1-t^{-\half} p^{i+1}q^{j+1} z }{1- t^{\half} p^i q^j z}\right) \left( \frac{1-t^{-\half} p^{i+1}q^{j+1} z^{-1} }{1- t^{\half} p^i q^j z^{-1}} \right) \ ,
\ee
where $\CI^V$ is the free $U(1)$ vector multiplet index. The integrand in the limit $t=qp^{N+1}$ can be simply written as in \eqref{eq:specialIndexHyp}. Also, if we take $p \to 1$, we simply get
\be \label{eq:sqedIdx}
 \CI_N(q)= {\CI^V_N} \oint \frac{dz}{2\pi i z} (q^{\half} z^{\pm}; q)^N \ .
\ee

The chiral algebra is simply given by that of the free fermions ($N>0$) or bosons ($N<0$) under the Gauss law constraint (along with the decoupled piece coming from the free vector multiplets)
\be
 \CA_N(\CT_{QED}) =
 \begin{cases}
 \CA_N(\CT_V) \otimes \left\{ \left\langle \Psi^i(z), \bar{\Psi}^i(z) \big| i=0, 1, \cdots, N  \right\rangle \right\} / U(1) & N>0,  \\
 \CA_N(\CT_V) \otimes  \left\{ \left\langle Q^i(z), \bar{Q}^i(z) \right\rangle \big| i=0, 1, \cdots, N \right\} / U(1) & N<0 .
 \end{cases}
\ee

\paragraph{Index for $N>0$.}
Let us compute the specialized indices $\CI_N$ (equivalently the character of the chiral algebra $\CA_N (\CT_{QED})$) for $N>0$.  
The integral \eqref{eq:sqedIdx} can be evaluated (with suitable insertion of operators already discussed to absorb the bosonic zero modes) to give
\be
 \CI_N (q) = \CI_N^{V}\times \frac{1}{(q)^N_\infty} \oint \frac{dz}{2\pi i z} \sum_{\vec{k} \in \IZ^{N}} q^{\half \sum_i k_i^2 } (-z)^{\sum_i k_i} = \CI_N^{V} \times\frac{1}{(q)^N_\infty}  \sum_{\vec{k} \in \IZ^{N-1}} q^{Q(\vec{k})}  ,
\ee
where
\be
Q(\vec{k})=\sum_i k_i^2+\sum_{i<j}k_ik_j.
\ee
By general theory of integral (Tits) quadratic forms \cite{ringel1984tame}, $Q(k_i)$ is $\mathbb{Z}$--equivalent to $\ell_i C_{ij}\ell_j/2$ where $C_{ij}$ is the $A_{N-1}$ Cartan matrix, i.e.\! there is a transformation $\ell_i=S_{ij}k_j$, with $S\in SL(N-1,\mathbb{Z})$ such that $Q(S^{-1}_{ij}\ell_j)\equiv \ell_i C_{ij}\ell_j/2$; explicitly, $\ell_i=k_i-k_{i-1}$ with the convention $k_0=0$. Hence
\be
\sum_{\vec{k} \in \IZ^{N-1}} q^{Q(\vec{k})} = \sum_{\ell_i\in\mathbb{Z}^{N-1}}q^{\ell_i C_{ij}\ell_j/2}\equiv \Theta_{SU(N)}(q) \ .
\ee
Therefore, we get
\be
\CI_{N}(q) = \CI^V_{N} \times \frac{\Theta_{SU(N)} (q)}{(q; q)^N_\infty} = \frac{\Theta_{SU(N)} (q)}{(q; q)^{3N}_\infty} \, .
\ee
The central charge is $c^{2d} = 2N + N = 3N$, where the first $2N$ is coming from the decoupled vector multiplet, which becomes $N$ free complex bosons.

\paragraph{Index for $N<0$.}
When $N=-1$, we get the Schur index. We obtain
\be
 \CI_{-1}(q) = \CI^V_{-1} \times \frac{\Psi(q)}{(q; q)_\infty^2} = \Psi(q) \ ,
\ee
where $\Psi(q) = {1\over 2} \sum_{n \in \IZ} q^{\frac{n(n+1)}{2}}$.   Note that even though we needed no insertion for $N=-1$ this
is not a modular weight zero object.  This is consistent with our analysis that showed that only for superconformal theories we expect to get a modular
weight zero object (at $N=-1$).  Nevertheless the theta function is a modular object with weight.
The central charge for the chiral algebra is $c^{2d}_{\textrm{tot}} = -2+2 = 0$. The first term $-2$ is coming from the free vector and the latter $2$ is the one coming from the hypermultiplets under the Gauss law constraint.
For the case of $N=-2$, we find
\be
 \CI_{-2}(q) = \CI^V_{-2} \times \frac{1}{(q; q)_\infty^4} \sum_{n \ge 0} (2n+1) q^{n^2 + n+1} \ .
\ee
We get the central charge as $c^{2d}_{\textrm{total}} = -4 + 4 = 0$ if we also include the vector multiplet.

%%%%%%%%%%%%%%%%%%%%%%%%%%%%

For general $N <0$,
with the help of identities in appendix
\ref{ap:partialthetas},
we get
\be
\mathcal{I}_{N}(q)&=& \mathcal{I}^V_{-|N|}\times \frac{1}{(q;q)^{2|N|}_\infty}\;\sum_{m_i\in\mathbb{Z}^{|N|-1}}
q^{(|\sum_im_i|+\sum_i |m_i|)/2}\psi(-q^{|\sum_im_i|},q)\prod_{i=1}^{|N|-1}\psi(-q^{|m_i|},q) \nn \\
 &=& \mathcal{I}^V_{-|N|}\times \frac{\Psi_{|N|}(q)}{(q;q)^{2|N|}_\infty}\equiv \Psi_{|N|}(q) \ ,
\ee
where
\be\label{rampartial}
\psi(x,q)=\sum_{m\geq0} q^{m(m+1)/2}x^m
\ee
 is Ramanujan's partial theta--function.
The function $\Psi_{N}(q)$ is the sum of $2^{|N|-1}$ multiple partial theta-functions, or more compactly
\be
\Psi_N(q)=\sum_{n_i\in\mathbb{N}^N\atop
m_j\in \mathbb{Z}^{N-1}}(-1)^{\sum_{i=1}^N n_i}\;q^{Q(m_i,n_j)/2+\sum_i (|m_i|+ n_i)/2} \ ,
\ee
where $Q(m_i,n_j)$ is the quadrant--wise quadratic form
\be
Q(m_i,n_j)=\sum_{j=1}^Nn_j^2+2\sum_{i=1}^{N-1}n_i\big|m_i\big|+2\, n_N\left|\sum_{i=1}^{N-1} m_i\right| \ .
\ee
The sum is a (multiple) partial theta function. We get the central charge $c^{2d}_{\textrm{total}} = -2N + 2N = 0$, if we include the vector multiplet contribution.

\subsubsection{$U(1)$ theory with $N_f$ hypermultiplets}
Let us consider a slightly more general case. 

\paragraph{Index for $N \ge 0$.}
Suppose we have several hypers of integral charges $e_a$ and fugacities $y_a$ ($a=1,2,\dots, N_f$) (one redundant).
For $N\geq0$,
\be \label{eq:u1NfIdx}
\mathcal{I}_N (q;y_a)=\frac{1}{(q)^{N (N_f+2)}_\infty}\int \frac{dz}{2\pi i z} \prod_{a=1}^{N_f}\Theta(-y_az^{e_a};q)^N=\frac{\Theta_N(e_a,y_a;q)}{(q)^{N(N_f+2)}_\infty}
\ee
where $\Theta_N(e_a,y_a;q)$ is the theta--function of the rank $NN_f-1$ positive--definite sub--lattice
\be
\Lambda_N(\{e_a\})= \Big\{k_{a,i}\in\mathbb{Z}^{N_f}\times \mathbb{Z}^N\;\Big|\;\sum_{i,a}e_a\,k_{i,a}=0\Big\}\subset \mathbb{Z}^{nN_f},
\ee
endowed with the quadratic form $Q$ induced by the standard one
in $\mathbb{Z}^{NN_f}$,
$\sum_{i,a}k_{a,i}^2$, i.e.
\be
\Theta_N(e_a,y_a;q)=\sum_{k_{i,a}\in \Lambda_N(\{e_a\})} q^{\sum_{i,a}k_{a,i}^2/2}\prod_a(-y_a)^{\sum_i k_{i,a}}.
\ee
In particular, if all $e_a$ are equal, we get the $N_f-1$ variable specialization of the $SU(NN_f)$ theta--function induced by (and covariant under) the subgroup inclusion
\be
SU(N_f)\times SU(N)\subset SU(N_f N).
\ee
Of course, this is just the statement that supersymmetric quantum electrodynamics (SQED) with $N_f$ quarks of the same charge has a $SU(N_f)$ symmetry: indeed, via the replica trick, the integral \eqref{eq:u1NfIdx} is identified with the one entering in the $N=1$ index for SQED with $NN_f$ quarks which has a $SU(NN_f)$ flavor symmetry and therefore produces the $SU(NN_f)$ theta--function specialized to the locus in fugacity space which invariant under the $SU(N)$ replica symmetry. In the basic case that all $e_a$ are equal and $N=1$ we get the full $SU(N_f)$ theta--function depending on all its $N_f-1$ fugacities.

Even more generally, we may couple $k$ Abelian vectors to $N_f$ hypers the $a$--th hyper having (integral) charge $e_{a,\alpha}$ under the $\alpha$--th photon ($\alpha=1,\dots,k$). For $n\geq0$ we get
\be
\mathcal{I}_N(q;y_a)= \frac{\Theta_N(e_{a,\alpha},y_a;q)}{(q)^{N(N_f+2k)}_\infty},
\ee
where in the numerator we have the obvious flavor group covariant specialization of
the theta function for the rank $NN_f-k$ lattice
\be
\Lambda_N(\{e_{a,\alpha}\})= \Big\{k_{a,i}\in\mathbb{Z}^{N_f}\times \mathbb{Z}^N\;\Big|\;\sum_{i,a}e_{a,\alpha}\,k_{i,a}=0\ \text{for }\alpha=1,\dots,k\Big\}\subset \mathbb{Z}^{NN_f}.
\ee

\paragraph{Index for $N < 0$.}
When $N <0$, we have
\be
\mathcal{I}_{-N}(q;y_a)=\frac{1}{(q)^{2|N| (N_f-k)}_\infty}\int \prod_{\alpha=1}^k\frac{dz_\alpha}{2\pi i z_\alpha} \prod_{a=1}^{N_f}\Xi(y_az_\alpha^{e_{a,\alpha}};q)^{|N|}=\frac{\Psi_{|N|}(e_{a,\alpha},y_a;q)}{(q)^{2|N|(N_f-k)}_\infty} \ ,
\ee
where $\Xi(z;q)$ is the function
\be
\Xi(z;q)=\frac{(q)^2_\infty}{(q^{1/2}z;q)_\infty\,(q^{1/2}z^{-1};q)_\infty}
\ee
which may be written as the sum of partial theta--functions in the form (see appendix \ref{ap:partialthetas})
\be
\Xi(z;q)=\sum_{n\in\mathbb{Z}}q^{|n|/2}\,\psi(-q^{|n|},q)\,z^n\equiv \sum_{n\in\mathbb{Z}\atop
m\geq0}(-1)^m q^{m(m+1)/2+|n|(m+1/2)}\,z^n.
\ee
\footnote{The right-hand side of this expression has to be understood with care, because the summand is not absolutely convergent and the ordering is important. Here the sum over $m\ge0$ has to be taken first. We thank O. Warnaar for the comment.}
The function $\Psi_{N}(e_{a,\alpha},y_a;q)$ is the sum of $2^{NN_f-k}$ partial theta functions which is invariant under the $SU(N)$ replica symmetry (which acts on the replica index $i=1,\dots, N$)
\be
\Psi_{N}(e_{a,\alpha},y_a;q)=
\sum_{n_{a,i}\in\Lambda_{N}(\{e_{a,\alpha}\})}\sum_{m_{a,i}\in\mathbb{N}^{NN_f}}(-1)^{\sum_{a,i}m_{a,i}}\,q^{\frac{Q(m_{a,i},n_{a,i})}{2}+\sum_{a,i}\frac{m_{a,i}+|n_{a,i}|}{2}}\, \prod_a y_a^{\sum_i n_{a,i}} \nn
\ee
where $Q(m_{a,i},n_{a,i})$ is the quadrant--wise quadratic form
\be
Q(m_{a,i},n_{a,i})=\sum_{a,i} m_{a,i}^2+2\sum_{a,i} |m_{a,i}|\,n_{a,i}.
\ee

%%%%%%%%%%%%%%%%%%%%%%%%%%%%%%%%%%%%%%%%%%%%%%
\subsubsection{Pure $SU(2)$}
Let us consider the case for the pure $SU(2)$ Yang-Mills theory. This theory is not superconformal away from $g_{YM}=0$, and we restrict our attention to this point.

\paragraph{Specialized index at $N=-1$.}
At $N=-1$ the specialized index is the same as Schur index and for the $SU(2)$ YM it is given by 
\begin{align}\label{shurpuresu2}
\begin{split}
 \CI^{SYM}_{-1}(q) =& \oint \frac{dz}{2\pi i z} \frac{(1-z^{\pm 2})}{2} (q z^{\pm 2, 0}; q)_\infty^2 \\
 =&\sum_{n=0}^{\infty}q^{n(n+1)} 
 = 1+q^2+q^6+q^{12}+\cdots \ .
\end{split}
\end{align}
It can be easily proven by using the Jacobi triple product identity $(q; q)_\infty (y; q)_\infty  (q/y; q)_\infty  = \sum_{k\in \IZ} q^{k(k+1)/2} y^{-k}$ as
\begin{align}
\begin{split}
 \CI^{SYM}_{-1}(q) =& \oint \frac{dz}{2 \pi i z} (q;q)_\infty (z^2;q)_\infty (q/z^2;q)_\infty (q;q)_\infty (z^{-2};q)_\infty (q z^2;q)_\infty \\
 =& \sum_{m, n \in \IZ} q^{\half m(m+1)+ \half n(n+1)} \oint \frac{dz}{2 \pi i z}  z^{2(m-n)} = \sum_{n \ge 0} q^{n(n+1)} \ . 
\end{split}
\end{align}
This is the character of the chiral algebra
\be
 \CA_{-1}(\CT_{SU(2)})\big|_{g=0} = \left\{ \left\langle \partial b^{a}(z), \partial c^{a}(z) \right\rangle  \right\}/ SU(2) ,
\ee
where $a$ denotes the gauge index. This is the singlet sector of the $bc$-system in the adjoint representation of $SU(2)$.

\paragraph{Specialized index at $N=-2$.}
The integral formula with line operators inserted for $N=-2$ specialized index is given by
\be
\CI^{SYM}_{-2}(q) = \oint \frac{dz}{2\pi i z} \frac{(1-z^{\pm 2})}{2} (q z^{\pm 2, 0}; q)_\infty^4  .
\ee
Expanding in powers of $q$ we have,
\begin{equation}
\CI^{SYM}_{-2}(q)=1+6q^2-4q^3+3q^4+12 q^5 - 2 q^6 - 12 q^7 + 18 q^8 + 8 q^9 + 12 q^{10}+\cdots.
\end{equation}
The chiral algebra is built upon the gauge invariant combinations of two $bc$-systems in the adjoint representation of $SU(2)$ with the zero modes $b^{a}(0)$ and $c^{a}(0)$ removed,
\be
 \CA_{-2} (\CT_{SU(2)})\big|_{g=0} = \left\{ \left\langle \partial b^{a, 0}(z), \partial c^{a, 0}(z), \partial b^{a, 1}(z), \partial c^{a, 1}(z) \right\rangle  \right\}/ SU(2) .
\ee
The partition function counts the number of operators with sign. Let us illustrate this by explicitly counting operators for low orders in $q$. In the weak coupling limit, the $6q^2$ term counts six operators
\be
  \Tr_a\partial b^{i}\partial c^{j}, \quad \Tr_a\partial b^{1}\partial b^{2}, \quad \Tr_a\partial c^{1}\partial c^{2}, 
\ee
with $\Tr_a$ the trace over adjoint representation of $SU(2)$. The $-4q^3$ term counts $16$ bosonic operators 
\be
 \Tr_a\partial^2b^i\partial b^j, \quad \Tr_a\partial^2b^i\partial c^j, \quad \Tr_a\partial^2c^i\partial b^j , \quad \Tr_a\partial^2c^i\partial c^j , 
\ee
and $20$ fermionic operators
\be
\epsilon_{abc}\partial b^{a,i}\partial b^{b,j}\partial b^{c,k}, \quad \epsilon_{abc}\partial b^{a,i}\partial b^{b,j}\partial c^{c,k}, \quad \epsilon_{abc}\partial b^{a,i}\partial c^{b,j}\partial c^{c,k},  \quad \epsilon_{abc}\partial c^{a,i} \partial c^{b,j}\partial c^{c,k}. 
\ee
There is no a priori reason that in a non-conformal theory, such as pure $SU(2)$ the theory
is conformal away from $g=0$.  However, if the algebra continued to exist beyond $g=0$ we would have expected that the $16$ bosonic states
would pair up with $16$ of the fermionic states leaving us with 4 fermionic operators.

\paragraph{Specialized index at $N=1$.}
The integral formula with the line operator inserted for $N=1$ specialized index is given by
\be
\CI^{SYM}_{1}(q) = \oint \frac{dz}{2\pi i z} \frac{(1-z^{\pm 2})}{2} \frac{1}{(q z^{\pm 2, 0}; q)_\infty^2} .
\ee
Expanding in powers of $q$ we have,
\begin{equation}
\CI^{SYM}_{1}(q)=1 + 3 q^2 + 4 q^3 + 15 q^4 + 24 q^5+\cdots.
\end{equation}
Note that all the coefficients are positive integers. 
The chiral algebra is built upon the gauge invariant combinations of a complex chiral bosons $\Phi^{a}(z)$ and $\bar{\Phi^{a}}(z)$ in adjoint representation of $SU(2)$ with the zero modes $\Phi^{a}(0)$ removed,
\be
 \CA_{1} (\CT_{SU(2)})\big|_{g=0} = \left\{ \left\langle \partial \Phi^{a}(z), \partial \bar{\Phi}^{a}(z) \right\rangle  \right\}/ SU(2) .
\ee
For example in the weakly coupled limit, $3q^2$ term counts three operators $\Tr_a\partial \Phi\partial \Phi$, $\Tr_a\partial \Phi\partial \bar{\Phi}$ and $\Tr_a\partial \bar{\Phi}\partial \bar{\Phi}$. Since there is no fermionic operator, coefficients of $\CI_1(q)$ are always greater than equal to zero.

%%%%%%%%%%%%%%%%%%%%%%%%%%%%%%%%%%%%%%%

\subsubsection{$SU(2)$ with $N_f=4$}

\paragraph{Specialized index at $N=-1$.}

This is the same as Schur index. In the unrefined limit where we turn off all the flavor fugacities, we get
\begin{equation}
 \CI_{-1}(q) =\oint \frac{dz}{2\pi i z} \frac{(1-z^{\pm 2})}{2} \frac{(q z^{\pm 2, 0}; q)_\infty^2}{((q^{\half} z^{\pm 1}; q)_\infty^2)^4},
\end{equation}
with the expansion
\begin{equation}
\CI_{-1}=1 + 28 q + 329 q^2 + 2632 q^3 + 16380 q^4 + 85764 q^5+\cdots.
\end{equation}
This is the character of $SO(8)_{-2}$ according to \cite{Beem:2013sza}. The chiral algebra $\CA_{-1} (\CT_{SU(2), N_f=4}) = SO(8)_{-2}$ has the central charge equal to that of the Sugawara central charge $c^{2d} = -14$.

\paragraph{Specialized index at $N=1$.}

The specialized index at $N=1$ with line operator inserted (in the unrefined limit) is,
\begin{equation}
\CI_{1}(q) =\oint \frac{dz}{2\pi i z} \frac{(1-z^{\pm 2})}{2} \frac{((q^{\half} z^{\pm 1}; q)^2)_\infty^4}{(q z^{\pm 2, 0}; q)_\infty^2},
\end{equation}
and its expansion in $q$ is
\begin{equation}
\CI_{1}(q)=1 + 36 q + 459 q^2 + 3700 q^3 + 23403 q^4 + 125232 q^5+\cdots.
\end{equation}
The chiral algebra is built upon the gauge invariant combinations of one complex boson $\Phi(z)$ in adjoint representation of $SU(2)$ and four complex bifundamental fermion $\Psi(z)$,
\be
 \CA_{1} (\CT_{SU(2),\,N_F=4}) = \left\{ \left\langle \partial \Phi^{a}(z), \partial \bar{\Phi}^{a}(z), \Psi(z), \bar{\Psi}(z) \right\rangle   \right\}/ SU(2),
\ee
where we remove the zero modes $\Phi^a(0)$ and $\bar{\Phi}^a(0)$. Here $\Psi^i(z)$ is in the $(2,4)$ representation of $SU(2)_{\CG}\times U(4)_{F}$ and $\bar{\Psi}^i(z)$ is in the $(2,\bar{4})$ representation of $SU(2)_{\CG}\times U(4)_{F}$. 

The term $36q$ in the index comes from the gauge invariant combination of $\Psi^{i}(0)\Psi^{j}(0)$, $\Psi^{i}(0)\bar\Psi^{j}(0)$ and $\bar\Psi^{i}(0)\bar\Psi^{j}(0)$, which form the adjoint representation of $USp(8)$ having dimension $36$. Moreover, it has correct decomposition under the flavor symmetry $SU(4)\times U(1) \subset USp(8)$. It shows that there is an $USp(8)$ subalgebra.
For $N>1$, we have $N$ copies $\Psi(z)$ and $\bar\Psi(z)$. We find that the coefficient of $q$ in the index $\CI_{N}(q)$ is given by the dimension of the adjoint representation of $USp(8N)$, which is $4N(8N+1)$.  

\paragraph{Specialized index at $N=-2$}

Let us consider the case with $N=-2$. The specialized index at $N=-2$ is given by
\be
 \CI_{-2}(q) = \oint \frac{dz}{2\pi i z} \frac{(1-z^{\pm 2})}{2} \frac{(q z^{\pm 2, 0}; q)_\infty^4}{(q^\half z^{\pm}; q)_\infty^4)^4} , 
\ee
where its expansion in $q$ is
\be
 \CI_{-2}(q) = 1 + 120 q + 5158 q^2 +124644 q^3 +2065459 q^4+26107916 q^5 + \cdots. 
\ee
The chiral algebra is constructed from the two copies of the $bc$-system ($b^{a, i=0, 1}$, $c^{a, i=0, 1}$) in the adjoint representation of $SU(2)_\CG$ and two copies of the symplectic bosons $Q^{i=0, 1}$ and $\bar{Q}^{i=0, 1}$ in the $(2, 4)$ and $(2, \bar{4})$ representations of $SU(2)_\CF \times U(4)_F$. Then we impose the Gauss-law constraint to build
\be
 \CA_{-2} (\CT_{SU(2),\,N_F=4}) = \left\{ \left\langle \partial b^{a, i}(z), \partial c^{a, i}(z), Q^{i}(z), \bar{Q}^{i}(z) \big| i=0, 1 \right\rangle   \right\}/ SU(2) , 
\ee
where we removed the zero modes in the $bc$-system. 

The term $120q$ in the index comes from the gauge invariant combinations of $Q^i(0)\bar{Q}^j(0)$, $Q^i(0) \bar{Q}^j(0)$ and $\bar{Q}^i(0)\bar{Q}^j (0)$. They form the adjoint representation of $SO(16)$ having dimension $120$. We see that for general $N<0$, the coefficient of the $q$ term is given by the dimension of the adjoint representation of $SO(-8 N)$, which is $4|N|(8|N|-1)$.  

%%%%%%%%%%%%%%%%%%%%%%%%%%%%%%%%%%%%%%%%

%%%%%%%%%%%%%%%%%%%%%%%%%%%%%%%%%%%%%%%%
%%%%%%%%%%%%%%%%%%%%%%%%%%%%%%%%%%%%%%%%%%%%%%%%%%%%%%%
\section{$T^2 \times S^2$ compactifications} \label{sec:T2S2}

Up to now we have discussed a specialization of the index and how to compute it in Lagrangian theories.  It turns out, somewhat
surprisingly, that almost the same result arises from a suitable compactification of the same theory on $S^2$ and considering its partition function on $T^2$.
We will find that this gives the same integrands as the ones appearing in the index computation, even though the contour prescriptions for the integrals
are slightly different.  The nice aspect of this correspondence is that in this context the $T^2$ is the relevant physical space upon taking
the limit where $S^2$ goes to zero size, and not just
part of the $S^3\times S^1$ geometry which is somewhat harder to visualize.  Moreover in this set up we can see more clearly the meaning of fermionic
zero modes for $N<-1$ and bosonic zero modes for $N>0$.  In this context it turns out that one can add a chemical potential associated
with the $U(1)$ rotation of $S^2$ which gets rid of zero modes.
We stress that this construction describes a full physical theory, and not just the supersymmetry protected states that contribute to the elliptic genus.
In a sense we get a (0,2) SCFT in the infrared (IR), whose elliptic genus we are computing.

\subsection{2d $\CN=(0, 2)$ theory from twisted compactification on $S^2$}
Let us consider putting the 4d $\CN=2$ superconformal theory on $\IR^2 \times S^2$. In order to preserve any amount of supersymmetry, we need to perform a partial topological twist \cite{Bershadsky:1995vm}.  The bosonic subgroup of the 4d $\CN=2$ superconformal group includes $SO(4) \times SU(2)_R \times U(1)_r$ where $SO(4)$ is the (Euclidean) Lorentz group acting on the spacetime. Let us consider the subgroup $SO(2)_E \times SO(2)_{S^2} \subset SO(4)$. The first factor acts on $\IR^2$ and the second factor acts on $S^2$. We can twist the theory by considering a linear combination of $SU(2)_R$ and $U(1)_r$
\be \label{eq:twist}
 J_{34}^{(N)} = J_{34} +\left(1+\half N\right) R - \half N r = J_{34} + R + \half N (R-r) \ ,
\ee
where $J_{34}$, $R$, $r$ are the generators of $SO(2)_{S^2}$, $SU(2)_R$, and $U(1)_r$ respectively. When $N=0$, the twisted theory preserves $\CN=(2, 2)$ SUSY in 2d. When $N=-2$, it preserves $\CN=(0, 4)$ SUSY in 2d \cite{Kapustin:2006hi} and it has been considered in \cite{Putrov:2015jpa}. See appendix \ref{app:twisting} for details.

Upon twisting and dimensional reduction, we obtain an effective 2d $\CN=(0, 2)$ theory on $\IR^2$. We can be very explicit for the free theory of hypermultiplet or vector multiplet. The 4d vector multiplet becomes $\CN=(0, 2)$ vector multiplet and $N+1$ chiral multiplets when $N>0$ and becomes vector and $|N|-1$ Fermi multiplets for $N<0$. A 4d hypermultiplet becomes $N$ Fermi multiplets for $N>0$ and $N$ chiral multiplets when $N<0$.  The astute reader may have noticed
that our twisting for odd values of $N$ is somewhat problematic because some fields end up having fractional spin and so we would need a further twist
by some global symmetry to make sense of them.
For odd $N$, this can only be done if we have an additional $U(1)$ global symmetry which distinguish $Q$ and $\widetilde{Q}$ in a hypermultiplet. In this case, we need to further twist the theory by this $U(1)$ \cite{Gadde:2015wta}. This can be always done for a Lagrangian CFT by breaking some of the global symmetry in 4d.
\begin{table}
\centering
\begin{tabular}{|c|c|}
\hline
4d $\CN=2$ multiplet & 2d $\CN=(0, 2)$ multiplets \\
\hline \hline
vector multiplet & 1 vector and
$ \begin{cases}
 N+1~~~ \textrm{chiral multiplets}& (N>0) \\  -(N+1)=|N|-1~ \textrm{Fermi multiplets}& (N<0)
 \end{cases} $ \\
\hline
hypermultiplet &
$\begin{cases}
N~~~ \textrm{Fermi multiplets} &(N>0) \\
 |N|~~ \textrm{chiral multiplets} &(N<0)
\end{cases}$ \\
\hline
\end{tabular}
\caption{4d $\CN=2$ matter multiplets in terms of 2d $\CN=(0, 2)$ multiplets after twisting}
\label{table:twist}
\end{table}

Let us start from a 4d $\CN=2$ SCFT realized as a gauge theory given by gauge group $\CG$ with hypermultiplets in some representation $\CR_i$ of $\CG \times \CF$ where $\CF$ is the flavor symmetry group.\footnote{ Such $\CN=2$ SCFTs are classified in the paper \cite{Bhardwaj:2013qia}.} For this class of theories, the gauge couplings are exactly marginal, so we can continuously deform the theory to the zero coupling limit. At this point, we perform partial topological twisting and shrink the size of $S^2$ to zero. Then we obtain the matter content in table \ref{table:twist}.

When $N<0$, we get $\CN=(0, 2)$ gauge theory with the same gauge group $\CG$ but $(|N|-1)$ copies of the Fermi multiplets ($\Theta$) in the adjoint of $\CG$ and $N/2$ copies of chiral multiplets ($Q, \tilde{Q}$) in the $\CR_i $ and also its conjugate $\bar{\CR}_i$ representation of $\CG \times \CF$. In the parenthesis, we write $\CN=(0, 2)$ superfields for each multiplet. We have both $\CR_i$ and $\bar{\CR}_i$ because the original 4d theory had chiral multiplets in the same representations. When $N$ is odd, we only have $\CR_i$ representation or its conjugate by further twisting by the baryonic charge. In addition, we have a $J$-type superpotential interaction term inherited from 4d $\CN=2$ given by
\be \label{eq:W2d}
 J_{\Theta} = \tilde{Q} Q  \quad \to \quad W = \Tr (\tilde{Q} \Theta Q) \ ,
\ee
where we suppressed indices of the $|N|/2$ copies of $Q, \tilde{Q}$ and $|N|-1$ copies of $\Theta$. One can impose $SU(|N|/2)$ symmetry rotating among the $\Phi$ along with $Q$, by tuning the coupling, but we will not enforce it in this discussion. Note that when $N=-2$, this is exactly the superpotential and matter content to preserve $\CN=(0, 4)$ supersymmetry discussed in \cite{Putrov:2015jpa}.

When $N>0$, we basically reverse the Fermi and chiral multiplets. We get $N+1$ copies of the chiral multiplets ($\Phi$) in the adjoint representation of $\CG$, and $N/2$ copies of chiral multiplets ($\G, \tilde{\G}$) in the $\CR_i $ and also its conjugate $\bar{\CR}_i$ representation of $\CG \times \CF$. As before, when $N$ is odd, we keep $N$ copies of Fermi multiplets in the $\CR_i$ representation. There is no gauge invariant $E$ or $J$ type interaction we can write.

\begin{figure}[t]
	\centering
	\begin{subfigure}[b]{2.9in}
	\centering
	\includegraphics[width=2.3in]{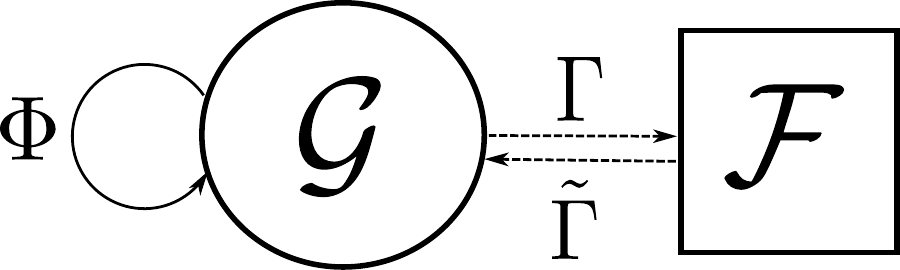}
	\caption{$N>0$: There are $N+1$ copies of $\Phi$ and $N/2$ copies of $(\G, \tilde{\G})$.}
	\end{subfigure}
	\quad
	\begin{subfigure}[b]{2.9in}
	\centering
	\includegraphics[width=2.3in]{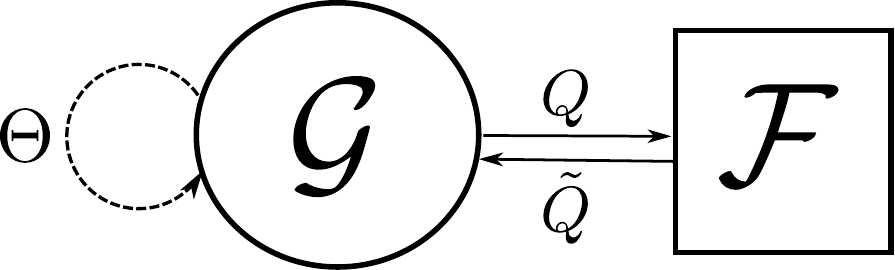}
	\caption{$N<0$: There are $|N|-1$ copies of $\Theta$ and $|N|/2$ copies of $(Q, \tilde{Q})$.}
	\end{subfigure}
	\caption{Schematic matter content of the 2d gauge theory obtained by twisted reduction on $S^2$. $\CG$ denotes the gauge group and $\CF$ denotes the flavor group.}
	\label{fig:2dgaugeN}
\end{figure}
Since 2d theories we obtain are chiral, we should worry about the potential gauge anomalies. When the gauge group $\CG$ has no $U(1)$ factor, we simply need to compute the trace anomaly $\tr(\g^3 \CG \CG)$. For simplicity, let us assume $\CG$ is simple, and write the representation of the hypermultiplets to be $(\CR_{\CG, i}, \CR_{\CF, i})$ where the first and second factor denotes that of gauge group and flavor group respectively. Then the gauge anomaly can be written as
\be
 \tr (\g^3 \CG \CG) = (-1) h^\vee_\CG + (N+1) h^\vee_\CG - N \sum_i \textrm{dim}(\CR_{\CF, i}) c_2(\CR_{\CG, i})  \ ,
\ee
where $h^\vee$ is the dual coxeter number of the gauge group and $c_2$ is the index of the representation. Here the first and second term comes from the gauge multiplet and the adjoint chiral or Fermi. The last term comes from the matter multiplets. Note that the expression works for both signs of $N$. This expression is nothing but  $N$ times the beta function of the 4d $\CN=2$ theory before dimensional reduction. As long as we start from a 4d conformal theory (which is crucial to perform $U(1)_r$ twist), gauge anomalies are absent in 2d.
As we have seen in section \ref{subsec:modular}, this condition is tantamount to the modularity condition of the integrand of the specialized 4d index $\CI_N$.

\subsection{Central charge}
The central charge of the 2d $\CN=(0, 2)$ gauge theory can be computed easily from the 't Hooft anomalies of the R-symmetry as
\be
 c_R = 3\, \tr(\g^3 R^2) , \qquad c_R-c_L = \tr\, \g^3 \ .
\ee
Once we know the $R$-charge of the 2d multiplets, we can compute the 2d central charges.

From the 2d viewpoint, there is no fixed choice of $R$-charge in general, because the theory has extra global symmetry (originating from the $U(1)$ isometry of the $S^2$) that can be mixed with $R$-symmetry. For even $N<0$, one can assign $U(1)_R$ charge for the $\Theta$ to be $1-2\a$ and $Q, \tilde{Q}$ to be $\a$ for arbitrary $\a$. One canonical choice is $\a=0$.  This is because there is a classical moduli space parametrized by $(Q, \tilde{Q})$ and in the semi-classical regime, where the field value is large, we expect there is a sigma-model description parametrized by the chiral multiplets \cite{Witten:1997yu}. When $N=-2$, this is the one studied in \cite{Putrov:2015jpa}. This choice gives central charges
\be
 c^0_L = -2N (n_h - n_v) , \qquad c^0_R = -3N (n_h - n_v) \ ,
\ee
which are positive for $N<0$, but negative for $N>0$.

When $N>0$, the canonical choice is to pick the $R$-charges for the chiral multiplets coming from the vector multiplets to be $0$, and those for the  Fermi multiplets coming from the hypermultiplet to be $0$.
We get
\be
 c^1_L = N(n_h + 2n_v) \ , \qquad c^1_R = 3N n_v \ .
\ee
This choice realizes the CFT on the Higgs branch parametrized by the scalar component of $\Phi$ field.

We find that the value $c_L$ gives the correct effective central charge $c_{\textrm{eff}}$ of $\Tr\, \CM(q)^N = \CI_N(q)$
\be \label{eq:2dcentral}
 c_{2d, N} = N (n_h + 2n_v) \ ,
\ee
for every value of $N$.
For any 4d $\CN=2$ SCFT, one can define effective number of hypermultiplets and vector multiplets by extrapolating the relation between conformal anomalies $(a^{4d}, c^{4d})$ and $(n_h, n_v)$ for Lagrangian theories
\be
 n_v = 4(2a^{4d}- c^{4d}) \ , \qquad n_h = 4(5c^{4d} - 4a^{4d}) \ .
\ee
We can write the 2d central charges in terms of the 4d conformal anomalies as
\be \label{eq:2dcentralc}
 c_{2d, N} = 12Nc^{4d} \ .
\ee
We note that when $N=-1$, the central charge for the left-movers is the same as the one given in \cite{Beem:2013sza}. For negative $N$, this value of central charge can be formally obtained by choosing the `wrong R-charge' $R[Q]=1$, $R[\Theta]=0$ for the matter multiplets. This is related to the fact that the integrand of the elliptic genus with the `wrong R-charge' assignment (for $N<0$) can be obtained from the specialization limit of the 4d index $\CI_{N}$. We discuss it in more detail in the next subsection.

\subsection{$T^2 \times S^2$ vs.\! $S^1 \times S^3$ partition functions}
Let us discuss the connection between our setup $S^1 \times S^3$ vs. $T^2 \times S^2$.
The partition function of a 2d $\CN=(0, 2)$ gauge theory on $T^2$, or elliptic genus, has been computed in \cite{Benini:2013nda,Benini:2013xpa, Gadde:2013dda}. The entire computation boils down to the integral over the moduli space of $\CG$-flat connections $\CM_{\textrm{flat}}(T^2, \CG)$ on $T^2$ with 1-loop determinant
\be
 Z_{ell} = \Tr (-1)^F q^{H_L} x^F = \int_{\CM_{\textrm{flat}}(T^2, \CG)} \prod_{i} du_i\, d\bar{u}_i\, Z_{\textrm{1-loop}} (u, \bar{u}) \ ,
\ee
where $\CG$ is the gauge group and $u, \bar{u} \in \CM_{\textrm{flat}}(T^2, \CG)$. The integrand has a number of poles, that are functions of various chemical potentials related to the global symmetry of the theory. By applying Stokes theorem carefully, the integral can be written in terms of a contour integral over certain holomorphic cycle of $\CM_{\textrm{flat}}$. This yields so-called the Jeffrey-Kirwan residue formula
\be
 Z_{ell} = \oint_{\textrm{JK}} \frac{dz}{z}\, Z_{\textrm{1-loop}}(z) \ ,
\ee
where the integrand is now holomorphic in $z=e^{2\pi i u}$. The integrand has to be elliptic (invariant under $z \to q z$) in order to be free of gauge anomalies.

The integrand for a $(0, 2)$ chiral multiplet and a Fermi multiplet are given by
\be
 Z_{\textrm{chiral}} (z; q) = \frac{(q)_\infty}{\theta(q^{\frac{R}{2}}z; q)} \ , \qquad
 Z_{\textrm{Fermi}} (z; q) = \frac{\theta(q^{\frac{R+1}{2}}z; q)}{(q)_\infty} \ ,
\ee
where $R$ is the R-charge of the multiplet and $\theta(z; q)$ is the theta function defined in equation \eqref{whattheta}
and $z$ is the fugacity for the $U(1)$ flavor associated to the matter multiplet. Here we used the integrand for the NS sector elliptic genus, which agrees with the RR elliptic genus under the spectral flow and multiplicative factor. The NS elliptic genus is more convenient to compare with the 4d index as we will see. For a vector multiplet, we have
\be\label{zvec}
 Z_{\textrm{vec}} (z; q) = (q)^{2r}_\infty \prod_{\a \in \Delta^{\pm}(\CG)} \vartheta(z^{\a}; q) \ ,
\ee
where $r$ is the rank of the gauge group $\CG$ and $\Delta^\pm (\CG)$ is the set of all (non-zero) roots of $\CG$. in eqn.\eqref{zvec} we use the short-hand notations $z^\alpha \equiv \prod_i z_i^{\a_i}$ and
\be
\vartheta(z;q)= \theta(z;q)/(q)_\infty.
\ee

Now, we write the elliptic genus of the $\CN=(0, 2)$ theory obtained from 4d $\CN=2$ theory on $S^2$ with the twist of \eqref{eq:twist}. Under the twisting and dimensional reduction, 4d vector multiplet gives
\be
  Z^{V}_{N>0} =  \frac{(q;q)^{2r}}{\vartheta(v q^{\frac{R_\Phi}{2}}; q)^{r(N+1)}} \prod_{\a \in \Delta^\pm (\CG)} \frac{\vartheta(z^{\a}; q)}{\vartheta(v q^{\frac{R_\Phi}{2}} z^{\alpha}; q)^{N+1}} \ .
\ee
The terms in the numerator come from the $\CN=(0, 2)$ vector and the denominator is coming from the chiral multiplets $\Phi$.
Here $v$ is introduced to regulate the zero modes of the chiral/Fermi multiplets. It is the fugacity of the $U(1)$ global symmetry associated to the rotation along $S^2$ direction. This is necessary because the theory is non-compact. As we have argued in the previous subsection, there is a CFT in the Higgs branch parametrized by $\Phi$. The correct R-charge for this case is $R_\Phi = 0$.

When $N<0$, we get
\be \label{eq:s2t2vecNneg}
  Z^{V}_{N<0} =  (q;q)^{2r} \vartheta( v q^{\frac{R_{\Theta}+1}{2}}; q)^{2r(|N|-1)} \prod_{\a \in \Delta^\pm (\CG)} \vartheta(z^{\a}; q)\,\vartheta( v q^{\frac{R_{\Theta}+1}{2}} z^{\alpha}; q)^{|N|-1} \ ,
\ee
coming from the vector multiplet and adjoint Fermi multiplets $\Theta$.

When we choose the R-charge for the adjoint chiral $R_\Phi$ to be $0$ and adjoint Fermi $R_\Theta$ to be $-1$, we get a formula
\be \label{eq:s2t2vec}
 Z^{V}_N = \frac{(q; q)^{2r}}{\vartheta(v; q)^{r(N+1)}} \prod_{\a \in \Delta^\pm (\CG)} \frac{\vartheta(z^\a; q)}{\vartheta(v z^\a; q)^{N+1}} ,
\ee
which works for all $N$. This choice of R-charge $R_\Phi = 0$ for $N>0$ is physical and gives the elliptic genus of a CFT on the Higgs branch parametrized by the scalar component of $\Phi$. On the other hand, the choice of $R_\Theta = -1$ for $N<-1$ seems to be unphysical. But we find that the equation \eqref{eq:s2t2vec} is exactly the same one we get from the $S^1 \times S^3$ partition function (superconformal index) upon the specialization limit $t=qp^{N+1}$ (when Haar measure is also included in the vector multiplet index) $\CI^V_N$, with $v=1$.

For the hypermultiplet, we get
\be \label{eq:s2t2hypN}
 Z_{N>0}^{H} = \vartheta (q^{\frac{R_\Gamma+1}{2}} z; q)^N \ , \qquad Z_{N<0}^H = \vartheta(q^{\frac{R_Q}{2}}; q)^{-|N|} \ ,
\ee
which comes from $|N|$ copies of Fermi or chiral multiplets. The physical choice is to set $R_\G = 0$ for $N>0$ and $R_Q = 0$ for $N<0$, which gives the CFT on the Higgs branch. On the other hand, if we pick $R_{\G}=0$ while $R_Q = 1$, we get
\be \label{eq:s2t2hyp}
 Z_N^{H} = \vartheta(q^{\half} z; q)^N \ .
\ee
Note that, this formula is valid regardless of whether $N$ is positive or negative. As before, choosing `unphysical' R-charge assignment for $N<0$ gives the integrand to be the same as the specialization limit of the 4d superconformal index $\CI^H_N$.

With the integrand given as above, we need to evaluate Jeffrey-Kirwan residues to obtain the partition function on $T^2 \times S^2$.\footnote{ Precisely speaking, we also need to sum over different flux sectors of the gauge fields on $S^2$ \cite{Benini:2015noa}, which will yield multiple copies of the 2d theory with different charges. Here we only focus on the zero flux sector. }
We note that the integrand \eqref{eq:s2t2vec} and \eqref{eq:s2t2hyp} is exactly the same one we get from the $S^1 \times S^3$ partition function upon specialization $t=qp^{N+1}$:
\be
 Z(T^2 \times S^2) = \oint_{\rm JK} \frac{dz}{z}\, Z^{(N)}_{\textrm{1-loop}} \ , \quad Z(S^1\times S^3) \big{|}_{t=qp^{N+1}} = \oint_{|z|=1} \frac{dz}{z}\, Z^{(N)}_{\textrm{1-loop}} \ .
\ee
The crucial difference here is that unlike the case of $T^2 \times S^2$, the integration contour is simply given by a unit circle. This integration contour is coming from the $\CG$--flat connections on a circle $S^1$ instead of a torus $T^2$. One can consider both geometries as locally the same $T^2 \times S^2$, where in one case $S^1 \subset T^2$ is fibered over $S^2$ non-trivially. This fibration removes one of the circle directions in the integral, so that we get an integration over the flat connections on a circle.  Another difference is that the $T^2 \times S^2$ partition function is guaranteed to be modular (Jacobi form of weight $0$) whenever the underlying theory is conformal, whereas the $S^1 \times S^3$ partition function is not invariant, as we discussed in section \ref{subsec:modular}.

Let us remark that when $N>0$, the above partition function $Z(T^2 \times S^2)$,  using the integrand \eqref{eq:s2t2vec} and \eqref{eq:s2t2hyp}, gives a valid elliptic genus of the 2d $\CN=(0, 2)$ CFT on the Higgs branch (along with the vector bundle) of the 2d theory (which is not the same as the Higgs branch of 4d theory) with central charge given in \eqref{eq:2dcentralc}. In order to actually evaluate the integral, we need to consider fully refined fugacities for the extra global symmetry rotating the $N+1$ copies of the adjoint chiral multiplets, otherwise the elliptic genus diverge. This is the case because our target space is non-compact. For $N<0$, we should use the integrand \eqref{eq:s2t2vecNneg} and \eqref{eq:s2t2hypN} with $R_Q = 0$, $R_\Theta = 1$ and with all the flavor fugacities turned on to get a proper elliptic genus on the Higgs branch (with vector bundle).  This is reminiscent of the fact that in the specializations of the index we had to insert operators to absorb fermionic/bosonic zero modes (for $N\not= -1$).

%%%%%%%%%%%%%%%%%%%%%%%%%%%%%%%%%%%%%%%%%%%%%%%%%%%%%%%%%%%%%%%%%%
%%%%%%%%%%%%%%%%%%%%%%%%%%%%%%%%%%%%%%%%
\section{BPS monodromy and the specialized index}
\label{sec:BPSandSIndex}

The quantum Kontsevitch--Soibelman (KS) wall-crossing formula for 4d $\mathcal{N}=2$ field theories states\footnote{ This version of the quantum KS formula holds under two genericity assumptions: 1) the only massless states are IR--free photons, and 2) massive BPS states with equal BPS phases, $\arg Z_\gamma=\arg Z_{\gamma^\prime}$, are mutually local i.e.\!
$\langle\gamma,\gamma^\prime\rangle=0$.
} that the
phase--ordered product\footnote{ The sign $\pm$ in the argument of the $q$--Pochammer symbol stands for the \emph{quadratic refinement}  of \cite{Gaiotto:2008cd}.} \cite{Kontsevich:2008fj}
\be\label{mmm1}
M(q)=\prod^\circlearrowright_{\text{massive}\hfill\atop\text{BPS states}} \big(\pm q^{s+1/2}X_{\gamma};\,q\big)_\infty^{(-1)^{2s}}
\ee
taken over the full spectrum of massive BPS states (of charge $\gamma\in\Gamma$
and spin $s$) in the clockwise order $\circlearrowright$ with respect to the BPS phase\footnote{ Here $Z_\gamma$ is the $\CN=2$ SUSY central charge associated to the charge $\gamma$; one has $Z_{\gamma+\gamma^\prime}=Z_\gamma+Z_{\gamma^\prime}$.} $\arg Z_\gamma$, is independent of the particular BPS chamber up to conjugacy in the quantum torus algebra $\mathbb{T}$ with multiplication table
\be\label{multx}
X_\gamma X_{\gamma^\prime}=q^{\langle \gamma,\gamma^\prime\rangle/2}\, X_{\gamma+\gamma^\prime}\qquad\gamma,\gamma^\prime\in\Gamma.
\ee
Hence the traces $\mathrm{Tr}\,M(q)^N$
 are \emph{absolute} wall-crossing invariants for all $N\in\mathbb{Z}$ and $q\in\mathbb{C}$ for which they are well defined.

 Suppose we start from a UV SCFT
 $\mathcal{T}$, and mass--deform it in a generic way by going to the Coulomb branch. In the IR we end up with a theory
 $\mathcal{T}_\text{IR}$ which contains $r$ massless photons ($r$ being the dimension of the Coulomb branch) and a non--trivial massive BPS spectrum. From this spectrum
we compute the set of traces $\{\mathrm{Tr}\,M(q)^N\}$ which are well defined for $\CT_\text{IR}$. The Kontsevitch-Soibelman theory implies that the $\{\mathrm{Tr}\,M(q)^N\}$ depend only on the UV fixed point $\mathcal{T}$ and not
 on the particular mass--deformation $\mathcal{T}_\text{IR}$. Hence
 the $\{\mathrm{Tr}\,M(q)^N\}$ are to be thought of as invariant
 properties of the UV fixed point, in perfect analogy with the 2d (2,2) situation \cite{Cecotti:1992rm,Cecotti:2010qn} reviewed in section \ref{sec:review2d}. The monodromy traces $\{\mathrm{Tr}\,M(q)^N\}$ then should be related to the natural SCFT invariants of $\mathcal{T}$, in particular to its specialized indices $\CI_N(q)$. In order to connect the monodromy traces and usual SCFT invariants, one has to treat all BPS states of $\mathcal{T}_\text{IR}$ on the same footing, including the massless ones which are not taken into account in the KS product $M(q)$. This gives an explanation of the prescription suggested in  \cite{Cordova:2015nma}.
 Then we redefine the monodromy operator by inserting the factors which correspond to the $r$ massless photons, seen as chargeless BPS vector multiplets,
 \be
 \mathcal{M}(q)= \frac{1}{(q)^{2r}_\infty}\,M(q),
 \ee
 and take the objects $\{\mathrm{Tr}\,\mathcal{M}(q)^N\}$ (to the extent such quantities are well defined) as the natural invariants to compare with the SCFT ones
 \be\label{?map}
 \big\{\;\mathrm{Tr}\,\mathcal{M}(q)^N\;\big\} \ \overset{?}{\longleftrightarrow}\ \big\{\CI_N(q)\big\}.
 \ee
For $N=-1$ this map is expected to be especially simple:
$\mathrm{Tr}\,\mathcal{M}(q)^{-1}$ gets identified with the SCFT index in the
Schur limit which is independent of $p$. Strong evidence for this identification
was given in \cite{Cordova:2015nma}.  However, even for the Schur case we encounter a difficulty because
the $\mathrm{Tr}\,\mathcal{M}(q)^{-1}$ is not always well defined, as computed from the BPS spectrum.  For example
for $SU(2)$ with $N_f=4$, even if one uses the chamber where there are only a finite number of BPS particles, it turns out
that the naive computation of $\mathrm{Tr}\,\mathcal{M}(q)^{-1}$ leads to divergencies which need to be regularized.
We do not know of any general prescription for how to do this.  In the absence of a general such prescription to regularize
such computations we instead resort to cases which is free from these ambiguities as well as some general properties
which do not depend on how the regularization is performed.  There is an additional issue for $\mathrm{Tr}\,\mathcal{M}(q)^N$ for
$N\not= -1$, because as we have explained we need to absorb extra zero modes to get a finite non-vanishing answer.
Therefore the comparison of the corresponding computation with the UV computation becomes more difficult due to mapping
operators from UV to IR.  Nevertheless the general properties which are regularization independent (such
as the effective 2d central charge) as well as the example of Argyres-Douglas theories for which the characters
seem to describe natural 2d chiral algebras, support the picture we have found in the previous sections.

For the case of Lagrangian theories we have stronger results.  First we show that the computation of the specialized index and the monodromy computation
are {\it identical} for free theories (see also  \cite{Cordova:2015nma}).  Moreover for interacting theories since the computation is independent
of where we are in the moduli space, we can take the limit of $g_{YM}\rightarrow 0$.   In this limit the BPS spectrum becomes
that of the free fields in the Lagrangian plus an infinite tower of dyonic BPS states whose mass $m\rightarrow \infty$.  Thus
the decoupling arguments for physical computations would lead us to consider the collection only of electric states and
project to the gauge invariant subsector.  This is identical to what one does both in the index computation and the $\mathrm{Tr}$ computation
of the BPS particles and so the computations are identical.\footnote{This argument also extends to the case of insertion of line operators.} The only point which is not entirely trivial is to argue the more or less
`obvious' physical fact that infinitely massive states should decouple from the computation.  An explicit proof of this
turns out to be difficult because of the ambiguities noted above in defining the trace. Instead we resort to highly non-trivial
consistency checks that shows we need to set the product of the contributions of BPS states coming from the dyonic towers to 1, i.e.
we can ignore them in the extreme weak coupling limit.

The organization of this section is as follows:  We first talk about the free case and then discuss in detail the argument why
the tower of dyonic states should not contribute.  We then talk about some general aspects of the trace of the monodromy
which applies to all theories, and not just the Lagrangian ones.  In the following section we show how these general expectations
are borne out in the context of Argyres-Douglas theories.

\subsection{Free hypers and hypers coupled to abelian gauge groups}\label{ss:freehypers}

There is a case in which the map
\eqref{?map} is obvious, namely when
the BPS particles are all mutually local.
Then the theory has a unique BPS chamber, and all BPS factors in the KS product belong to a commutative sub--algebra of the quantum torus algebra $\mathbb{T}$. We may think of $\mathcal{M}(q)^N$ as ordinary
functions of the fugacities for the local Noether charges (electric and flavor). The traces $\{\mathrm{Tr}\,\mathcal{M}(q)^N\}$ are then the
integrals of these functions with respect to the electric fugacities.

Examples of this situation are the free theories of hypermultiplets and vector multiplets, as well as the Abelian gauge theories coupled to
hypermultiplets considered in section \ref{ss:abelianmatter}.
(These last theories are however not UV complete,
so our arguments should be considered purely formal in this case).

In the free theory case, each BPS factor in the KS product \eqref{mmm1} is equal to the inverse of the corresponding factor in the Schur index, and $\mathcal{M}(q)^N$ (which in the free case coincides with its trace) is by definition equal to the index $\CI_N$ described in section\,\ref{ss:Lagrangianfreehypers}.

In the SQED case,  $\mathcal{M}(q)^N$, as a function, is by construction equal to the integrand of the specialized index $\CI_N$ (cfr.\! section\,\ref{ss:abelianmatter}),
while taking their trace amounts to integrating along the unit circle all the non--flavor fugacities (cfr.\! equation \eqref{mm12a}). Since the integrand depends only on the electric and flavor fugacities, this reproduces the same prescription we gave in section\,\ref{ss:abelianmatter} to compute the specialized
index $\CI_N$. For instance, for SQED with one quark we have
\begin{align}
\begin{split}
 \Tr\, \CM(q)^N &= (q)^{2}_\infty\,\Tr \!\left( \prod_{n=0}^\infty (1-q^{n+1/2}X)^{N} (1-q^{n+1/2} X^{-1})^{N} \right) \\
  &\equiv (q)^2_\infty\oint \frac{dz}{2\pi i z} (1-q^{n+1/2} z)^N (1-q^{n+1/2} z^{-1})^N = \CI_N(q)\, .
\end{split}
\end{align}
Note here that the comparison between the trace computation and index computation involves the integral
representation of the monodromy trace.

These simple examples corroborate the idea of a direct identification in \eqref{?map}.

\subsection{Physical arguments for models having Lagrangian formulation}

The next class of models is the $\mathcal{N}=2$ non--Abelian gauge theories coupled to hypermultiplets, which coincide with the $\mathcal{N}=2$ models with a Lagrangian description.

From the superconformal index side, the situation
looks rather similar to the SQED one.
The specialized indices are expressed as integrals over the electric fugacities dual to the Cartan charges of the non--Abelian electric gauge group
(with the appropriate gauge invariant measure),
while the integrand is a product of
factors associated to the quarks states,
which have the same form as in the Abelian case, and new factors for the charged $W$ bosons which are essentially identical to the KS monodromy factors
for BPS vector multiplets with the $W$ boson quantum numbers,
see equation \eqref{mmm1}.

From the quantum monodromy side the situation looks far more complicated.
In any chamber where the Yang--Mills coupling $g_\text{YM}$ is small, the BPS spectrum of $\mathcal{N}=2$ supersymmetric quantum chromodynamics (SQCD) contains two kinds of states:
\begin{itemize}
\item[a)] the perturbative spectrum, consisting of mutually--local particles with zero magnetic charges, whose masses remain  bounded as $g_\mathrm{YM}\to 0$. There are only finitely many such \emph{light} particles, and they have spin $\leq 1$, so belong to either hypermultiplets (quarks)
or vector multiplets (photons and charged $W$ bosons);
\item[b)] infinite towers of \emph{heavy} dyonic states, with non--zero magnetic charges, whose masses
are of order $1/g^2_\mathrm{YM}$ as $g_\text{YM}\to 0$ and whose spins are not necessarily bounded.
\end{itemize}
Thus, in a weakly coupled chamber, the KS product \eqref{mmm1} contains the factors $$(\pm q^{s+1/2}X_\gamma;q)_\infty^{(-1)^{2s}}$$
of all these infinitely many BPS states. The KS factors associated to the light states a) are essentially equal to the corresponding factors in the specialized index integrand; but
in addition we have infinitely many other factors
associated to the heavy dyons.
Since taking the trace is the same as integrating over the non--flavor fugacities, up to possible subtleties with the integration measure,
the only difference between the definitions of the superconformal
index $\CI_N(q)$ and $\mathrm{Tr}\,\mathcal{M}(q)^N$ is the insertion in the latter of the factors
associated to the heavy dyonic towers.
Were we allowed to forget these dyonic factors, the two sets of SCFT invariants in equation \eqref{?map} would be equal on the nose, as they are for free theories and SQED.

The idea that at weak coupling we may simply neglect the dyonic factors sounds physically plausible. Indeed, the monodromy traces are independent of the Yang--Mills coupling $g_\text{YM}$, and they may be safely computed at parametrically small $g_\text{YM}$.
In this limit the dyon masses go to infinity; physical intuition says that they decouple completely. Hence, at least heuristically,
at extreme weak coupling we may identify $\mathrm{Tr}\,\mathcal{M}(q)^N$ with the corresponding specialized index for all $\mathcal{N}=2$ models having a weakly coupled Lagrangian description.

The $\{\mathrm{Tr}\,\mathcal{M}(q)^N\}$ are chamber independent (as long as they are well defined). Many interesting $\mathcal{N}=2$ non--Abelian gauge theories have strongly--coupled chambers with \emph{finite} BPS spectra consisting of just $h$ hypermultiplets. In this situation $\{\mathrm{Tr}\,\mathcal{M}(q)^N\}$ may be computed directly from the strongly coupled phase. In this way one may check
the above physical picture at extreme weak coupling against
an independent computation of the monodromy traces. Such a check was performed in \cite{Cordova:2015nma} for
SQCD with gauge group $SU(2)$ and $N_f\leq 3$: It was shown that, for these models,
 $\mathrm{Tr}\,\mathcal{M}(q)^{-1}$ as computed from the minimal BPS chamber with $h=N_f+2$ agrees with the extreme weak coupling answer, i.e.\! with its computation at weak coupling neglecting the dyons.
For instance, for pure $SU(2)$ SYM
$\mathrm{Tr}\,\CM(q)^{-1}$, computed as an integral over the electric and magnetic fugacities
with all dyonic factors inserted in the integrand, is (appendix \ref{a:toolbox}),
\be
\mathrm{Tr}\,\CM(q)^{-1}=
\int \frac{d\theta\,d\phi}{(2\pi)^2}\left|\frac{(q)_\infty\, (qe^{i(\theta+\phi)})_\infty}{(q^{1/2}e^{i\theta})_\infty\,(q^{1/2}e^{i\phi})_\infty} \right|^2=\psi(1,q^2),
\ee
which agrees with the Schur index \eqref{shurpuresu2} given by the corresponding integral
without the dyonic contributions \cite{Cordova:2015nma}.
The case $N=-1$ is especially
 simple since there is no subtlety with the integration measure (involving the absorbtion of bosonic/fermionic zero modes). Indeed,
consider $\mathcal{N}=2$ SQCD with gauge group $\CG$ of rank $r_{\CG}$ and let $\Delta^+$ be the set of its positive roots. As a function of the electric fugacities $e^{iH}$, $H\in\text{(Cartan subalgebra)}$, the light BPS factors in the KS product $\mathcal{M}(q)^{-1}$ are
\begin{multline}
(q)^{2r_{\cal G}}_\infty\!
\left(\prod_{\alpha\in\Delta^+}
\big(qe^{i\alpha(H)};q\big)_\infty\,
\big(e^{i\alpha(H)};q\big)_\infty\,
\big(e^{-i\alpha(H)}\big)_\infty
\big(qe^{-i\alpha(H)}
\big)_\infty\right)\!
\Big(\text{quark factors}\Big)\equiv\\
\equiv \overbrace{\left(\prod_{\alpha\in\Delta^+}(1-e^{i\alpha(H)})(1-e^{-i\alpha(H)})\right)}^{\text{Weyl measure}}\;
\mathsf{PE}\!\left[-\frac{2q}{1-q}\,\chi_\mathrm{adj}(H)\right]
\Big(\text{quark factors}\Big),
\label{weak}
\end{multline}
so that the $W$ boson KS factors correctly reproduce the gauge invariant Weyl measure as well as the proper factor from the vectors fields in the Schur index integrand.

Ignoring the dyonic towers --- \textit{if legitimate} --- would be a major simplification for the computation of monodromy traces of $\mathcal{N}=2$ models with a Lagrangian formulation and would make their equivalence obvious. In view of this,
it is interesting to see to what extend this procedure may be justified.  In the next couple of subsections we argue why
the natural regularization of the contribution of the infinite tower of dyonic states should set their contribution to the monodromy operator to 1.
In particular we show that setting their contribution to 1 leads to the action of monodromy operator on the line operator which is consistent with what one
expects that action of the monodromy operator be on the line operators from other viewpoints.

\subsubsection{The physical picture at extreme weak coupling}\label{ss:ewc}

Let us make the physical picture more precise.
For $q\sim 1$ the quantum torus becomes classical and the $\{X_\gamma\}$ may be identified with the Darboux coordinates $\{\mathcal{X}_\gamma\}$ on the hyperK\"ahler target space of the $\sigma$--model dual to the 3d compactification of the 4d $\mathcal{N}=2$ theory \cite{Gaiotto:2008cd}. In other words, the $\{X_\gamma\}$ and the $\{\mathcal{X}_\gamma\}$ have the same transformation under the monodromy as $\tau\to0$. The $\mathcal{X}_\gamma$ has the physical interpretation of expectation value of the supersymmetric
Wilson--'t Hooft line operator of electro--magnetic charge $\gamma$.
In a given quantum phase of the non--Abelian gauge theory, one distinguishes two classes of
such line operators, ``light'' and ``heavy''
\cite{tHooft:1977hy,tHooft:1979uj,tHooft:1981ht}. ``Heavy'' lines may mix with ``light'' ones, but ``light'' ones may only mix between themselves: This is just a rephrasing in the physical language of the mathematics of the Stokes phenomenon for the Darboux coordinates $\mathcal{X}_\gamma$ \cite{Gaiotto:2008cd}. In the Coulomb phase, the ``light'' lines are the electric ones while the dyonic lines are ``heavy''. It follows that under monodromy the electric lines go into themselves while the magnetic ones may mix with the electric ones. The mixing is dictated by
the Witten effect \cite{Witten:1979ey}. In the $\mathcal{N}=2$ case these general non--perturbative arguments may be made more explicit as we are going to discuss.

\paragraph{$SU(2)$ SQCD with $N_f$ quarks.} For definiteness, we consider the case of $\mathcal{N}=2$ SQCD with gauge group $SU(2)$ and $N_f$ flavors of quark
\cite{Seiberg:1994aj}; the general Lagrangian case is similar. Since the $U(1)_R$ symmetry is anomalous, a
$U(1)_R$ rotation by $2\pi$ is equivalent to
 a shift of the Yang--Mills angle $\theta$ by a multiple of $2\pi$; then the effect of the
 monodromy on the electric $e$ and magnetic charge $m$ is
 \begin{equation}
 \big(e,m\big)\longrightarrow \big(e+2(4-N_f),m\big),
 \end{equation}
 that is, in the weakly coupled phase the electric and magnetic lines transform under monodromy as
 \be
 \mathcal{X}_e \to \mathcal{X}_e,\qquad\quad
 \mathcal{X}_m\to \mathcal{X}_{m+2(4-N_f)e}.
 \ee
 Indeed, in general one has \cite{Gaiotto:2008cd}
 \be
 \mathcal{X}_\gamma= \exp\!\Big(R a_\gamma \zeta+ i\theta_\gamma+ R\overline{a}_\gamma/\zeta+\cdots\Big)
 \ee
where $a_\gamma\equiv Z_\gamma$ and $\cdots$ stands for terms exponentially suppressed as $R\to\infty$. Since \cite{Seiberg:1994aj}
\begin{equation}\label{SW}a_m\approx -\frac{4-N_f}{\pi i}\,a_e\,\log(a_e/\Lambda),\quad\text{for  } |a_e|\ggg \Lambda,\end{equation}
while the monodromy acts on the Lagrangian fields as $a_e\to e^{2\pi i}a_e$, we have the expected result \eqref{exrrea}. Thus, as $q\to 1$
the action of the monodromy is simply
\be\label{exrrea}
X_e \to X_e,\qquad X_m\to X_{m+2(4-N_f)e}.
\ee

How is the physical answer
\eqref{exrrea} related to the quantum monodromy $\mathcal{M}(q)$ written as the KS product of BPS factors \eqref{mmm1}?
To compare the two pictures of quantum monodromy, we choose the mass--deformations so that the BPS phases of the quark states are aligned with those of the $W$ boson (since $W$ bosons and quarks are mutually local, this is still a `generic' situation according to our definition).
Identifying BPS particles with representations of the corresponding BPS quiver\footnote{In this paper when we refer
to quiver, we mean the quiver quantum mechanics which captures the BPS spectrum of the 4d theory and not the quiver
theory describing the Lagrangian degrees of freedom of the 4d gauge theory.} $Q$, and using the conventions which are standard in that context
\cite{Alim:2011kw,Cecotti:2012va},
the BPS phase of the light BPS states ($W$--bosons and quarks) is set to $\pi/2$
while the phase of the monopole is $0$.
Thus states whose central charge belong to the
upper (lower) half--plane have positive (negative) electric charge, while states with central charges in the right (left) half--plane have
positive (negative) magnetic charge.
The monodromy $\mathcal{M}(q)$ computed from the chamber at strong coupling with finite number of BPS states is the monodromy which starts at phase zero (the monopole phase) makes a $2\pi$ rotation and comes back to zero.
Instead the monodromy $\mathcal{M}(q)^\text{el}$ which preserves the light electric lines, eqn.\eqref{exrrea}, starts and ends at $\pi/2$. Thus, as operators,
\be\label{conj}
\mathcal{M}(q)^\text{el}=KS(0,\pi/2)\,\mathcal{M}(q)\, KS(0,\pi/2)^{-1},
\ee
where $KS(\alpha,\beta)$ is the KS ordered product \eqref{mmm1} taken over BPS states with $\alpha\leq \arg Z_\gamma <\beta$.
Thus the monodromy operators which are relevant at strong coupling and weak coupling are different operators in the same conjugacy class (recall that monodromy is defined only up to conjugacy). Their intertwinner, $KS(0,2\pi)$, is a quite complicated object: it is an \emph{infinite} product over the BPS dyonic towers with positive magnetic charge and non--negative electric charge. Relating monodromy computations at strong and weak coupling requires to give a definite meaning to the formal infinite non--commutative product $KS(0,2\pi)$.

From \eqref{exrrea}, we see that $\mathcal{M}(q)^\text{el}$ commutes with $X_e$. Under a mild regularity assumption, one concludes that $\mathcal{M}(q)^\text{el}$ does not contain $X_m$, i.e.\! that it is a function only of $X_e$ and the $SO(2N_f)$ flavor fugacities $y_a$
\be
\mathcal{M}(q)^\text{el}= m(X_e, y_a;q).
\ee
The function $m(X_e,y_a;q)$ is determined
from its adjoint action on the magnetic lines, eqn.\eqref{exrrea},
\be\label{exrrea2}
\CM(q)^\text{el}\,X_m\,(\CM(q)^\text{el})^{-1}=X_{m+2(4-N_f)e}.
\ee
If we know the singularity structure of the function $m(X_e,y_a;q)$, we may use this equation to determine it
 up to multiplication by a $c$--number function of $q$ and $y_a$. In facts, eqn.\eqref{exrrea2} translates into a functional equation for $m(X_e, y_a;q)$
related to the Yang--Baxter equation studied in \cite{fad}.
As a preparation, consider the $\theta$--function
\be\label{thetadef}\theta(z;q)=\sum_{n\in\Z}q^{n(n-1)/2}(-z)^n=(q;q)_\infty\, (z;q)_\infty\,(q/z;q)_\infty.\ee
It satisfies the functional equation
\be\label{zzz12}\theta(qz)=-z^{-1}\,\theta(z)\quad\Rightarrow\quad
\theta(q^2z)=(qz^2)^{-1}\theta(z).\ee
Then for operators satisfying $X_eX_m=qX_mX_e$ we have
\cite{fad}
\begin{equation}
\theta(q^aX_{2e};q)^{-1}X_m\,\theta(q^aX_{2e};q)=X_m\,
\theta(q^{a+2}X_{2e};q)^{-1}X_m\,\theta(q^aX_{2e};q)=
q^{2a+1}X_mX_{4e}.
\end{equation}

All singularities of the monodromy function $m(X_e,y_a;q)$ should
have a physical origin.
The only possible mechanism to generate a singularity in the monodromy is that
some generically massive BPS particles become massless,
which implies a discontinuity in the KS product over \emph{massive}
BPS states. Since in the weak coupling chamber the $W$ boson is nowhere massless, for pure $SU(2)$ it is natural to look for a
regular solution to\eqref{exrrea2}.
We compare the adjoint action \eqref{exrrea2} with the following ``Yang--Baxter identity''
\be\begin{split}
&[\theta(q^aX_{2e};q)\,\theta(q^bX_{2e};q)]^{-1}X_m
[\theta(q^aX_{2e};q)\,\theta(q^bX_{2e};q)]=\\
&\hskip1cm =q^{2(a+b)+2}X_mX_{8e}
=q^{2(a+b-1)}X_{m+8e}.
\end{split}
\ee
Then the general \emph{regular} solution to \eqref{exrrea} is
\begin{equation}\label{oureeewc}
\begin{split}
\big(\mathcal{M}(q)^\text{el}\big)^{-1}&= f(q)\,\theta(qX_{2e};q)\,\theta(X_{2e};q)=\\
&=f(q)\,(q;q)^2_\infty
(qX_{2e};q)_\infty\,(X_{2e};q)_\infty\,(qX^{-1}_{2e};q)_\infty\,(X^{-1}_{2e};q)_\infty,
\end{split}
\end{equation}
with $f(q)$ undetermined.
Choosing $f(q)=1$ we get the answer that we expected from the heuristic argument that infinitely heavy particles (the dyons at extreme weak coupling) have no effect on the physics.

\subparagraph{$SU(2)$ SQCD with $N_f$ quarks.} The argument is easily generalized to $N_f\neq0$. However in this case it is possible for the quarks to get massless while keeping $g_\text{YM}$ parametrically small.
Hence we expect the solution $m(X_e,y_a;q)$ to \eqref{exrrea2}
to have singularities as a function of the quark fugacities $y_a$ of the appropriate form.
Then, defining the
theta function
\be\tilde\theta(z;q)=\sum_{n=0}^\infty q^{n^2/2}z^n,\ee we solve the equation \eqref{exrrea2} in the form
\begin{equation}\label{ewcnf}
\big(\mathcal{M}(q)^\text{el}\big)^{-1}=f(q;y_a)\;\frac{\theta(qX_{2e};q)\,\theta(X_{2e};q)}{\prod_{i=1}^{N_f}\big(\tilde\theta(y_aX_{e};q)\,\tilde\theta(y_a^{-1}X_e;q)\big)}.
\end{equation}
 Setting the undetermined function $f(q;y_a)=(q)_\infty^{2N_f}$
we recover the physically expected answer. In particular, the form of the singularity in flavor fugacity is consistent with its physical interpretation in terms of BPS states becoming massless.
Additional (and stronger) arguments for equation \eqref{ewcnf} are presented in the following subsections.
\medskip

From this analysis\footnote{ We warn the reader that the `physical' argument suffers from a minor ambiguity. In facts, we deduced \eqref{exrrea2} from the $\tau\equiv \log q/2\pi i\to 0$ limit, and hence the actual answer may differ by a factor which is trivial in this limit. In practice, viewing the KS product as a partition function for BPS states, the answer may be off by \emph{finitely many} modes. An example is given by eqn.\eqref{weak} where the monodromy differs from the index integrand by the zero--mode factors which produce the correct gauge invariant measure. This will not affect robust quantities like $c_\text{eff}$.} we learn that
the physical picture of the quantum monodromy
at weak coupling where we simply forget the heavy dyons, and the strong coupling picture
where $\mathcal{M}(q)$ is a complicated element of the quantum torus algebra depending non--trivially on both $X_e$ and $X_m$ are not in contradiction since the two monodromies are different elements in the same conjugacy class,
eqn.\eqref{conj}. Their intertwinner, $KS(0,\pi/2)$, is a formal operator whose precise definition is hard to pinpoint. Different prescriptions may lead to different answers, so we must expect a degree of ambiguity in computations which require explicit use of the strong--weak intertwinner
$KS(0,\pi/2)$.  We can interpret the results of
\cite{Cordova:2015nma}, for the Schur case, as a confirmation of the general arguments presented here.

\subsubsection{Decoupling of infinitely massive dyons and S-duality}

The result of the physical argument
in section\,\ref{ss:ewc}, say equation \eqref{ewcnf} for $SU(2)$ SQCD with $N_f$ flavors, implies strong (and somehow paradoxical) properties of the KS products
in $\mathcal{N}=2$ non--Abelian gauge theories. First of all, the mathematical legitimacy of neglecting the heavy dyonic states in the KS product
requires that the KS products over the dyons, as operators, satisfy remarkable identities. We shall check these identities in the next subsection.

In addition, the physical picture leads to apparent paradoxes, since at first sight it seems to clash with $S$--duality.
Consider $SU(2)$ with $N_f=4$.
From the physical picture advocated above, it is natural to conclude that the monodromy has the form
\eqref{ewcnf} with $N_f=4$: This is the result of `forgetting' the heavy dyons. But which `heavy' dyons are we supposed to forget? The notion of a BPS state to be a dyon depends on the $S$--duality frame we use, and its being `heavy' or not depends on the value of the Yang--Mills coupling $g_\text{YM}$. But the KS monodromy does not depend on $g_\text{YM}$.
Our physical picture would describe
the monodromy in one $S$--frame
as the  function \eqref{ewcnf} of $X_e$, $\CM(q)=m(X_e,y_a;q)$, and in a different frame as the same function but of
different arguments, $\CM(q)=m(X_{pe+qm}, y^\prime_a;q)$, (here $(p,q)$ are coprime integers). It seems we got a paradox.

The solution of the paradox is that the monodromy $\mathcal{M}(q)$ is not unique, only its conjugacy class is an absolute invariant. The physical picture is consistent provided the two
candidate monodromies are conjugate by the appropriate KS operator i.e.
\be
m(X_{pe+qm},y_a^\prime,q)= K\!S(\arg Z_{pe+qm},\arg Z_e)^{-1}\; m(X_e,y_a;q)\;K\!S(\arg Z_{pe+qm},\arg Z_e),
\ee
where $K\!S(\theta,\theta^\prime)$
is the KS product over all BPS
states with charges $\gamma$ satisfying
\be
\theta^\prime < \arg Z_\gamma < \theta.
\ee
That $m(X_{pe+qm},y_a^\prime,q)$ and  $m(X_e,y_a;q)$ are conjugate in the quantum torus algebra is obvious (at least when $y_a\equiv 1$);
that the required action of $SL(2,\mathbb{Z})$ is generated by the KS products in the appropriate angular sectors will be shown, for the $SU(2)$ $N_f=4$ example, in sect.\,\ref{ss:nf4su2} starting from the monodromy as computed from the strongly--coupled finite BPS chamber.

\subsubsection{Product identities for dyonic towers}

The `physical' picture implies some remarkable identities for the KS products over dyonic towers.
Here we focus on pure $SU(2)$, but the results apply with minor modifications to all Lagrangian theories.

The notion of the quantum monodromy arises
from considering a closed path in parameter space along which the phase of the central charge increases by $2\pi$
$$Z\to e^{2\pi i\,t}Z\qquad t\in[0,1].$$ At $t\to1$ we go back to the original theory, but the line operators $\mathcal{X}_\gamma$ do not return to themselves. The map $\{\mathcal{X}_\gamma\}_{t=0}\to \{\mathcal{X}_\gamma\}_{t=1}$ defines the adjoint action of the quantum monodromy $\mathcal{M}(q)$. However, the theory has come  back to itself already at $t=1/2$,
since $\gamma\to -\gamma$ corresponds to the action of the $SU(2)$ Weyl symmetry, which is part of the gauge group. Hence the argument of the previous subsection should hold also for the KS product of BPS states whose central charges belong to a half--plane. In our conventions that the BPS phase of the $W$--boson is $\pi/2$,
the previous argument would imply that the product over all BPS states with central charge in the positive half--plane, i.e.\! over all dyons with positive magnetic charge, is just $1$
\be
\prod_{\text{BPS states with}\atop
-\pi/2< \arg Z_\gamma <\pi/2}^\curvearrowright \big(q^{1/2}X_\gamma;q\big)_\infty =1,
\ee
or, explicitly,
\be\label{qqqarw}
\prod^\leftarrow_{j\in\mathbb{Z}}(q^{1/2}X_{m+2je};q)_\infty=1,
\ee
where the notation $\prod\limits^\leftarrow$ means that the factor $(q^{1/2}X_{m+2je};q)_\infty$
is on the left of the factors $(q^{1/2}X_{m+2\ell e};q)_\infty$ with $\ell <j$, and $X_m$, $X_{2e}$ satisfy the commutation relations
$X_{2e}X_m=q^2 X_mX_{2e}$.
Showing that the identity \eqref{qqqarw} holds (in the appropriate sense) is a important check on the full picture.

Formally, we may expand the inverse of the LHS of
\eqref{qqqarw} into terms of definite magnetic charge $M$
\be\label{connmmea}
\prod^\rightarrow_{j\in\mathbb{Z}}(q^{1/2}X_{m+2je};q)_\infty^{-1}=
\sum_{M=0}^\infty X_m^M\, A_M(X_{2e};q).
\ee
To write the $A_M(X_{2e};q)$ in compact
form, we introduce the following notation:
For $M\in\mathbb{N}$, we write $\mathbb{A}(M)$ for the set
of all maps $\boldsymbol{k}\colon \mathbb{Z}\to\mathbb{N}$, $j\mapsto k_j$, such that $\sum_j k_j=M$ (in particular, at most $M$ $k_j$'s are non zero). Then we put
\be
|\boldsymbol{k}|=\sum_{j\in\mathbb{Z}}k_j,\qquad [\boldsymbol{k}]=\sum_{j\in\mathbb{Z}} j k_j,\qquad (q)_{\boldsymbol{k}}=\prod_j(q)_{k_j},
\ee
and consider the quadratic form
\begin{equation}\label{wQ}
Q(\boldsymbol{k})=\sum_{i,j} \min(i,j)k_ik_j
\end{equation}
which is an inverse of the $A^\infty_\infty$ Cartan matrix
\be\sum_j\big(2\delta_{ij}-\delta_{i,j-1}-\delta_{i,j+1}\big)\min(j,k) =\delta_{i,k}.\ee
Then formally
\begin{equation}\label{wTm}A_M(X_{2e};q)=q^{M/2}\sum_{\boldsymbol{k}\in \mathbb{A}(M)}
\frac{q^{Q(\boldsymbol{k})}}{(q)_{\boldsymbol{k}}}\,
X_{2e}^{[\boldsymbol{k}]}.\end{equation}
To give a concrete meaning to the operator in the LHS of \eqref{connmmea}, we have to make
sense of the infinite sums $\{A_M(X_{2e};q)\}_{M\in\mathbb{N}}$. For $M=0$ we have simply
$A_M(X_{2e};q)=1$. For $M>0$ we focus on the set of functional equations satisfied by the would be sums.
The group $\Z$ acts on the set $\mathbb{A}(M)$ by the shift operators
 $[n]\colon\mathbb{A}(M)\to\mathbb{A}(M)$ given by
$\boldsymbol{k}[n]_j=k_{j-n}$.
Under the unit shift $[1]$ we have
\be|\boldsymbol{k}|\to |\boldsymbol{k}|,\qquad
[\boldsymbol{k}]\to[\boldsymbol{k}]+|\boldsymbol{k}|,\qquad
Q(\boldsymbol{k})\to Q(\boldsymbol{k})+ |\boldsymbol{k}|^2,\qquad
(q)_{\boldsymbol{k}}\to(q)_{\boldsymbol{k}}.\ee
Summing over the shifted maps $\boldsymbol{k}[1]\in\mathbb{A}(M)$, we get
\begin{equation}
\label{firstrel}
\begin{split}A_M(X_{2e};q)&=
q^{M/2}\sum_{\boldsymbol{k}[1]\in \mathbb{A}(M)}
\frac{q^{Q(\boldsymbol{k}[1])}}{(q)_{\boldsymbol{k}[1]}}\,
X_{2e}^{[\boldsymbol{k}[1]]}
=\\
&=\big(q^{M^2}X_{2e}^M\big)q^{M/2}
\sum_{\boldsymbol{k}\in \mathbb{A}(M)}
\frac{q^{Q(\boldsymbol{k})}}{(q)_{\boldsymbol{k}}}\,X_\delta^{[\boldsymbol{k}]}\equiv q^{M^2}\,X_{2e}^M\,A_M(X_{2e};q),
\end{split}
\end{equation}
so that $A_M(X_{2e};q)$ satisfies
the functional equation
\be
\Big(1- q^{M^2}\,X_{2e}^M\Big)A_M(X_{2e};q)=0.
\ee
For $M>0$ the only \emph{regular} solution to this equation is $A_M(X_{2e};q)=0$.
The regularity condition we use here reflects the regularity condition we used to pinpoint a unique solution for the Yang--Baxter equation of pure $SU(2)$ in the previous subsection.
Then we conclude that
$A_M(X_{2e};q)=\delta_{M,0}$, that is
\be
\prod^\rightarrow_{j\in\mathbb{Z}}(q^{1/2}X_{m+2je};q)_\infty^{-1}=1,
\ee
in agreement with the decoupling of infinitely massive dyons.

\subsection{$SU(2)$ SQCD with $N_f=4$ and $S$--duality}\label{ss:nf4su2}

To further illustrate and corroborate the above physical picture of the quantum monodromy $\mathcal{M}(q)$, we present some
 non--trivial monodromy computations in
 SCFT examples.\medskip

The simplest non--trivial $\mathcal{N}=2$ SCFT with a Lagrangian description is SQCD with gauge group $SU(2)$ and four fundamental quarks.
We choose its BPS quiver in the form
\cite{Cecotti:2011gu,Alim:2011ae,Cecotti:2013sza}
\begin{equation}\label{Qwz}\begin{gathered}\xymatrix{&& c_1\ar[dr]\ar[drr]\\
a_1\ar[urr] &a_2\ar[ur]&& b_1\ar[dl] &b_2\ar[dll]\\
&&c_2\ar[ull]\ar[ul] }
\end{gathered}
\end{equation}
which has the $\Z_2$ symmetry $\iota$
\begin{equation}
(a_1,a_2,c_1,b_1,b_2,c_2)\overset{\iota}{\longleftrightarrow}
(b_1,b_2,c_2,a_1,a_2,c_1),
\end{equation}
which is a simple instance of Galois symmetry in the sense of \cite{Cecotti:2015qha} which, when present, is the most powerful tool to compute the quantum monodromy. Since $\iota$ is an involutive automorphism of the quiver, for all $\gamma,\gamma^\prime\in\Gamma$,
\be\label{mutlocal}
\langle \iota(\gamma),\gamma^\prime\rangle= \langle \gamma,\iota(\gamma^\prime)\rangle
\quad\Rightarrow\quad
\langle \gamma,\iota(\gamma)\rangle=0
\ee
i.e.\! BPS states in the same orbit of the
Galois symmetry are automatically mutually--local.

$SU(2)$ with $N_f=4$ has a strongly--coupled finite BPS chamber with 12 hypermultiplets \cite{Cecotti:2011gu,Alim:2011ae,Cecotti:2013sza}, which is invariant under the $\Z_2$ symmetry
$\iota$ in the sense that if $\gamma\in\Gamma$ is the charge of a stable hypermultiplet so is $\iota(\gamma)$.
Indeed the set of the charge vectors of the 12  hypermultiplets is given by the union $\Delta^+\amalg \Delta^+$ of the positive roots  $\Delta^+$ of the
two $A_3$ Dynkin full subquivers over the nodes $\{a_1,a_2,c_1\}$ and $\{b_1,b_2,c_2\}$ which are interchanged by $\iota$ \cite{Cecotti:2011gu,Cecotti:2013sza,Cecotti:2015qha}. Inside this strongly--coupled chamber there is a locus where the central charge is
also $\iota$--symmetric, i.e.\!
$Z_{\iota(\gamma)}=Z_\gamma$.
Finally the quadratic refinement is also $\iota$--invariant.
Then the monodromy KS product, as computed at the $\iota$--invariant locus in the strongly--coupled finite chamber, takes the form \cite{Cecotti:2015qha}
\be\label{galoism}
\mathcal{M}(q)= (q)^{-2}_\infty\;\prod^\circlearrowright_{\text{Galois} \atop\text{orbits}} \Big\{\big(\pm q^{1/2}X_\gamma;q)_\infty\,(\pm q^{1/2}X_{\iota(\gamma)};q)_\infty\Big\},
\ee
where the two factors inside each
curly bracket commute between themselves by eqn.\eqref{mutlocal}.

This model has an $SO(8)$ flavor symmetry; correspondingly the exchange matrix $B_{ij}$
of the quiver \eqref{Qwz} has rank $2$, and we may parametrize its quantum torus algebra $\mathbb{T}_Q$ in terms of four commuting flavor fugacities $y_a$ and two operators $X_{e_1}$, $X_{e_2}$ satisfying the $\mathbb{T}_{\vec A_2}$ quantum torus algebra  $X_{e_1}X_{e_2}=q\,X_{e_2}X_{e_1}$ as follows
\be\label{toruspar}
\begin{aligned}
X_{c_1}&=y_1X_{e_1}, &X_{c_2}&=y_1\, X_{e_1}^{-1},\\
X_{b_1}&=y_2y_3 \,X_{e_2},\quad
X_{b_2}=y_2^{-1}y_3 \,X_{e_2}
&X_{a_1}&=y_4y_3\, X_{e_2}^{-1},\quad
X_{a_1}=y_4^{-1}y_3\, X_{e_2}^{-1}.
\end{aligned}
\ee
The monodromy $\mathcal{M}(q)$ may then be seen as an element of the
$\mathbb{T}_{\vec A_2}$ algebra
which depends on the $y_a$ parameters.
Although we could be more general, since the flavor fugacity dependence plays a secondary role in our discussion, for simplicity we set $y_a=1$. At this special point in flavor fugacity space eqn.\eqref{toruspar} implies
\be
X_{\iota(\gamma)}= X_\gamma^{-1},
\ee
and hence each curly bracket in
 eqn.\eqref{galoism} takes the form
 \be
 \Big\{\big(\pm q^{1/2}X_\gamma;q)_\infty\,(\pm q^{1/2}X_{\iota(\gamma)};q)_\infty\Big\}\longrightarrow \frac{\theta(\pm q^{1/2}X_\gamma;q)}{(q)_\infty}
 \ee
 where $\theta(z;q)$ is the function defined in eqn.\eqref{thetadef}.
 Hence we may write,
\be\label{rrrq123}
\mathcal{M}(q)\Big|_{y_a=1}=
 (q)^{-12 c}_\infty\; \Big(L_4^2\,L_3\,L_2\,L_1\Big)^2,
\ee
where $c=14/12$ is the 4d conformal central charge, and the $L_i$, $(i=1,2,3,4)$ are Kontsevich--Soibelman products in suitably angular sectors containing two (or four) mutually--local stable hypermultiplets. Explicitly,
\be
\begin{aligned}
L_1&=\theta(q^{1/2} X_{e_1};q) &L_2&=\theta(-q^{1/2}X_{e_2-e_1};q)^2\\
L_3&=\theta(q^{1/2}X_{2e_2-e_1};q) &L_4&=\theta(q^{1/2}X_{e_2};q).
\end{aligned}
\ee
Using the functional equation
\eqref{zzz12} for $\theta$, we get
\be\label{www12z}
\begin{aligned}
L_1\begin{Bmatrix}X_{e_1}\\ X_{e_2}\end{Bmatrix}L_1^{-1}&=
\begin{Bmatrix}X_{e_1}\\ -X_{e_2-e_1}\end{Bmatrix},
& L_2\begin{Bmatrix}X_{e_1}\\ X_{e_2}\end{Bmatrix}L_2^{-1}&=
\begin{Bmatrix}X_{2e_2-e_1}\\ X_{3e_2-2e_1}\end{Bmatrix}\\
L_3\begin{Bmatrix}X_{e_1}\\ X_{e_2}\end{Bmatrix}L_3^{-1}&=
\begin{Bmatrix}X_{4e_2-e_1}\\ -X_{3e_2-e_1}\end{Bmatrix},
& L_4\begin{Bmatrix}X_{e_1}\\ X_{e_2}\end{Bmatrix}L_4^{-1}&=
\begin{Bmatrix}X_{e_1+e_2}\\ X_{e_2}\end{Bmatrix},
\end{aligned}
\ee
so that the half--monodromy acts as
\be
(L_4^2L_3L_2L_1)\begin{Bmatrix}X_{e_1}\\ X_{e_2}\end{Bmatrix}(L_4^2L_3L_2L_1)^{-1}=
\begin{Bmatrix}X_{-e_1}\\ X_{-e_2}\end{Bmatrix}\ee
and the full--monodromy $\mathcal{M}(q)$ as
\be \label{aaa123}
\mathcal{M}(q)\begin{Bmatrix}X_{e_1}\\ X_{e_2}\end{Bmatrix}\mathcal{M}(q)^{-1}=
\begin{Bmatrix}X_{e_1}\\ X_{e_2}\end{Bmatrix}
\ee
in agreement with eqn.\eqref{exrrea}.

We may fix our reference $S$--duality frame so that the operator $X_{ae_1+be_2}$ has electric/magnetic charges $(a,b)$. Then
 eqn.\eqref{www12z} may be rephrased as the statement that
the KS operators $L_i$ induce the following $SL(2,\mathbb{Z})$
$S$--duality transformations
on the electro--magnetic charges of the line operators
\be\label{wwwex}
\begin{aligned}
L_1&\to\begin{pmatrix}1 &0 \\
-1 & 1\end{pmatrix},
&L_2&\to\begin{pmatrix}-1 &2 \\
-2 & 3\end{pmatrix}\\
L_3&\to\begin{pmatrix}-1 &4 \\
-1 & 3\end{pmatrix},
&L_4&\to\begin{pmatrix}\;1\; &1 \\
\;0\; & 1\end{pmatrix}.
\end{aligned}
\ee
In particular $L_1,L_4$ suffice to generate the full $SL(2,\mathbb{Z})$ action on line operators.
We conclude that the KS product of
BPS factors associated to an angular sector is the same thing as
the telescoping operator\footnote{ These operators correspond to the mathematicians' telescopic (endo)functors in the corresponding derived category \cite{meltzer}. For a review in the present physical context, see \cite{Cecotti:2015hca}. In the categoric language, the BPS states of $SU(2)$ with $N_f$ quarks are the stable objects in the derived category $\mathsf{D}^b(\mathsf{coh}\,\mathbb{P}_{N_f})$ of coherent sheaves on the orbifold of $\mathbb{P}^1$ with $N_f$ double points. The quantum monodromy $\CM(q)$ is the autoequivalence
$(\tau^{-1}[1])^2$ of $\mathsf{D}^b(\mathsf{coh}\,\mathbb{P}_{N_f})$
\cite{Cecotti:2010fi}. The physical equations \eqref{exrrea} are just Serre duality on $\mathbb{P}_{N_f}$.  In particular, the monodromy is an autoequivalence of the Abelian category of objects with $m=0$, $e>0$, (the category of finite--length coherent sheaves) i.e.\! $\CM(q)X_e\CM(q)^{-1}=X_e$. If the orbifold $\mathbb{P}_{N_f}$ has zero Euler characteristic, i.e.\! for $N_f=4$, we have also
$\CM(q)X_m\CM(q)^{-1}=X_m$. } implementing the $S$--duality
by the corresponding angle.

Equation \eqref{aaa123} is the same as \eqref{exrrea2} for $N_f=4$. It has many solutions.
If we assume it to be a function only
of the `electric' fugacity $X_{e_1}$,
as suggested by the `oblique confinement' physical picture we wish to check, we have solutions of the general form
\be\label{wwz1P}
\mathcal{M}(q)= f(q)\,\prod_{a=1}^K \theta(q^{\ell_a}X_{e_1}^{m_a};q)^{n_a}
\ee
with $K$ even and
\be
\sum_a n_a m_a^2=0,\qquad \sum_a m_a n_a\left(\ell_a+ \frac{m_a-1}{2}\right)=0,\qquad \sum_am_an_a=0\mod2.
\ee
Two particular solutions to these
Diophantine equations correspond to the extreme weak coupling description of the monodromy for the two Lagrangian SCFT with one--dimensional Coulomb branch, i.e.\!
$SU(2)$ $\mathcal{N}=2^*$ and
$N_f=4$ respectively
\begin{align}
&\bullet\ \mathcal{N}=2^*\hskip 1cm
(\ell_a,m_a,n_a)=\overbrace{(1,2,1),(0,2,1)}^{W\ \text{boson}},\overbrace{(1/2,2,-1)(1/2,2,-1)}^{\text{adj. quark}}\\
&\bullet\ N_f=4\hskip 1cm
(\ell_a,m_a,n_a)=\overbrace{(1,2,1),(0,2,1)}^{W\ \text{boson}},\overbrace{(1/2,2,-1)(1/2,2,-1)\cdots (1/2,1,-1)}^{\text{(8 times) quarks}}
\end{align}
These are (essentially) the unique solutions if we require the singularities to be of the respective `right' form (i.e.\! singularities allowed only if they may arise from quarks becoming massless).

General KS theory implies, in particular, that the monodromies written with respect to two different $S$--duality frames are equal modulo conjugacy, more precisely that they differ by the adjoint action of an operator which is a KS product of BPS factors. Then consider the monodromy written with respect to the frame in which the weakly coupled $W$ boson has charge $2(pe_1+qe_2)$ with
$p$, $q$ coprime. If the monodromy in the original $S$--frame was given by equation \eqref{wwz1P}, the monodromy in the new frame should be simply
\be\label{wwz1P2}
\mathcal{M}(q)^\text{new}= f(q)\,\prod_{a=1}^K \theta(q^{\ell_a}X_{pe_1+qe_2}^{m_a};q)^{n_a}
\ee
i.e.\! the same function where we replaced $X_{e_1}$ by $X_{pe_1+qe_2}$.  Consistency of the physical picture of monodromy
proposed in sect.\! 6.2 with the Kontsevitch--Soibelman formula
then requires that there exists an operator $W\in\mathbb{T}$ such that
\be\label{newold}
\mathcal{M}(q)^\text{new}= W\,\mathcal{M}(q)^\text{old}\,W^{-1},
\ee
and moreover that $W$ can be chosen to be equal to the product of BPS factors with phases in the range $\arg Z_{e_1} <\theta <\arg Z_{pe_1+qe_2}$. The existence of a $W$ with the prescribed properties for all pair of coprime integers $(p,q)$ is a significant check on the physical scenario.

The existence of $W$ follows from the previous monodromy computation in the strong coupling chamber.
For each $(p,q)$ coprime there is
an $A\in SL(2,\mathbb{Z})$
which transforms $e_1$ into $pe_1+qe_2$. Since $L_1, L_4$ generates $SL(2,\mathbb{Z})$,
there is a word $W$ in $L_1$, $L_2$
equal to $A$. Then
\be
W X_{e_1} W^{-1}= X_{ae_1+be_2},
\ee
and eqn.\eqref{newold} is satisfied.
By construction, $W$ is a product of BPS factors $(\pm q^{1/2}X_\gamma,q)_\infty$. So the physical picture passes this check of consistency with KS theory.

\subsection{A non--Lagrangian example}\label{s:e6exam}

The above results for $SU(2)$
 with $N_f=4$ generalize to all $\mathcal{N}=2$ SCFTs with a finite chamber such that all operators have integral $U(1)_R$ charge \cite{Cecotti:2010fi}, even if they do not have a Lagrangian description.
When the Coulomb branch has dimension 1, one gets a $SL(2,\mathbb{Z})$ action on the line operators $X_\gamma$ implemented by the KS products which is similar to the one we described for $SU(2)$
with $N_f=4$. If the model has no Lagrangian formulation, the physical implications of this $SL(2,\mathbb{Z})$ auto--equivalence is less obvious since we have no weak coupling intuition.
We illustrate these facts in the simplest non--Lagrangian model of this class.

\begin{figure}
\begin{equation*}\begin{gathered}\xymatrix{&&&& c_1\ar[dr]\ar[drr]\ar[drrrr]\\
a_1\ar@<0.35ex>[urrrr] &a_2\ar[urrr] &\cdots &a_r\ar[ur]&& b_1\ar[dl] &b_2\ar[dll]&\cdots &b_s\ar[dllll]\\
&&&&c_2\ar[ullll]\ar[ulll]\ar[ul] }
\end{gathered}
\end{equation*}
\begin{equation*}\begin{gathered}\xymatrix{&&&& \gamma_2\ar@{-}[d]\\
\alpha_{r-1}\ar@{-}[r]& \alpha_{r-2}
\ar@{-}[r] &\cdots\ar@{-}[r] & \alpha_1 \ar@{-}[r] &\gamma_1 \ar@{-}[r] & \beta_1
\ar@{-}[r] &\beta_2 \ar@{-}[r] &\cdots
\ar@{-}[r]& \ar@{-}[r] &\beta_{s-1}}
\end{gathered}
\end{equation*}
\caption{The $Q(r,s)\equiv Q(s,r)$ quiver $(r+s)>0$. The corresponding $\mathcal{N}=2$ theory has one--dimensional Coulomb branch and  rank $(r+s)$ flavor symmetry group $G_F$ with the Dynkin graph in the lower graph.}\label{qrsquiver}
\end{figure}
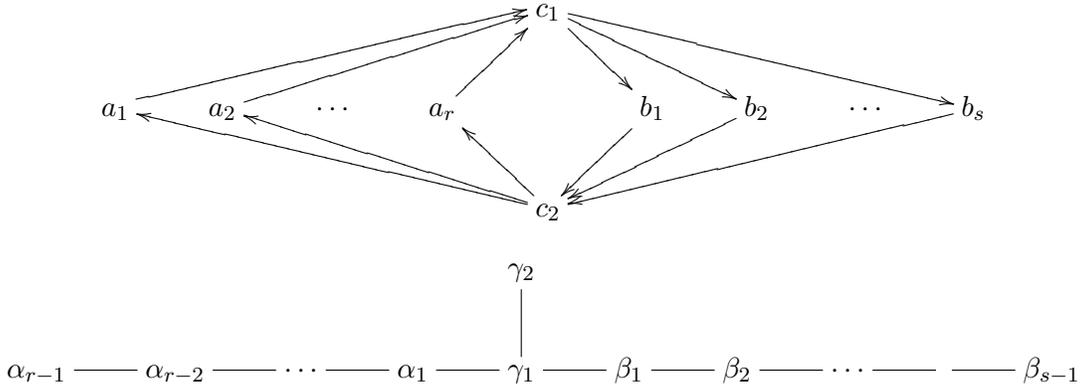

$SU(2)$ with $N_f=4$ is a special instance of a sequence of $\mathcal{N}=2$ models with one--dimensional Coulomb branch and large flavor symmetry. They are described by the class of the $Q(r,s)$ quivers in figure
\ref{qrsquiver} where $r$, $s$ are non--negative integers not both zero.
These models have a non--Abelian flavor symmetry of rank $r+s$ described by the Dynkin graph in the lower part of the figure. Since the flavor symmetry group of a meaningful QFT must be compact, only pairs $(r,s)$ such that the lower graph is a Dynkin graph of finite--type correspond to a $\mathcal{N}=2$ model.
In particular, $Q(2,N_f)$ is the quiver of $SU(2)$ with $N_f$ quarks, and
$Q(3,3)$, $Q(3,4)$ and $Q(3,5)$ are the quiver of the Minahan--Nemeschansky theories with flavor symmetry $E_{r+s}$.

The SCFTs described by a $Q(r,s)$ quiver --- which are known as $H_1,H_2,D_4,E_6,E_7,E_8$ ---
as well as many others like the Argyres--Douglas theories of type $D$, have in common some remarkable properties that we shall use in section \ref{s:monodromiesAD}
as a guiding principle to compute in an efficient way their monodromy traces. From the viewpoint of the 2d/4d correspondence of \cite{Beem:2013sza} the property may be stated as follows: these SCFT have a
flavor group $G_F$ with a 4d level $k$ and a conformal central charge $c$ to which there correspond the 2d CFT quantities $c_{2d}=-12c$ and $k_{2d}=-k/2$. These models have the special property that
the 2d Sugawara energy--momentum tensor of the flavor
symmetry $G_F$ saturates the Virasoro central charge, i.e.\! we have the equality
\be\label{sugsaturation}
c_{2d}=\frac{k_{2d}\cdot \dim G_F}{h_{G_F}+k_{2d}}.
\ee
The same SCFTs were shown in \cite{Cecotti:2013sza} to have
a finite BPS chamber which is $c$--saturating, i.e.\! such that the
conformal central charge $c$ computed from the BPS spectrum pretending it consists of \emph{free} particles is, for that particular chamber, equal to its actual value.
The $c$--saturating chambers lead to very convenient expressions for the monodromy which greatly simplify the task of computing their traces.  A first example of $c$--saturating is the 12 hyper chamber
of $SU(2)$ with $N_f=4$:
in eqn.\eqref{rrrq123}
the monodromy, as written in the $c$--saturating chamber, is equal to
$(q)^{-12c}_\infty$ times a product of theta functions. This formula is true
(for an appropriate product of theta functions) for all rank 1
SCFT models in the above list (again switching off flavor fugacities).
In particular, experience suggests that in all SCFTs for which  \eqref{sugsaturation} holds,
for all $N>0$ such that $\mathrm{Tr}\,\CM(q)^N$ makes sense, we have
\be
\mathrm{Tr}\,\CM(q)^N = \frac{\boldsymbol{\Theta}_N(y_a;q)}{(q)^{12Nc}_\infty}
\ee
where $c$ is the 4d SCFT central charge and $\boldsymbol{\Theta}_N(y_a;q)$ is the theta--function of some lattice. In section
\ref{s:monodromiesAD} we shall find many instance of this phenomenon so helpful in computing the traces.

For brevity we discuss only the first Minahan--Nemeschansky model
with $G_F=E_6$ .
Its quiver $Q(3,3)$ also has a $\mathbb{Z}_2$ Galois automorphism $\iota$
\be
a_i\overset{\iota}{\longleftrightarrow}b_i\qquad
c_1\overset{\iota}{\longleftrightarrow}c_2,
\ee
which is a symmetry of the BPS spectrum in the finite $c$--saturating 24 hypers chamber. Since the Coulomb branch is one--dimensional, we may parametrize the quantum torus algebra $\mathbb{T}_{Q(3,3)}$ in terms of the algebra $\mathbb{T}_{\vec A_2}$, $X_{e_1}X_{e_2}=qX_{e_2}X_{e_1}$ and six flavor fugacities $y_a$. Setting the $E_6$ fugacities to $1$, we again get
$X_{\iota(\gamma)}=X_\gamma^{-1}$ and, by the same argument we used for $SU(2)$ with $N_f=4$, the monodromy reads
\be
\mathcal{M}(q)=(q)_\infty^{-12c}\,
\Big(L_6L_5L_4L_3L_2L_1\Big)^2
\ee
where $c=13/6$ is the model's conformal central charge and
\be
\begin{aligned}
L_1&=\theta(q^{1/2}X_{e_1};q) &L_2&=\theta(-q^{1/2}X_{e_2-e_1};q)^3
& L_3&=\theta(q^{1/2}X_{3e_2-2e_1};q)\\
L_4&=\theta(q^{1/2}X_{2e_2-e_1};q)^3
&L_5&=\theta(-q^{1/2}X_{3e_2-e_1};q)
&L_6&=\theta(q^{1/2}X_{e_2};q)^3.
\end{aligned}
\ee
Using
\eqref{zzz12}, we get
\be\label{www12zII}
\begin{aligned}
L_1\begin{Bmatrix}X_{e_1}\\ X_{e_2}\end{Bmatrix}L_1^{-1}&=
\begin{Bmatrix}X_{e_1}\\ -X_{e_2-e_1}\end{Bmatrix},
& L_2\begin{Bmatrix}X_{e_1}\\ X_{e_2}\end{Bmatrix}L_2^{-1}&=
\begin{Bmatrix}-X_{3e_2-2e_1}\\ -X_{4e_2-3e_1}\end{Bmatrix}\\
L_3\begin{Bmatrix}X_{e_1}\\ X_{e_2}\end{Bmatrix}L_3^{-1}&=
\begin{Bmatrix}-X_{9e_2-5e_1}\\ -X_{7e_2-4e_1}\end{Bmatrix},
& L_4\begin{Bmatrix}X_{e_1}\\ X_{e_2}\end{Bmatrix}L_4^{-1}&=
\begin{Bmatrix}X_{12e_2-5e_1}\\ -X_{7e_2-3e_1}\end{Bmatrix},\\
L_5\begin{Bmatrix}X_{e_1}\\ X_{e_2}\end{Bmatrix}L_5^{-1}&=
\begin{Bmatrix}-X_{9e_2-2e_1}\\ -X_{4e_2-e_1}\end{Bmatrix},
& L_6\begin{Bmatrix}X_{e_1}\\ X_{e_2}\end{Bmatrix}L_6^{-1}&=
\begin{Bmatrix}-X_{e_1+3e_2}\\ X_{e_2}\end{Bmatrix}
\end{aligned}
\ee
so that the adjoint action of each $L_i$, and hence of their products, acts (a part for the signs from quadratic refinement) is just a $SL(2,\mathbb{Z})$ transformation
\be
L_i X_\gamma L_i^{-1}= X_{A_i(\gamma)}\qquad A_i\in SL(2,\mathbb{Z}).
\ee
However now the half--monodromy acts as the identity
\be
(L_6L_5\cdots L_1)\begin{Bmatrix}X_{e_1}\\ X_{e_2}\end{Bmatrix}(L_6L_5\cdots L_1)^{-1}=\begin{Bmatrix}X_{e_1}\\ X_{e_2}\end{Bmatrix}.
\ee
The situation for $E_7$ and $E_8$ Minahan--Nemeschansky is similar,
but the monodromy has a less compact expression since these models have no Galois symmetry.
The analysis of their monodromy in the classical limit $q\to1$ (but general flavor fugacities)
may be found in \cite{Cecotti:2014zga}.

\subsection{Generalities on monodromy traces}\label{S:tracegeneralities}

Having discussed the Lagragian cases and some extensions of them, in this section we turn to some general properties of the monodromy traces which follow
from their formal  structure. Appendix \ref{a:toolbox} contains a survey of the main techniques to compute monodromy traces. The interested reader can find more details and explicit expressions
for all quantities related to the traces that we will need.
\medskip

The methods of \cite{Cecotti:2010fi} work for models having a \emph{finite} BPS chamber i.e.\! a chamber whose BPS spectrum consists of finitely many hypermultiplets; then the KS product $M(q)^N$ contains
finitely many factors $(q^{1/2}X_\gamma;q)_\infty$.
One expands each  such factor in the basis $\{X_\gamma\}_{\gamma\in\Gamma}$ of the quantum torus algebra $\mathbb{T}$ and then
multiplies them with the rule \eqref{multx} to get a (formal) expression of the form
\be\label{mN1a}
\mathcal{M}(q)^N=\sum_{\gamma\in \Gamma} \mu(N;q)_\gamma\,X_\gamma,
\ee
for certain coefficient functions $\{\mu(N;q)_\gamma\}_{\gamma\in\Gamma}$ given by multiple $q$--hypergeometric sums. The trace on the quantum torus algebra $\mathbb{T}$ is defined by the rule\footnote{ This is slightly different from the notion of `canonical trace' used in \cite{Cecotti:2010fi} where the trace
included a projection also to flavor neutral sector.  The relation between the two notions of trace is as follows:
$$\mathrm{Tr}\,\mathcal{M}(q)^N\bigg|_{\text{CNV}\hfill\atop
\text{sense \cite{Cecotti:2010fi}}}=\oint \mathrm{Tr}\,\mathcal{M}(q)^N\bigg|_{\text{present}\atop
\text{sense}\hfill}\,\prod_a\frac{dy_a}{2\pi i\, y_a}.$$}
\be\label{trtr1a}
\mathrm{Tr}\,X_\gamma=\begin{cases}
y_\gamma & \gamma\ \text{is a flavor charge}\\
0 & \text{otherwise}.
\end{cases}
\ee
A charge $\gamma$ is a \emph{flavor} charge iff it belongs to the radical of the Dirac form, i.e.\!
$\langle\gamma, \nu\rangle=0$ for all $\nu\in\Gamma$. The flavor charges form a sublattice $\Gamma_f\subset\Gamma$ of the charge lattice.
 From \eqref{mN1a}\eqref{trtr1a} the monodromy traces, as functions of $q$ and the flavor fugacities $y_a$, are then
\be
\mathrm{Tr}\,\mathcal{M}(q)^N=\sum_{\phi\in\Gamma_f} \mu(N;q)_\phi\;y_\phi.
\ee
The computation of $\mathrm{Tr}\,\CM(q)^N$ is thus reduced to the evaluation of $q$--hypergeometric sums $\{\mu(N;q)_\phi\}_{\phi\in\Gamma_f}$ which have the general form \cite{Cecotti:2010fi}
\be\label{qhypera}
\sum_{n_1,\dots, n_s\geq0} \frac{q^{Q(n_i)/2}\, \prod_i z_i^{n_i}}{\prod_i(q)_{n_i}}\prod_a \delta_{b_{ai}n_i,t_a},
\ee
where $Q(n_i)\equiv n_iA_{ij}n_j$ is an integral quadratic form
and the Kronecker deltas enforce the proper restriction on the summation range;
see appendix \ref{a:toolbox} for full details, including explicit expressions for all elements appearing in \eqref{qhypera}. $q$--sums of this form
are familiar from the thermodynamical Bethe ansatz (TBA) and related $Y$--systems, see e.g.\! \cite{Nahm:2004ch}. Ideas and techniques developed in those contexts may be applied also to monodromy traces. Depending on $N$ and the particular model at hand,
the sum \eqref{qhypera} may or may not be absolutely convergent; if not an appropriate prescription is required to define
$\mathrm{Tr}\,\CM(q)^N$.

\paragraph{Moyal functions.}
Let $\{e_s\}$ be a set of generators of the charge lattice $\Gamma$,
and set $B_{s,t}=\langle e_s,e_t\rangle$. The quantum torus algebra $\mathbb{T}$, which has
generators $\{X_{e_s}\}$ and commutation relations
$X_{e_s}X_{e_t}=q^{B_{st}}X_{e_t}X_{e_s}$, is isomorphic to the algebra of holomorphic functions on the classical torus $T=(\mathbb{C}^*)^{\text{rank}\,\Gamma}$, with coordinates $u_s\in\mathbb{C}^*$, endowed with the Moyal product $\ast$ defined by the $2$--vector $B_{ts}$, i.e.\!
\be
f\ast g(u_s)= \exp\!\big(\pi i \tau B_{st}\, x_sy_t\,\partial_{x_s}\partial_{y_t}\big)f(u_s+x_s)g(u_s+y_s)\Big|_{x=y=0}\quad\text{where }q=e^{2\pi i\tau}.
\ee
 To the charge $\gamma=\sum_sn_se_s\in\Gamma$ we associate the function $u_\gamma=\prod_s u_s^{n_s}$ on $T$. If $\gamma$ is a flavor charge, the corresponding function $u_\gamma$ is called
a flavor fugacity.
In the basis $\{X_\gamma\}$ of $\mathbb{T}$ the
isomorphism between $\mathbb{T}$ and the Moyal functions is simply
\begin{equation}X_\gamma \leftrightarrow u_\gamma.\end{equation}
Comparing with \eqref{mN1a},
we see that the Moyal function
$\mu(u_s;q)^{(N)}$ on $T$ which corresponds to the
operator $\CM(q)\in\mathbb{T}$
is
\be
\mu(u_s;q)^{(N)}=\sum_\gamma \mu(N;q)_\gamma\, u_\gamma.\ee
Then
\be\label{mm12a}
\mathrm{Tr}\,\CM(q)^N= \oint \frac{dv_s}{2\pi i v_s}\, \mu(u_s;q)^{(N)}\,,
\ee
where the integral is on the unit
circle $|u_s|=1$ at fixed values of the flavor fugacities $y_a$.
The Moyal formalism has two  advantages
with respect to the quantum torus algebra: First the quantum cluster mutations \cite{Cecotti:2010fi} may be rephrased as functional equations for the Moyal functions which are often easier to solve.
Second, it is usually simpler to find a prescription to make convergent an integral like \eqref{mm12a} than a $q$--series.

In appendix \ref{a:toolbox} it is shown that
the integral \eqref{mm12a} may be recast in the form ($q=e^{2\pi i\tau}$)
\begin{equation}\label{Sinte}
\int_\CC \exp\!\left(-\frac{S(t_i,y_a;\tau)^{(N)}}{2\pi i \tau}\right)\prod_i dt_i,
\end{equation}
where $S(t_i,y_a;\tau)^{(N)}$ is a function which is regular as $\tau\to 0$ (see appendix \ref{a:toolbox} for its explicit form). The integral \eqref{mm12a} belongs to the well known class of ``oscillatory integrals'', which may be interpreted
as brane amplitudes $\int_\CC dX_i\, \exp(-W(X_i)/\zeta)$ for 2d (2,2) SCFTs \cite{Hori:2000ck}.
 For such an integral, giving the appropriate convergence prescription
amounts to specifying the correct integration contour $\CC$.
We stress that, while the  contour $\CC$ (and hence the proper definition of $\mathrm{Tr}\,\CM(q)^N$)
may be rather subtle,
the function $S(t_i,y_a;\tau)^{(N)}$ contains information on the quantum traces $\mathrm{Tr}\,\CM(q)^N$ which is both very valuable and totally non--ambiguous.

\paragraph{The effective central charge $c_\text{eff}$.}
In particular the function $S(t_i,y_a;\tau)^{(N)}$
contains the  ``2d CFT data'',
$c_\text{eff}$ and $\{h_i\}$,
associated with the monodromy trace $\mathrm{Tr}\,\CM(q)^N$.
These data are defined as follows: View $\mathrm{Tr}\,\CM(q)^N$ as a power series in $q$
\be\label{qqser}
\mathrm{Tr}\,\CM(q)^N=\sum_{n\geq0} a_n(y)\,q^n\ee
whose coefficients $a_n(y)$ are Laurent polynomials in the flavor fugacities $y_a$ with integral coefficients. The \emph{effective central charge}
$c_\text{eff}$
of the power series $\sum_{n\geq0} a_n\,q^n$
is defined as
\be\label{ceffy}
c_\text{eff}=\frac{3}{2\pi^2} \lim_{n\to\infty}\frac{(\log a_n)^2}{n},
\ee
or equivalently (setting $q=e^{2\pi i \tau}$) by its $\tau\to0$ asymptotic behavior
\be
\sum_{n\geq0} a_n\, q^n \simeq
 \exp\!\left(\frac{2\pi i}{\tau}\,\frac{c_\text{eff}}{24}+O(1)\right)\quad\text{as }\tau\to 0.
\ee
The name `effective central charge' stems from the fact that, when the $q$--series \eqref{qqser} is a conformal block
of a (not necessarily unitary) 2d CFT, one has
\begin{equation}
c_\text{eff}\equiv c-24\min_i(h_i)
\end{equation}
 with $h_i$, $c$ the conformal weights and Virasoro central charge of the CFT. In particular, for unitary CFTs, $c_\text{eff}\equiv c$.
If there are flavor symmetries, from eqn.\eqref{ceffy} we get a function of the flavor fugacities, $c_\text{eff}(y_a)$; in this case, we define the effective
central charge $c_\text{eff}$ as the value of this function at a (suitable) critical point, $\partial_{y_a} c_\text{eff}(y)=0$.

To compute $c_\text{eff}$, it suffices to evaluate the integral \eqref{Sinte} in the limit $\tau\to0$. This can be done by standard saddle point techniques: $c_\text{eff}$ for $\mathrm{Tr}\,\CM(q)^N$ is essentially the value of  the function $S(t_i,y_a;\tau)^{(N)}$ at its dominating critical point.
The saddle point equations may be solved explicitly by adapting  standard TBA methods \cite{Nahm:2004ch} to our situation. The details may be found in appendix \ref{a:toolbox}. Here we quote the final result:
$c_\text{eff}$ for $\mathrm{Tr}\,\CM(q)^N$ (as computed from a  chamber with $h$ hypers)
is given by
\be\label{bottom1}
c_\text{eff}= 2rN+2|N|\sum_{i=1}^h \frac{6}{\pi^2}\,L(z_i),
\ee
 where $L(z)$ is the Roger dilogarithm
 \be
 L(z)= \mathrm{Li}_2(z)+\frac{1}{2}\log z\,\log(1-z),
 \ee
 and the $z_i$ are the solutions to the
Nahm--like equations
 \begin{equation}\label{nahma}
 z_i^2=\left\{
 \begin{aligned}
 &(1-z_j)^{C_{ij}}, &&N>0\\
 &(1-z_j)^{2\,\delta_{ij}-C_{ij}} &&N<0
 \end{aligned}\right.
 \end{equation}
where $C_{ij}$ is the integral symmetric $h\times h$ matrix specified in appendix \ref{a:toolbox}.

Below we will need to recall that $c^{4d},a^{4d}$ for AD $(A,A')$ theory:
$$c^{4d}={\dim A\dim A'-r_Ar_{A'}\over 12(h_A+h_{A'})}$$
$$a^{4d}={4\dim A\dim A'-5r_Ar_{A'}\over 48(h_A+h_{A'})}$$

\paragraph{Example 1:
$(G,A_1)$ models with $N>0$.}
In this case the equations \eqref{nahma} take the form
\be\label{zzz15a}
z_i^2=(1-z_j)^{C_{ij}}
\ee
where $C_{ij}$ the Cartan matrix of the simply--laced Lie algebra $G$. Writing $z_i=w_j^{-C_{ij}}$, the equations take the form
\be\label{nahm12a}
1+w_j^{2\delta_{ij}-C_{ij}}=w_i^2.
\ee
The equations \eqref{zzz15a} have a unique solution with $0<z_i<1$ which corresponds to the vacuum character of the
2d coset  CFT $G_{(2)}/U(1)^{r_G}$ \cite{Nahm:2004ch,Keegan:2011ci}.
Then $6\sum_i L(z_i)/\pi^2$ is just the central charge of the  $G_{(2)}/U(1)^{r_G}$ coset CFT, and
\be\label{ceffNposa}
c_\text{eff}=2N\left(r+\frac{r_G\, h_G}{h_G+2}\right),
\ee
where $r$ is the rank of the Coulomb branch of the AD theory, and  $r_G$, $h_G$ are the rank and the Coxeter number of the Lie algebra $G$
(related to its dimension by the Coxeter formula $\dim G=r_G(h_G+1)$). $2r$ is equal to $r_G$ minus the multiplicity of $h_G/2$ as an exponent of $G$.
For $(A_1,G)$ Argyres-Douglas theories the $12 N c^{4d}$ is expected to be given by the above formula for $c_\text{eff}$ and this is in agreement with the expected
answer $(A_1,A_{r_G})$.

\paragraph{Example 2: $(G,A_1)$ models
with $N<0$.} In this case the equations \eqref{nahma} take the form
\begin{equation}
z^2_i=(1-z_j)^{2\delta_{ij}-C_{ij}}.
\end{equation}
Setting $z_i=-w_j^{2\delta_{ij}-C_{ij}}$ we get back the Nahm equations \eqref{nahm12a}.
For $N<0$ we are interested in a \emph{different} solution of these equation. We may find it by the same Lie theoretic methods as for $N>0$ \cite{Nahm:2004ch}.
For instance,  for the $(A_1,A_{2\ell})$ model with $N<0$
one finds (appendix \ref{a:toolbox})
\be\label{effccaa}
c_\text{eff}=\frac{2|N|\ell}{2\ell+3}.
\ee
Note that this agree with the expected answer $-48N(c^{4d}-a^{4d})$ for the AD theory when $G=A_{2l}$.

The effective central charge of the $(p,q)$ Virasoro
minimal model is
\be\label{effeN-1a}
c_\text{eff}(p,q)=c-24\min_{r,s} h_{r,s}=1-\frac{6}{pq}.
\ee
Taking $p=2$ and $q=2\ell+3$, we get the effective central charge \eqref{effccaa} for $N=-1$.
Note that this agrees with the general prediction that $h_{min}=(-1/2)(5c^{4d}-4a^{4d})$.

\paragraph{Example 3:
$(G,G^\prime)$ models with $N>0$.}
Again we reduce the saddle point conditions to systems of algebraic equations already studied in a related context by
Nahm \cite{Nahm:2004ch}. Using his results, one obtains the following formula
\be\label{eccceff}
c^{2d}=c_\text{eff}= 2Nr+N\,\frac{r_Gr_{G^\prime}h_Gh_{G^\prime}}{h_G+h_{G^\prime}}=12Nc^{4d},
\ee
which has been checked explicitly for $(G,A_2)$.  Using this and the $r$-charges for the Coulomb branch
of the $(G,G')$ models which are known, one can also deduce $a^{4d}$ for all $(G,G')$ AD theories from the relation \cite{Shapere:2008zf}
\be
 4(2a^{4d} - c^{2d}) = \sum_{i} (2r_i - 1) \ , 
\ee
where the sum is over all the Coulomb branch operators with dimension given by $r_i$.

%%%%%%%%%%%%%%%%%%%%%%%%%%%%%%%%%

%%%%%%%%%%%%%%%%%%%%%%%%%%%%%%%%%

\section{Monodromies and $2d$ Chiral Algebras of Argyres-Douglas Theories}\label{s:monodromiesAD}
\subsection{Insertion of line operators and characters of chiral algebras}

The Argyres-Douglas (AD) theories are isolated fixed points of 4d $\CN=2$ supersymmetric field theories with finite number of BPS states in all chambers \cite{Cecotti:2011rv}. Therefore to write down $q$--series expressing $\tr \,\CM(q)^N$ is straightforward. (But, for large $|N|$, the resulting $q$--series are not absolutely convergent, and suitable prescriptions are required to make sense out of them). On the other hand, the computation of the supersymmetric indices $\CI_N(q)$ of AD theories is difficult even in principle, because most of them do not have a Lagrangian description. In view of the proposed correspondence
\eqref{?map}, the ``easy'' computation of the monodromy traces
may be used as a prediction of superconformal indices in the limit $t=qp^{N+1}$ and $p\rightarrow {\rm exp}(2\pi i)$.

More generally,
we may consider monodromy traces
with line operator insertions
\be
\langle X_\gamma\rangle_N=\mathrm{Tr}\big[\CM(q)^N\,X_\gamma\big]\qquad \gamma\in\Gamma.
\ee
We stress that the $\langle X_\gamma\rangle_N$ are not absolute invariants, since $\CM(q)$ is unique only up to conjugacy. $\CM(q)$ depends on the choice of a reference ray in the $Z$--plane, which conventionally we take to be the positive real axis. $\langle X_\gamma\rangle_N$ is invariant only under the deformations of the parameters such that no central charge $Z_\gamma$ of a BPS particle crosses the real axis. In short, in the standard quiver formulation
\cite{Alim:2011kw}  the $\langle X_\gamma\rangle_N$ depend on the choice of a quiver in the mutation class, but not on the point in parameter space, as long as it is covered by the given quiver.

Fixing the quiver $Q$, the $\langle X_\gamma\rangle_N$ may be read directly from the
expansion of the operator $\CM(q)^N$ in the standard basis of the quantum torus algebra $\mathbb{T}$ (see eqn.\eqref{mN1a})
\be\label{fourcoeff}
\mathcal{M}(q)^N=\sum_{\gamma\in\Gamma}\langle X_\gamma\rangle_N\; X_{-\gamma}.
\ee
The $\langle X_\gamma\rangle_N$ may be used to compute traces of powers of $\CM(q)^N$. For instance
\be\label{2ntrace}
\mathrm{Tr}\,\CM(q)^{2N}=\sum_{\gamma\in\Gamma}\langle X_\gamma\rangle_N\,\langle X_{-\gamma}\rangle_N.
\ee
In this section we choose the quiver
to have the Dynkin form with the linear orientation.

\subsection{$(A_1,A_2)$ AD theory}

\begin{figure}
  \centering
  \begin{tikzpicture}
			\node[W,red] (1) at (0,0){1};
		 	\node[W,blue] (2) at (2,0) {2};
			\draw[->] (1)--(2);
  \end{tikzpicture}
  \caption{\label{fig:A1A2quiver}
  BPS quiver}
\end{figure}
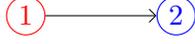

Let us start the discussion with the AD theory of the $(A_1, A_2)$ singularity. The BPS quiver of this theory is shown in figure \ref{fig:A1A2quiver}.

\paragraph{Trace of $\mathcal{M}(q)^{-1}$.}

The trace of the inverse monodromy for the $(A_1, A_2)$ theory was first computed in \cite{Cecotti:2010fi}. The result is (see appendix \ref{mminusonea2})
\begin{equation}
\Tr\, \mathcal{M}(q)^{-1}=H(q),
\end{equation}
where $H(q)$ is the Rogers-Ramanujan function,
\begin{equation}\label{rogra1}
H(q)=\sum_{l=0}^{\infty}\frac{q^{l^2+l}}{(q)_l}=\prod_{n=1}^{\infty}\frac{1}{(1-q^{5n-2})(1-q^{5n-3})}.
\end{equation}
It is known that $q^{11/60}\,H(q)$ is the vacuum character $\chi_{1,1}$ of the $(2,5)$ Virasoro minimal model. The central charge of the $(2,5)$ minimal model is $-22/5$, with matches our prediction in table \ref{table:ADcentrals} of appendix B (see also equation \eqref{effcc}). The central charge of the 2d chiral algebra is $-12$ times the central charge of the 4d theory as noted in \cite{Beem:2013sza,Beem:2014zpa, Cordova:2015nma}.

\paragraph{Trace of $\CM(q)$.}

The monodromy trace $\Tr\, \CM(q)$ is equal to \cite{Cecotti:2010fi}
\begin{equation}
\Tr\, \mathcal{M}(q)=\frac{G(q)}{(q)_{\infty}^4},
\end{equation}
where $G(q)$ is the other Rogers-Ramanujan function,
\begin{equation}\label{rogra2}
G(q)=\prod_{k=1}^{\infty}\frac{1}{(1-q^{5n-1})(1-q^{5n-4})}.
\end{equation}
The effective central charge
of $\Tr\, \CM(q)$
is $22/5$ as shown in table \ref{table:ADcentrals} (cfr.\! euqation \eqref{ceffNpos}).
Indeed, $q^{-1/60}\,(q)^4_\infty\,\Tr \,\CM(q)$ is the character $\chi_{1,3}$  of the $(2,5)$ minimal model which has an effective central charge
\be
c_\text{eff}=1-\frac{6}{2\cdot 5}\equiv \frac{22}{5}-4.\ee

We have provided evidence in the Lagrangian cases that at least in the extreme weak limit
we have a chiral algebra, whose character is computed by the trace of the monodromy.  We now
can test this for the $\Tr\, \CM(q)$.  Which chiral algebra has the character $G(q)/\eta(q)^4$?
Note that $G(q)$ is a character of a representation of the $(2,5)$ model but it is not the character of the
corresponding algebra (i.e. of the vacuum module).  So our general conjecture {\it anticipates} that there
should be nevertheless another chiral algebra whose vacuum character gives $G(q)/\eta(q)^4$.  The
$1/\eta^4$ is simply the chiral algebra associate to 4 free boson (with zero modes deleted).  So the question
is whether we can view $G(q)$ as the character of a chiral algebra.

\paragraph{The chiral algebra $\CA_1$ associated to $\mathrm{Tr}\,\CM(q)$.} As shown in \cite{Feigin:1993qr}, $G(q)$ is the vacuum character of the subalgebra of $SU(2)$
at level $k=1$ generated by $J^+(z)=J^1(z)+i J^2(z)$.  For level $1$ this algebra is simple and has only one relation given by
\begin{equation}
\label{eq:relA1A2}J^+(z)J^+(z)=0.
\end{equation}
Let us see why the characters of this algebra is given by $G(q)$.
To see this we rewrite $G(q)$ as
\begin{equation}
G(q)=\sum_{n=0}^{\infty}\frac{q^{n^2}}{\prod_{i=1}^n (1-q^i)}=1+\frac{q}{1-q}+\frac{q^4}{(1-q)(1-q^2)}+\cdots.
\end{equation}
Clearly $J^+(z)$ and its derivatives gives the module $q/(1-q)$. There is no $q^2$ and $q^3$ term because of the relation $\partial(J^+(z)J^+(z))=0$ and $J^+(z)\partial J^+(z)=\partial(J^+(z)J^+(z))=0$. The module $q^4/((1-q)(1-q^2))$ comes from $\partial J^+(z)\partial J^+(z)$ and its derivatives, the denominator is $(1-q)(1-q^2)$ instead of $(1-q)^2$ is due to the symmetry between two $\partial J^+(z)$'s. The algebra behind positive monodromy $\Tr\mathcal{M}(q)$ is just the aforementioned $J^+$ subalgebra of $SU(2)_1$ together with four free scalars. One interesting fact is that $\Tr\mathcal{M}^{-1}(q)=H(q)$ which is the character
of the module for $N=-1$ is the character of the other module of this $J^+$ subalgebra of $SU(2)_1$, which has as the vacuum
the fundamental representation.  Moreover one can get to $H(q)$ for $\Tr\, \CM(q)$ by inserting  a line operator (as was shown in  \cite{Cecotti:2010fi} and we will now review).

\paragraph{Insertion of line operators.} With reference to the
quiver in figure \ref{fig:A1A2quiver},
the $N=+1$  traces $\langle X_\gamma\rangle_{+1}$ were computed\footnote{ In the conventions of \cite{Cecotti:2010fi} the reference ray was $\theta=\pi$ instead of $\theta=0$. Hence
$\langle X_\gamma\rangle\big|_{\text{\cite{Cecotti:2010fi}}}=\langle X_{-\gamma}\rangle\big|_{\text{this paper}}$.
}
in \cite{Cecotti:2010fi}; there it was shown that they satisfy a three--term recursion in the charge $\gamma$ of the inserted line operator. This recursion relation is a manifestation of
 the Verlinde algebra of the underlying 2d CFT. Explicitly \cite{Cecotti:2010fi}
\be
\langle X_{me_1+ne_2}\rangle_{N=+1}= \frac{(-1)^{m+n}}{(q)^4_\infty}\,q^{(m^2+n^2-mn)/2}\,G_{m-n}(q)
\ee
where $G_\ell(q)$ is the Rogers--Ramunjan function
\be
G_\ell(q):= \sum_{k\geq0}\frac{q^{k^2+\ell k}}{(q)_k}
\ee
which for $\ell\geq0$, $G_\ell(q)$ may be written as a sum of (at most) $\ell$ theta--functions\footnote{ $\theta$ is defined as in eqn.\eqref{thetadef}.}\be
G_\ell(q)=
\frac{1}{(q)_\infty}\sum_{s=0}^\ell
\begin{bmatrix}\;\ell\; \\
\;s\;\end{bmatrix}_q\, q^{2s(s-\ell)}\;\theta\big(q^{3+4s-2\ell};q^5\big),\qquad \ell\geq0.
\ee
In particular, $G_0(q)\equiv G(q)$ and  $G_1(q)\equiv H(q)$.
The Garrett--Ismail--Stanton theorem\footnote{ We thank Ole Warnaar for pointing out the GIS theorem to us.} \cite{GIS}
solves the Verlinde algebra three--terms recursion for $G_\ell(q)$
in terms of the Schur polynomials $e_\ell(z)$ and $d_\ell(z)$ ($\ell\in\mathbb{Z}$).
This allows to write
$G_\ell(q)$ for $\ell\in\mathbb{Z}$
in a form which generalizes the RHS of eqns.\eqref{rogra1}\eqref{rogra2}
\be
G_\ell(q)=\frac{e_{-\ell}(1/q)}{(q,q^4;q^5)_\infty}+\frac{d_{-\ell}(1/q)}{(q^2,q^3;q^5)_\infty}
\ee
where the Schur polynomials $e_{\ell}(x)$, $d_\ell(x)$ are
 defined for $\ell\geq0$ as
\be
d_{\ell-1}(q)=\sum_j q^{j^2+j}\begin{bmatrix}\ell-j-2\\
j\end{bmatrix}_q\qquad
e_{\ell-1}(q)=\sum_j q^{j^2}\begin{bmatrix}\ell-j-1\\
j\end{bmatrix}_q,
\ee
and for $\ell<0$
\be
e_{-\ell}(1/q)=(-1)^\ell\,q^{-{\ell\choose 2}}\,d_{\ell-1}(q),\qquad
d_{-\ell}(1/q)=-(-1)^\ell\, q^{-{\ell \choose 2}}\,e_{\ell-1}(q).
\ee

The $N=-1$ trace with line operators insertions is computed in appendix \ref{mminusonea2}. Using the identities in appendix \ref{ap:partialthetas} we get
an expression in terms of  partial theta functions $\psi(z,q)$
\begin{multline}
\langle X_{me_1+ne_2}\rangle_{N=-1}=\\
= \frac{q^{mn/2}}{(q)^2_\infty}\sum_{k\geq0} (-1)^k\,
\frac{q^{(k^2+k+|m-k|+|n+k|)/2}}{(q)_k}\;\psi(-q^{|m-k|},q)\,\psi(-q^{|n+k|},q).
\end{multline}
This sum may be explicitly evaluated in each quadrant of $\mathbb{Z}^2$  in terms of Rogers--Ramanujan functions $G_\ell(q)$. The simplest case is $m\geq 0$ and $n\leq0$ where one gets (see appendix \ref{mminusonea2})
\be
\langle X_{me_1+ne_2}\rangle_{N=-1}=
q^{(m-n-nm)/2}\,G_{m-n+1}(q)\qquad \text{for }m\geq 0,\ n\leq0,
\ee
which, in particular, reduces to $\mathrm{Tr}\, \CM(q)^{-1}=H(q)$
for $m=n=0$. In the other three quadrants
one gets the same formula up to a finite sum
\be
\langle X_{me_1+ne_2}\rangle_{N=-1}=
q^{(m-n-nm)/2}\bigg(G_{m-n+1}(q)+\text{a \textit{finite} $q$--sum}\bigg).
\ee
See appendix \ref{mminusonea2} for explicit expressions in the various cases and additional details.

\paragraph{The traces $\mathrm{Tr}\,\CM(q)^{\pm 2}$.}
For any $\CN=2$ model and any $N$ one has the identity
\be
\mathrm{Tr}\,\CM(q)^{2N}= \sum_{\gamma\in\Gamma} \langle X_\gamma\rangle_N\;
\langle X_{-\gamma}\rangle_N,
\ee
where $\langle X_\gamma\rangle_N$ is defined as in \eqref{fourcoeff}
(we may see
the monodromy traces as norms, see sect.\! \ref{hardynorms}).
Thus for the $(A_1,A_2)$ model we may write $\mathrm{Tr}\, \CM(q)^{\pm 2}$ as a bilinear sum of Rogers--Ramanujan functions $G_\ell(q)$. Note that as $q\to 0$
\be
\begin{aligned}
\langle X_{me_1+ne_2}\rangle_{N=+1}\;\langle X_{-me_1-ne_2}\rangle_{N=+1}&= O(q^{m^2+n^2-mn+[|n-m|/2]^2})
\\ \langle X_{me_1+ne_2}\rangle_{N=-1}\;\langle X_{-me_1-ne_2}\rangle_{N=-1}&= O(q^{|m|+|n|}),
\end{aligned}\ee
so both traces $\mathrm{Tr}\,\CM(q)^{\pm2}$ have a nice expansion in integral non--negative powers of $q$.
Checking the $q$--expansion of $\mathrm{Tr}\,\CM(q)^2$, one is lead to the following expression
\begin{equation}
\Tr \CM(q)^2=\left(\frac{G(q)}{(q)_{\infty}^4}\right)^2\!\!
\left(1-10\sum_{k=1}^{\infty}\left(\frac{q^{5k+1}}{1-q^{5k+1}}-2\frac{q^{5k+2}}{1-q^{5k+2}}+2\frac{q^{5(k+1)-2}}{1-q^{5(k+1)-2}}-\frac{q^{5(k+1)-1}}{1-q^{5(k+1)-1}}\right)\!\right).
\end{equation}
which can be rewritten
\begin{equation}
\Tr\, \CM(q)^2=\left(\frac{G(q)}{(q)_{\infty}^4}\right)^2
\left(1-10\frac{\partial}{\partial\ln a}\theta(a,q^5)|_{a\rightarrow q}+20\frac{\partial}{\partial\ln a}\theta(a,q^5)|_{a\rightarrow q^2}\right).
\end{equation}
Therefore we see $\Tr\, \CM^2$ is $(\Tr\, \CM)^2$ multiplied by a modular function.  This again suggests
that there may be a chiral algebra associated with it.

\subsection{$(A_1, A_{2n})$ AD theories}

\paragraph{Trace of $\CM(q)^{-1}$.}

The trace of the inverse monodromy $\Tr \CM(q)^{-1}$ for the model
$(A_1,A_{2n})$ (as computed from the minimal chamber) is
\cite{Cordova:2015nma}
\be\mathrm{Tr}\,\CM(q)^{-1}=
(q)_\infty^{2n}
\sum_{\ell_1,\cdots,\ell_{2n}\geq0}
\frac{q^{\sum_i (\ell_i\ell_{i-1}+\ell_i)}}{\prod_i (q)^2_{\ell_i}}.
\ee
The sum in the RHS is explicitly evaluated  in appendix \ref{ap:andrews}; one gets
\be\label{ooguri}
 \tr\, \CM(q)^{-1} = \frac{1}{\prod_{i=0}^{n-1}(q^{2i+2}; q^{2n+3})(q^{2n+1-2i}; q^{2n+3})} \ .
\ee
It can be also written as
\be
 \tr \,\CM(q)^{-1} = \frac{1}{\prod_{i=1}^{r} \theta (Q^{R_i/2}; Q=q^{2n+3})} \ ,
\ee
where $r$ is the dimension of the Coulomb branch and $R_i$ are the dimension of the Coulomb branch operators.

The RHS of \eqref{ooguri} is the Feigin--Nakanishi--Ooguri formula for the vacuum character
of the $(2,2n+3)$ Virasoro minimal model
\cite{Feigin:1991wv,Nahm:1992sx}.
This confirms the claim of
 \cite{ Cordova:2015nma} that $\Tr\, \CM(q)^{-1}$ is equal to the vacuum character of the minimal model. The central charge of $(2, 2n+3)$ minimal model is
\begin{equation}
c_{2,2n+2}=-\frac{2n(6n+5)}{2n+3}.
\end{equation}
Its effective central charge is then
\be
c_\text{eff}=\frac{2n}{2n+3}
\ee
in agreement with eqn.\eqref{effeN-1a}. The 2d central charge is again $-12$ times the 4d central charge, implying that this the the correct 2d chiral algebra constructed in \cite{Beem:2013sza,Beem:2014zpa}. It also agrees with the TQFT computation of the Schur index \cite{Song:2015wta}.

\paragraph{Trace of $\CM(q)$.}
The monodromy traces $\Tr\, \CM(q)$ for $(A_1,A_{2n})$ were computed in \cite{Cecotti:2010fi}; with the present normalizations
they are
\begin{equation}\label{traceA2n}
\Tr\, \CM(q)=\frac{1}{(q)^{4n}_\infty}\sum_{\boldsymbol{l}\in\mathbb{Z}^{2n-1}_+}\frac{q^{\boldsymbol{l}\cdot C_{2n-1}\cdot \boldsymbol{l}/2}}{(q)_{l_1}\cdots(q)_{l_{2n-1}}},
\end{equation}
where $C_{2n-1}$ stands for  the Cartan matrix of $A_{2n-1}$. The sum in the RHS is the Nahm sum associated to the pair of graphs $(T_1,A_{2n-1})$ \cite{Nahm:2004ch}. Then its effective central charge is
\begin{equation}
c_{(A_1, A_{2n})}=\frac{2n(6n+5)}{2n+3}.
\end{equation}
in agreement with eqn.\eqref{ceffNposa} with $r=n$, $r_G=2n$, and $h_G=2n+1$.
The sum in the RHS of \eqref{traceA2n} is evaluated in
\cite{etafunctions} \textbf{Theorem 1.2}, where they are related to
the Macdonald identities for the twisted Kac--Moody algebra $A^{(2)}$ as well as to the
Feigin--Stoyanovsky theorem \cite{Feigin:1993qr} we are going to discuss.

\paragraph{The chiral algebra associated to
$\CM(q)$ for the $(A_1,A_{2n})$ model.}
The Feigin and Stoyanovsky theorem \cite{Feigin:1993qr} describes\footnote{ In fact, at least conjecturally, their results are expected to describe the chiral algebra of all $\mathcal{N}=2$ models of the class $(A_k,G)$, with $G$ a simply laced Lie algebra. The relation level/rank in their theorem is the same one as in the monodromy trace case, see \cite{Cecotti:2010fi}.
 } a class of $q$--sum
of the form
\begin{equation}
\Phi_{\mathfrak{g}}(q):=\sum_{\boldsymbol{l}\in\Z_+^r}\frac{q^{\boldsymbol{l}\cdot C_{\mathfrak{g}}\cdot \boldsymbol{l}/2}}{(q)_{l_1}\cdots(q)_{l_{2n-1}}},
\end{equation}
where $C_\mathfrak{g}$ is the Cartan matrix of a simply--laced Lie algebra $\mathfrak{g}\in ADE$ of rank $r$. They consider a Cartan splitting $\mathfrak{g}=\mathfrak{n}_-\oplus \mathfrak{h}\oplus\mathfrak{n_+}$ and consider the loop algebra
$\widehat{\mathfrak{n}}_+$ of the nilpotent Lie algebra $\mathfrak{n}$ seen as a subalgebra of the affine $\widehat{\mathfrak{g}}_{(1)}$ Lie algebra of type $\mathfrak{g}$ and level 1.
One takes the vacuum integrable highest weight module $\CV$ of $\widehat{\mathfrak{g}}_{(1)}$ with vacuum vector $|v\rangle$ and focus on the cyclic submodule
\be
U(\widehat{\mathfrak{n}}_+)|v\rangle\in \CV,
\ee
where $U(\cdot)$ stands for the universal covering algebra. The theorem states that $\Phi_\mathfrak{g}(q)$ is just the character restricted to the submodule
\begin{equation}
\mathrm{Tr}\, q^{L_0}\bigg|_{U(\widehat{\mathfrak{n}}_+)|v\rangle}=\Phi_\mathfrak{g}(q).
\end{equation}
In other words, the construction described around eqn.\eqref{eq:relA1A2} for  $\mathfrak{g}=\mathfrak{su}(2)$,
in relation with the $(A_1,A_2)$ AD model, extends to all simply--laced Lie algebras $\mathfrak{g}$
(and, in fact, to other situations as well). For all $\mathfrak{g}$ the function $\Phi_\mathfrak{g}(q)$ is the vacuum character
of the chiral algebra $U(\widehat{\mathfrak{n}}_+)$.

In particular, for the
 $(A_1, A_{2n})$ AD model
 the chiral algebra $\CA_1$, whose partition function is
 $\mathrm{Tr}\,\CM(q)$,   is generated by the upper triangular $\widehat{SU}(2n)$ currents at level $1$.  Again this is a highly non-trivial
 realization of our conjecture that the relevant characters can be viewed as characters of a 2d chiral algebra.

\subsection{$(A_1, A_{2n+1})$ AD theories}

The models $(A_1,A_{2n+1})$ have a $U(1)$ flavor symmetry, enhanced to $SU(2)$ for $n=1$, and a Coulomb branch of dimension $n$.
The monodromy traces are function of $q$ and the flavor fugacity $y$.

\paragraph{Trace of $\CM(q)^{-1}$.}
The minimal chamber expression for the trace of the inverse monodromy of the $(A_1,A_{2n+3})$ model, as a function of $q$ and the flavor fugacity $y$, is given by
\be\begin{split}
&\mathrm{Tr}\,\CM(q)^{-1}\equiv \sum_{m\in\mathbb{Z}} y^m\,M(q)_m^{(n)}= (q)^{2n}_\infty \sum_{m\in\mathbb{Z}}y^m\times\\
&\times \sum_{k_1,\cdots k_{2n+1}\geq0\atop \ell_1,\cdots,\ell_{2n+1}\geq0} \frac{q^{(\sum_ik_i(k_{i+1}-\ell_{i+1}+\ell_i+1)+\sum_i \ell_i(\ell_{i+1}+1))/2}}{\prod_i (q)_{k_i} (q)_{\ell_i}}\,\prod_{j=1}^{n+1}
\delta_{k_{2j-1},\ell_{2j-1}+|m|}\;\delta_{k_{2j},\ell_{2j}}
\end{split}
\ee
(where we used the convention $k_{2n+2}=\ell_{2n+2}=0$).
The multiple sum in the second line of the RHS may be evaluated analytically, see appendix
\ref{app:a2n+1}. The coefficient
$M(q)^{(n)}_m$ of $y^m$ has the form
\be
M(q)^{(n)}_m= \frac{q^{|m|}}{(q)_\infty\,(q^{m+1})_\infty}\,\sum_{\ell\geq0} q^{(2n+1)\ell(\ell+m+1)}\,\alpha(|m|)_\ell,
\ee
where the Bailey coefficients $\alpha(|m|)_\ell$ are given by finite sums, see appendix
\ref{app:a2n+1} for explicit formulae and alternative expressions.

\paragraph{Trace of $\CM(q)$.}

The argument in sect.\,9.8.1 of \cite{Cecotti:2010fi} yields for the trace $\Tr\, \CM(q)$ of the $(A_1,A_{2n+1})$ model,  seen as a function of $q$ and $y$,
\begin{equation}\label{fff321}
\Tr\, \CM(q)=\frac{1}{(q)^{4n+1}_\infty}\sum_{k\in\mathbb{Z}}y^k q^{(n+1)k^2/2}\left(\sum_{\boldsymbol{l}\in\mathbb{Z}_+^{2n}}\frac{q^{\boldsymbol{l}\cdot C_{2n}\cdot \boldsymbol{l}/2 -k\sum_{i}(-1)^il_i}}{(q)_{l_1}\cdots(q)_{l_{2n}}}\right),
\end{equation}
with $C_{2n}$ the Cartan matrix of $A_{2n}$. Each sum in the large parenthesis belongs to the Nahm class for the pair of graphs $(T_1,A_{2n})$. For $k=0,1$ these sums have been evaluated
in \cite{etafunctions} (see their \textbf{Theorem 2.3}).
The effective central charge is then
\begin{equation}
c_{(A_1,A_{2n+1})}=\frac{2(3n^2+5n+1)}{n+2},
\end{equation}
in agreement with sect.\! \ref{S:tracegeneralities}.

As before, the chiral algebra $\CA_1$ for the $(A_1, A_{2n+1})$
model is generated by the upper triangular $\widehat{SU}(2n+1)$ currents at level $1$. Moreover, according to Fortin et al \cite{Fortin:2006dn} this is given by the character of $\widehat{Sp}(2n)$ at level $1$.

\subsubsection{The special $(A_1, A_3)$ model}\label{ss:a3}

The model $(A_1,A_3)$ is special in three (not unrelated) respects. First the Abelian flavor symmetry $U(1)$ in this case enhances to $SU(2)$.
Second, since $\mathfrak{su}(4)\simeq \mathfrak{so(4)}$ this model is equivalent to
$(A_1,D_3)$ AD, and hence enjoys the special properties which single out the type $D$ Argyres--Douglas models from their type $A$, $E$
brothers. Third, it it the $H_1$ model in the sequence of SCFT with one--dimensional Coulomb branch and maximal flavor symmetry compatible with the dimension $\Delta$ of the Coulomb branch field. Hence, as discussed inn sect.\,\ref{s:e6exam},  it has a $c$--saturating BPS chamber which implies the formula
\be\label{thetaprediction}
\mathrm{Tr}\, \CM(q)=\frac{\Theta(y,q)}{(q)^{12c}_\infty},
\ee
for some theta--function $\Theta(y,q)$ function. The 4d SCFT central charge for this model is $c=1/2$.
Let us show that the prediction  \eqref{thetaprediction} is correct. The simplest way is to use
the identity
\be
\begin{split}
\Phi_1(z;q)&:=\sum_{\ell_1,\ell_2\geq0} \frac{q^{\ell_1^2+\ell_2^2-\ell_1\ell_2} z^{\ell_2-\ell_1}}{(q)_{\ell_1}(q)_{\ell_2}}=\\
&= \sum_{k=-\infty}^{+\infty} q^{k^2} z^k \left(\sum_{\ell_1=\max\{0,-k\}}^\infty\frac{q^{\ell_1(\ell_1+k)}}{(q)_{\ell_1}\,(q)_{\ell_1+k}}\right)
\equiv \frac{1}{(q)_\infty}\sum_{k=-\infty}^{+\infty} q^{k^2}\,z^k,
\end{split}
\ee
where in the first equality we wrote $\ell_2=\ell_1+k$ and in the second one we used Cauchy's identity.
Indeed, from eqn.\eqref{fff321}
with $n=1$
\be
\mathrm{Tr}\,\CM(q)=\frac{1}{(q)^5_\infty}\sum_{k\in\mathbb{Z}}y^k\,q^{k^2}\,\Phi_1(q^{-k};q)=\frac{1}{(q)^6_\infty}\,\sum_{\boldsymbol{k}\in\mathbb{Z}^2} q^{\boldsymbol{k}\cdot C_2\cdot \boldsymbol{k}/2}\,y^{k_1},
\ee
i.e.\! we get \eqref{thetaprediction} with $\Theta(y,q)$ the following one--parameter specialization of the $SU(3)$ theta--function
\be
\begin{aligned}
&\Theta(y,q)=\theta_{SU(3)}(y,0,q)\\
&\text{where }\theta_{SU(3)}(y_1,y_2,q)=\sum_{\boldsymbol{k}\in\mathbb{Z}^2} q^{\boldsymbol{k}\cdot C_2\cdot \boldsymbol{k}/2}\, y_1^{k_1}\,y^{k_2}_2.
\end{aligned}
\ee

\subsection{$(A_1,D_{2n+1})$ AD theories}

The Argyres--Douglas models $(A_1,D_{2n+1})$ have a $SU(2)$ flavor symmetry with
level $k=8n/(2n+1)$; its conformal central charge is $c=n/2$.
The corresponding 2d quantities,
$c_{2d}=-12 c$ and $k_{2d}=-k/2$
then satisfy the $SU(2)$ Sugawara bound \eqref{sugsaturation}. The model has a $c$--saturating BPS chamber.

\paragraph{Trace of  $\CM(q)^{-1}$.}
The traces of the inverse monodromy for the $(A_1,D_{2n+1})$ AD models have been studied in \cite{ Cordova:2015nma} where they were found to agree with the
vacuum character of $\widehat{SU(2)}$ at level $-4n/(2n+1)$ as expected from the arguments of \cite{Beem:2013sza}
in view of the Sugawara saturation
of $c_{2d}$. Then
\be
\mathrm{Tr}\,\CM(q)^{-1}=
\frac{1}{(q)_\infty (y^2q)_\infty (y^{-2}q)_\infty}\sum_{m=0}^\infty q^{(2n+1)m(m+1)/2}\; \frac{z^{2m+1}-z^{-(2m+1)}}{z-z^{-1}}.
\ee
\medskip

\paragraph{Trace of $\CM(q)$.}
Since these models saturate the two--dimensional $SU(2)$ Sugawara bound, one expects that, as a function of $q$ and the flavor fugacity $z$,
\be\label{exD2n+1}
\mathrm{Tr}\,\CM(q)= \frac{\boldsymbol{\Theta}(z,q)}{(q)^{12c}_\infty}\,
\ee
for some theta function $\boldsymbol{\Theta}$. We check this prediction and identify
$\boldsymbol{\Theta}$
which turns out to the a one--parameter specialization of the
theta--function for the $SU(2n+1)$ root lattice.\medskip

\begin{figure}
\begin{equation*}
\begin{gathered}
\xymatrix{&&&&&1\\
r-1\ar@{-}[r] & r-2\ar@{-}[r] &\cdots \ar@{-}[r] & 4\ar@{-}[r] &3\ar@{-}[rr]\ar@{-}[ur]\ar@{-}[dr] && 0\ar@{..}[lu]
\ar@{..}[ld]
\\
&&&&& 2  }
\end{gathered}
\end{equation*}
\caption{\label{bigraph}The bi--graph of the ideal triangulation of the punctured disk with $r$ marks on the boundary having a self--folded triangle inside an \emph{internal} 2--gon.}
\end{figure}
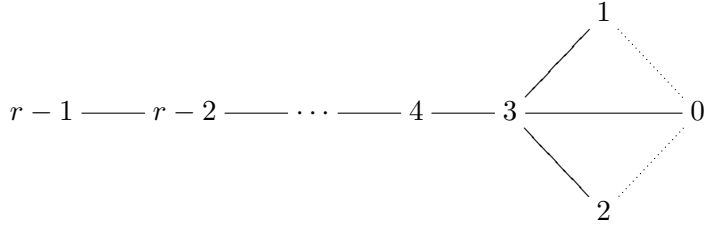

Following
sect.\,9.8.1 of \cite{Cecotti:2010fi}, for all $(A_,D_r)$ models we choose the quiver $Q$ in the form
\begin{equation}\label{dnquiv}
\begin{gathered}\xymatrix{&&&&& 1\\
r\ar[r] & r-1 \ar[r] &r-2\ar[r] &\cdots\ar[r] & 3\ar[ru]\ar[rd]\\
&&&&& 2}
\end{gathered}\end{equation}
Then in the minimal chamber we have
(setting $\Theta(z)\equiv \sum_{k\in\Z}q^{k^2/2}(-z)^k$)
\begin{equation}\label{mqzz}
\begin{split}
\CM(q)&=
\frac{1}{(q)^{r+2[(r-1)/2]}_\infty}\,
\Theta(X_{e_1};q)\,(q^{1/2}X_{e_2})_\infty\,
(q^{1/2}X_{e_1+e_2-e_3})_\infty\,
(-q^{1/2}X_{e_1-e_3})_\infty
(-q^{1/2}X_{e_2-e_3})_\infty\;\times\\
&\times\;\Theta(X_{e_3})\,(q^{1/2}X_{e_3-e_4})_\infty\,\Theta(X_{e_4})
(-q^{1/2}X_{e_4-e_5})_\infty\,
\cdots\, (-q^{1/2}X_{e_{r-1}-e_{r}})_\infty\,\Theta(X_{e_r}).
\end{split}
\end{equation}
The Dynkin quiver \eqref{dnquiv} is the incidence quiver of an ideal triangulation of the punctured disk with $r$ marks with a self--folded triangle inside a 2--gon having a side on the boundary \cite{Cecotti:2011rv}.
Flipping the internal arc of the 2--gon we get an equivalent quiver
whose Dynkin
bi--graph is represented in figure \ref{bigraph}. We recall that a Dynkin bi--graph encodes a generalized Cartan matrix $A_{ij}$: Along the main diagonal one sets $A_{ii}=2$, while for $i\neq j$
\be
A_{ij}=\#\{\text{dashed edges between $i$ and $j$}\}-\#\{\text{solid edges between $i$ and $j$}\},
\ee
The Tits form of a bi--graph is the quadratic form
$Q(\boldsymbol{m})= \boldsymbol{m}\cdot A\cdot \boldsymbol{m}/2$.
See \cite{Cecotti:2015qha} for the physical properties of the Tits forms arising from $\mathcal{N}=2$ BPS quivers. For the bi--graph in figure \ref{bigraph} we have
\begin{equation}\label{flippedTits}
2\,Q(\boldsymbol{m})=m_{r-1}^2+\sum_{j=4}^{r-1}(m_j-m_{j-1})^2+\frac{1}{2}(m_3-m_0-2m_1)^2+\frac{1}{2}(m_3-m_0-2m_2)^2+m_0^2,
\end{equation}
The $\mathbb{Z}$--equivalence class of the Tits form is a mutation--invariant \cite{Cecotti:2015qha}; hence the quadratic form
$Q(\boldsymbol{m})$ is $\Z$--equivalent to the standard Tits form of $D_r$. In particular it is positive definite.

\medskip

Now we specialize to the odd rank case, $r=2n+1$ (see below for $r$ even). Expanding \eqref{mqzz}
in the basis of $\mathbb{T}_Q$ and taking the trace one finds
 \be\label{ggrep}
 \mathrm{Tr}\,\CM(q)=\frac{1}{(q)^{4n+1}_\infty}\,\sum_{k\in\Z} q^{k^2} z^k
 \left(\sum_{m_0,m_1,\cdots,m_{2n}\geq0} \frac{q^{Q(\boldsymbol{m})+k(m_2-m_1)}}{(q)_{\boldsymbol{m}}}\right)
 \ee
 The sum inside the large parenthesis is a Nahm series \cite{Nahm:2004ch} for the positive definite `flipped' quadratic form $Q(\boldsymbol{m})$, eqn.\eqref{flippedTits}. More generally, we may consider the
 Nahm sum\footnote{ Here and below we use the compact notation
 $(q)_{\boldsymbol{m}}=\prod_{i=0}^{r-1}(q)_{m_i}$.}
 \be
 \Phi(x;q):=\sum_{m_0,m_1,\cdots,m_{2n}\geq0}
 \frac{q^{Q(\boldsymbol{m})}\,x^{m_1-m_2}}{(q)_{\boldsymbol{m}}}
 \ee
 Eqn.\eqref{ggrep} may then be written in the form
 \be
 \mathrm{Tr}\,\CM(q)=\frac{1}{(q)^{4n+1}_\infty}\,\sum_{k\in\Z} q^{k^2} z^k
 \;\Phi(q^{-k};q).
 \ee

The fact that for the $(A_1,D_{2n+1})$ model the 2d central charge $c_{2d}$ is saturated by the
Sugawara value for its $SU(2)$ current algebra, eqn.\eqref{sugsaturation}, strongly suggests that the
 following magical identity holds
 \be\label{conje2}
 \Phi(x;q)=\frac{1}{(q)^{2n-1}_\infty}
 \sum_{\boldsymbol{k}\in \Z^{2n-1}} q^{\boldsymbol{k}\cdot C_{2n-1}\cdot \boldsymbol{k}/2}\, x^{k_1},
 \ee
 where $C_{2n-1}$ is Cartan matrix of $A_{2n-1}$. In other words,
 $(q)^{2n-1}_\infty\,\Phi(z;q)$ \textit{is just a one--parameter specialization of the
 $SU(2n)$ theta--function.} Thus
 \be\label{trmdodd}
 \mathrm{Tr}\,\CM(q)=\frac{1}{(q)_\infty^{6n}}\sum_{\boldsymbol{k}\in\Z^{2n}}z^{k_0} q^{k_0^2+\boldsymbol{k}\cdot C_{2n-1}\cdot
 \boldsymbol{k}/2-k_0k_1}\equiv
 \frac{1}{(q)_\infty^{6n}}\sum_{\boldsymbol{k}\in\Z^{2n}}
 z^{k_1} q^{\boldsymbol{k}\cdot C_{2n}\cdot \boldsymbol{k}/2}
 \ee
 in agreement with the physical expectation in eqn.\eqref{exD2n+1}, where now we see that $\boldsymbol{\Theta}$ is the one--parameter specialization of the $SU(2n+1)$ theta--function
 in the RHS of \eqref{trmdodd}.

 Specialized to $z=1$, $\mathrm{Tr}\,\CM(q)$ is $(q)^{-2n}_\infty$ times the vacuum character of $\widehat{SU}(2n+1)_1$, while the CNV trace
 \be
 \mathrm{Tr}\,\CM(q)\Big|_\text{CNV}\equiv \oint \frac{dz}{2\pi iz} \mathrm{Tr}\,\CM(q),
 \ee
 is $(q)^{-(2n+1)}_\infty$ times the vacuum character of
 $\widehat{SU}(2n)_1$.\medskip

 The magical identity \eqref{conje2}, as well as its counterpart for $(A_1,D_r)$ with $r$ even (cfr.\! eqn.\eqref{conj3}), originally motivated by the Sugawara saturation and its implications for the monodromy $\CM(q)$, has recently been proved by O. Warnaar \cite{conwarnaar}.\footnote{ We thank Ole Warnaar for communicating his proof to us.}

\subsection{$(A_1,D_{2n})$ AD theories}

These Argyres--Douglas models
have $SU(2)\times U(1)$ flavor symmetry for $n>2$, which is further enhanced to
$SU(2)\times SU(2)$ for $n=1$ and to $SU(3)$ for $n=2$ \cite{Cecotti:2011rv}. Then SCFT charges
are
\be
c=(3n-2)/6,\quad \text{and}\quad k=2(2n-1)/n.\ee
 The corresponding 2d
quantities $c_{2d}=-12 c$ and $k_{2d}=-k/2$ saturates the Sugawara conditions for all\footnote{ Also for $n=1$, taking into account that the quoted value of $k$ refers to the diagonal $SU(2)$ subgroup.} $n\geq 2$
\be
\frac{\left(-\frac{2n-1}{n}\right)\cdot 3}{2+\left(-\frac{2n-1}{n}\right)}+1= -6n+4\equiv c_{2d}.
\ee
Note that the $+1$ in the above formula comes from the central charge of the chiral algebra of the current associate to the $U(1)$ global symmetry.
Correspondingly, we expect
\be\label{exp34}
\mathrm{Tr}\,\CM(q)= \frac{\boldsymbol{\Theta}(u,v;q)}{(q)^{12 c}_\infty}
\ee
for the theta--function $\boldsymbol{\Theta}(u,v;q)$ of some positive--definite lattice $\Lambda_n$,
specialized to two parameters i.e.\! the $SU(2)$ fugacity $v$ and $U(1)$ fugacity $u$. Below we shall show that this prediction is correct, and identify the lattice $\Lambda_n$
and the precise two variable specialization.

\paragraph{Traces of  $\CM(q)^{-1}$.} These traces were studied in \cite{ Cordova:2015nma} and shown to match the conjectural Schur index for these Argyres--Douglas models \cite{Buican:2015ina, Buican:2015tda}. We refer the reader to those papers for further details.

\paragraph{Traces of $\CM(q)$.}
Setting $r=2n$ in eqn.\eqref{mqzz},
and computing the trace,
eqn.\eqref{ggrep} gets replaced by
\begin{equation}
\Tr\, \CM(q)=
\frac{1}{(q)^{4n-2}_\infty}\sum_{k,\ell\in\mathbb{Z}}u^k v^\ell\, q^{k^2+(2n-1)\ell^2}\,\Phi(q^{-k},q^{-\ell};q),
\end{equation}
where $\Phi(w,z;q)$ is the following Nahm sum for the Tits form $Q(\boldsymbol{m})$ of the bi--graph in figure \ref{bigraph} (with $r=2n$)
\be
\Phi(w,z;q)=
\sum_{m_0,m_1,\cdots,m_{2n-1}\geq 0}\frac{q^{Q(\boldsymbol{m})}}{(q)_{\boldsymbol{m}}}\;w^{m_1-m_2}\;z^{m_1+m_2+2m_0+2\sum_{j=3}^{n-1}(-1)^j m_j}
\ee
The even rank counterpart to
the magic identity \eqref{conje2} is
the the following identity
\begin{equation}\label{conj3}
\Phi(w,z;q)=\frac{1}{(q)^{2n-2}_\infty}\;\sum_{\boldsymbol{m}\in\Z^{2n-2}} q^{\boldsymbol{m}\cdot C_{2n-2}\cdot \boldsymbol{m}/2}\; w^{m_1-m_2}\;z^{m_1+m_2+2\sum_{j\geq 3}(-1)^jm_j}
\end{equation}
where $C_{2n-2}$ is the Cartan matrix of $A_{2n-2}$.
In other words, $(q)^{2n-2}_\infty\,\Phi(w,z;q)$ \textit{is a two--variable specialization of the theta function for the $SU(2n-1)$ root lattice.}
Also this second magic identity is now a proven mathematical theorem \cite{conwarnaar}.

In conclusion, we get \eqref{exp34}
with
\begin{equation}
\boldsymbol{\Theta}(u,v;q)=
\sum_{(k,\ell,\boldsymbol{m})\in\Z^{2n}} q^{\boldsymbol{m}\cdot C_{2n-2}\cdot \boldsymbol{m}/2+k^2+(2n-1)\ell^2+k(m_2-m_1)-
\ell\{m_1+m_2+2\sum_{j\geq 3}(-1)^jm_j\}}\,u^k\,v^\ell.
\end{equation}
from which we read the lattice $\Lambda_n$. In particular we note that
\be
\begin{aligned}
\oint \frac{dv}{2\pi i\,v}\;\boldsymbol{\Theta}(u,v;q)&=
\text{theta--function for the $SU(2n-1)$ lattice}\\
\oint \frac{du\,dv}{(2\pi i)^2u\,v}\;\boldsymbol{\Theta}(u,v;q)&=
\text{theta--function for the $SU(2n-2)$ lattice.}
\end{aligned}
\ee

%%%%%%%%%%%%%%%%%%%%%%%%%%%%%%%%%%%%%%%%%%%%%%%%%%%%

%%%%%%%%%%%%%%%%%%%%%%%%%%%%%%%%%%%%%%%%%%%%%%%%%%%%%%%%%
\section{Concluding Remarks}\label{sec:conclusion}
In this paper we have shown that an integer sequence of specializations of the 4d superconformal index for $\CN=2$ theories
has interesting connections to 2d chiral algebras, and moreover the characters of this algebra are captured by
the traces of the powers of KS monodromy operator.

There are many things that can be done to extend the present work.  The most important unanswered question is
whether there always is a canonical unambiguous meaning to the trace of the powers of the monodromy operator.
Moreover it would be nice to make the corresponding 2d theory more physical.  There is a sense in which we have succeeded
in doing this:  Namely there is a closely related family of 2d CFT's whose characters give the same integrand as the
ones we get in chiral algebras.  Namely the 2d theories obtained by $\frac{N}{2} U(1)_r$ twisted KK reduction of the 4d theory.
It would be interesting to explore the connections between these two dual perspectives.

Finally, and perhaps most importantly, we have conjectured the existence of an algebra for all $N$, but
we only have circumstantial evidence for their existence (except for $N=-1$).  It would be important
to come up with an a priori definition of these algebras in general.  A starting point for this is to show
their existence in Lagrangian theories.  For this class we managed to prove that in the extreme weak limit,
there is a chiral algebra realizing the character.

%%%%%%%%%%%%%%%%%%%%%%%
\acknowledgments{We thank T. Dumitrescu, S. Gukov, A. Kirillov, K. Lee, P. Putrov, L. Rastelli, V. Stylianou, Y. Tachikawa, and O. Warnaar for useful discussions.
We wish to thank the hospitality of the Simons Center for Geometry and Physics where this work was initiated during the 2015 summer workshop in mathematics and physics.
SC would also like to thank the Bershadsky Visiting Fellowship while visiting the Department of Physics at Harvard University.  JS would also like to thank hospitality of the Department of Physics at Harvard University. 
The research of CV is supported in part by NSF grant PHY-1067976.
The research of JS is supported by the US Department of Energy under UCSD's contract de-sc0009919. 
The research of WY is supported by the Center for Mathematical Sciences and Applications at Harvard University.}

%%%%%%%%%%%%%%%%%%%%%%%%%%%%%%%%%%%%%%%%%%%%%%%%%%%%%%%%
\appendix

\section{Partial topological twisting and effective 2d central charges} \label{app:twisting}
Let us put 4d $\CN=2$ theory on a Riemann surface $\CC_g$ of genus $g$ and take the small volume limit to get an effective 2d theory. In order to preserve any supersymmetry, we need to perform topological twisting along $\CC_g$ \cite{Bershadsky:1995vm, Kapustin:2006hi}. The symmetry group of the 4d $\CN=2$ superconformal theory includes $SU(2)_1 \times SU(2)_2 \times SU(2)_R \times U(1)_r$, where $SU(2)_1 \times SU(2)_2 = SO(4)$ is the (Euclidean) Lorentz group and $SU(2)_R \times U(1)_r$ is the R-symmetry group. Upon dimensional reduction, the symmetry group becomes $SO(2)_E \times SO(2)_\CC \times SU(2)_R \times U(1)_r$, where $SO(2)_E$ and $SO(2)_\CC$ are the Lorentz group along the $\IR^2$ and $\CC_g$ respectively.

There are two linearly independent choices of twisting. We can twist with either $U(1)_r$ or $SU(2)_R$. If we twist by $U(1)_r$, we get $\CN=(0, 4)$ SUSY in two-dimension since $Q_-^1, Q_-^2, \tilde{Q}_-^1, \tilde{Q}_-^2$ are preserved in 2d. Note that they all have charge $-\half$ under $SO(2)_E$. If we twist with $SU(2)_R$, the conserved supercharges are $Q_-^1, Q_+^2, \tilde{Q}_+^1, \tilde{Q}_-^2$ so that we get $\CN=(2, 2)$. See the table \ref{table:N2susy}. Also, if we choose to do more general twisting by considering a linear combination of the two, we get $\CN=(0, 2)$ SUSY in 2d.\footnote{Closely related twisting in the context of M5-branes wrapped on 4-cycles has been also considered in \cite{Benini:2013cda,Gadde:2013sca, Bah:2015nva}.}

\begin{table}[h]
\be
\begin{array}{c|cc|cc|cc}
Q & U(1)_E & U(1)_\CC  & SU(2)_R & U(1)_r & U(1)_R^{(a, b)} & U(1)_\CC^{(a, b)} \\
\hline \hline
Q_-^1 & -\half & -\half & \half & \half & \half & 0 \\
Q_+^1& \half & \half  & \half & \half & \half & 1 \\
Q_-^2  & -\half & -\half & -\half & \half & \half -a & -a \\
Q_+^2  & \half & \half & -\half & \half & \half -a & 1-a \\
\hline
\tilde{Q}_-^1 & -\half & \half & \half & -\half & a - \half & a \\
\tilde{Q}_+^1 & \half & -\half & \half & -\half & a -\half & a-1 \\
\tilde{Q}_-^2 & -\half & \half & -\half & -\half & -\half & 0 \\
\tilde{Q}_+^2 & \half & -\half & -\half & -\half & -\half & -1
\end{array} \nn
\ee
\caption{Supercharges of the $d=4, \CN=2$ supersymmetry upon twisting. Here $U(1)_R^{(a, b)} = a R + b r$ with $a+b=1$, and $U(1)_E$ is the 2d Lorentz group, and $U(1)_\CC$ is the Lorentz group on $\CC$ before twisting. We see that for general twisting, only 2 supercharges $Q_-^1, \tilde{Q}_-^2$ are preserved, which are both right-moving. }
\label{table:N2susy}
\end{table}

\subsection{Twisting of the free theory}\footnote{We thank Vasilis Stylianou for the discussions and sharing his note related to this section.}  \label{sec:02twist}
Let us consider the effect of twisting for the free hypermultiplet and vector multiplet. Aspects of this twisting has been already discussed in \cite{Putrov:2015jpa}, and what we do here is simply to consider a linear combination of the two twists $U(1)_r$ and $SU(2)_R$.

One can consider a linear combination of the $SU(2)_R$ twist and $U(1)_r$ twist,
\be
 \CC^{(a, b)} = \CC + a R + b r \ ,
\ee
which yields $\CN=(0, 2)$ theory. When $(a, b)=(1, 0)$, we get $\CN=(2, 2)$ theory and for $(a, b)=(0, 1)$ we get $\CN=(0, 4)$ theory. We also need to have $a+b=1$.

Upon the general twisting of a hypermultiplet, we get the charges as in the table \ref{table:Hyper02}.
\begin{table}[h]
\be
\begin{array}{c|cc|cc|cc}
 & U(1)_E & U(1)_\CC & SU(2)_R & U(1)_r &U(1)_R^{(a, b)} & U(1)_\CC^{(a, b)} \\
\hline \hline
\psi_\pm & \pm \half & \pm \half & 0 & - \half & \frac{a-1}{2} & (\frac{a}{2}, \frac{a}{2}-1) \\
\tilde{\psi}_{\dot{\pm}}^\dagger & \pm \half & \mp \half & 0 & \half & \frac{1-a}{2} & (-\frac{a}{2}, 1 - \frac{a}{2})\\
\psi^\dagger_{\dot{\pm}} & \pm \half & \mp \half & 0 & \half & \frac{1-a}{2} & (-\frac{a}{2}, 1 - \frac{a}{2}) \\
\tilde{\psi}_{\pm} & \pm \half & \pm \half & 0 & -\half & \frac{a-1}{2} & (\frac{a}{2}, \frac{a}{2}-1) \\
\hline
q & 0 & 0 & \half & 0 & \frac{a}{2} & \frac{a}{2} \\
\tilde{q}^\dagger & 0 & 0 & -\half & 0 & -\frac{a}{2} & -\frac{a}{2} \\
q^\dagger & 0 & 0 & -\half & 0 & -\frac{a}{2} & -\frac{a}{2} \\
\tilde{q} & 0 & 0 & \half & 0 & \frac{a}{2} & \frac{a}{2}
\end{array} \nn
\ee
\caption{Twisting hypermultiplets with both $SU(2)_R$ and $U(1)_r$. }
\label{table:Hyper02}
\end{table}
We see that $(q, \psi_+)$ and $(\tilde{q}, \tilde{\psi}_+)$ form $\CN=(0, 2)$ chiral multiplets, and become sections of $K^{\frac{a}{2}}$, where $K$ is the canonical bundle over the Riemann surface $\CC_g$. Also, we get Fermi multiplets from $(\psi_{\dot -}^\dagger)$ and $(\tilde{\psi}_{\dot{-}}^\dagger)$, which are sections of $K^{1-\frac{a}{2}}$. We expect that the twisted Lorentz group on the Riemann surface $U(1)_\CC^{(a, b)}$ becomes a global symmetry of the 2d theory, and $U(1)_R^{(a, b)}$ to become $R$-symmetry of the theory.

\begin{table}[h]
\be
\begin{array}{c|cc|cc|cc}
 & U(1)_E & U(1)_\CC & SU(2)_R & U(1)_r & U(1)_R^{(a, b)} & U(1)_\CC^{(a, b)} \\
\hline \hline
A_{\a \dot{\b}} & (1, -1, 0, 0) & (0, 0, 1, -1) & 0 & 0 & 0 & (0, 0, 1, -1)\\
\lambda_\pm & \pm \half & \pm \half & \half & \half & \half & (1, 0) \\
\tilde{\lambda}_\pm & \pm \half & \pm \half & -\half & \half & \half -a & (1-a, -a)\\
\lambda^\dagger_{\dot{\pm}} &  \pm \half & \mp \half & -\half & -\half & -\half & (-1, 0) \\
\tilde{\lambda}^\dagger_{\dot{\pm}} & \pm \half & \mp \half & \half & -\half & a - \half & (a-1, a) \\
\phi & 0 & 0 & 0 & 1 & 1-a & 1-a \\
\phi^\dagger & 0 & 0 & 0 & -1 & a-1 & a-1
\end{array} \nn
\ee
\caption{Twisting vector multiplets with both $SU(2)_R$ and $U(1)_r$.}
\label{table:Vector02}
\end{table}
For the case of vector multiplet, we get an $\CN=(0, 2)$ vector from $(A_{+\dot{+}}, \lambda_-)$, and chiral multiplets from $(A_{+\dot{-}}, \lambda_+)$ in the section $\G(\CC_g, K)$. Also, we get another chiral multiplets from $(\phi, \tilde{\lambda}_-)$ in $\G(\CC_g, K^b)$ and Fermi multiplets from $(\lambda_-^\dagger)$ in $\G(\CC_g, K^a)$. We summarize this in the table \ref{table:Matter02twist}.

\begin{table}[h]
\centering
\begin{tabular}{|c|c|cc|cc|c|c|}
\hline
superfield & components & $SU(2)_R$ & $U(1)_r$ & $U(1)_{R_0}^{(a, b)}$ & $U(1)_R$ & multiplicity & $\CC_g = \IP^1$ \\
\hline \hline
vector $U$ & $(A_{+\dot{+}}, \lambda_-)$ & $(0, \half)$ & $(0, \half)$ & $0$ & $0$ & $h^0(\CC_g, \CO) = 1$ & $1$ \\
chiral $\S$ & $(A_{-\dot{+}}, \lambda_{\dot +}^\dagger)$ & $(0, -\half)$ & $(0, -\half)$ & $0$ & $0$ & $h^0(\CC_g, K) =g$ & $0$ \\
chiral $\Phi$ & $(\phi, \tilde{\lambda}_+)$ & $(0, -\half)$& $(1, \half)$ & $2-2a$ & $2-2\a$ & $h^0(\CC_g, K^b)$ & $2a-1$\\
Fermi $\Theta$ & $(\tilde{\lambda}_{-})$ & $-\half$ & $\half$ & $2b-1$ & $1-2\a$ & $h^0(\CC_g, K^a)$ & $2b-1$ \\
\hline
chiral $Q$ & $(q, \psi_+)$ & $(\half, 0)$ & $(0, -\half)$ & $1-b$ & $\a$ & $h^0(\CC_g, K^{\frac{a}{2}})$ & $b$ \\
chiral $\tilde{Q}$ & $(\tilde{q}, \tilde{\psi}_+)$ & $(\half, 0)$ & $(0, -\half)$ & $1-b$ & $\a$ & $h^0(\CC_g, K^{\frac{a}{2}})$ & $b$ \\
Fermi $\Gamma$ & $(\psi_{\dot{-}}^\dagger)$ & $0$ & $\half$ & $1-a$ & $1-\a$ & $h^0 (\CC_g, K^{1-\frac{a}{2}})$ & $a-1$ \\
Fermi $\tilde{\Gamma}$ & $(\tilde{\psi}_{\dot{-}}^\dagger)$ & $0$ & $\half$ & $1-a$ & $1-\a$ & $h^0 (\CC_g, K^{1-\frac{a}{2}})$ & $a-1$ \\
\hline
\end{tabular}
\caption{Summary of general $\CN=(0,2)$ twist of 4d $\CN=2$ multiplets. $R_0 = 2(a R + b r)$, $R=2(\a R + \b r)$ with $a+b=1, \a+\b=1$. }
\label{table:Matter02twist}
\end{table}

We see that when $a=1, b=0$, $(U, \Phi)$ forms an $\CN=(2, 2)$ vector multiplet and $(\S, \Theta)$ forms an $\CN=(2, 2)$ chiral multiplet in the adjoint representation, and $(Q, \G)$ and $(\tilde{Q}, \tilde{\G})$ form $\CN=(2, 2)$ chiral multiplets in the conjugate representations. When $a=0, b=1$, $(U, \Theta)$, $(\S, \Phi)$ form $\CN=(0, 4)$ vector, twisted hypermultiplet respectively. Also, $(Q, \tilde{Q})$ forms an hypermultiplet.

The Riemann-Roch theorem for $\CC_g$ tells us that
\be \label{eq:RRthm}
 h^0 (\CC_g, K^a) - h^0(\CC_g, K^{1-a}) = (2a-1)(g-1) \ ,
\ee
from the fact that the degree of $K$ is $2g-2$. This can be used to determine the number of each multiplets in 2d. Especially, when $a=1$, this is nothing but Poincare duality between 1-forms and vectors.
When $a>1, g>1$, we have $h^0(\CC_g, K^{1-a})=0$. Hence we have
\be \label{eq:h0Ka}
h^0 (\CC_g, K^a) = (2a-1)(g-1) \quad \qquad (a>1, g >1) \ .
\ee
For the case of $g=0$, we have $h^0(\IP^1, \CO(n)) = n+1$ for $n\geq0$, and $0$ for otherwise. Since $K_{\IP^1}=\CO(-2)$, we have
\be
 h^0 (\IP^1, K^a) = h^0(\IP^1, \CO(-2a)) = 1-2a = 2b-1 \qquad (a\leq 0, b \geq 1) .
\ee

Generally, one cannot choose $a$ or $b$ to be half-integer because there is no square root of the spinor bundle. But, we can further twist the hypermultiplet by $U(1)$ baryonic symmetry to split $Q$ and $\tilde{Q}$ or $\G$ and $\tilde{\G}$. This makes the table effectively correct with half integers by picking up only one of the chiral or Fermi multiplets between the pair.

\subsection{Central charge of the 2d effective theory}
The central charge of the 2d $\CN=(0, 2)$ gauge theory can be computed easily from the 't Hooft anomalies of the R-symmetry as
\be
 c_R = 3 \tr \g^3 R^2 , \qquad c_R-c_L = \tr \g^3 \ .
\ee
Once we know the effective number of 4d hypermultiplets and vector multiplets, we can compute the central charges of the 2d theory using the above formula and the table \ref{table:Matter02twist}. Now, for the twist parameter $N \equiv -2b=2-2a$, we get
\be
 c_R &=& 3n_v ((2\a - 1)^2 (2a - 1) -1) + 6 n_h (1-\a)^2 (1-a) \ , \\
 c_L &=& c_R - 2(a-1) (n_v - n_h) \ ,
\ee
where we have not specified the R-charge $\a$ of the chiral multiplet $Q, \tilde{Q}$ yet. If we pick $\a=0$, we get a central charge for the CFT on the Higgs branch of the theory. If we pick $\a=1$, we set the chiral adjoint $\Phi$ to have R-charge $0$, therefore we are looking at a CFT on the Coulomb branch. For $\a=0$, we get
\be \label{eq:2dCCa0}
 c_L^0 = -2N (n_h - n_v) \ , \qquad c_R^0 = -3N (n_h - n_v) \ ,
\ee
and for $\a=1$, we get
\be \label{eq:2dCCa1}
 c_L^1 = N(n_h + 2n_v) \ , \qquad c_R^1 = 3N n_v \ .
\ee

\subsubsection*{Argyres-Douglas theories}
The 4d central charge $(a, c)$ of the generalized Argyres-Douglas theory of type $(A_{k-1}, A_{n-1})$ is given as \cite{Shapere:2008zf,Xie:2012hs,Xie:2013jc}
\be
 a(k, n) &=& \frac{R_A (k, n)}{4} + \frac{R_B(k, n)}{6} + \frac{5r_{k, n}}{24} \ , \\
 c(k, n) &=& \frac{R_B(k, n)}{3} + \frac{r_{k, n}}{6} \ ,
\ee
where
\be
 R_A(k, n) &=& \sum_{i=m+2+\lfloor j(l-1)/k \rfloor}^{l m} \left( \frac{i k - l j}{k+km-j}-1 \right) \ , \\
 R_B(k, n) &=& \frac{(k-1)(n-1)nk}{4(n+k)} \ ,
\ee
with $n=km+j$ and $r(k, n)$ being the dimension of the Coulomb branch of the theory.
When $k$ and $n$ are coprime, the central charges are given by
\be
 a(k, n) &=& \frac{(n-1)(k-1)(4kn+4n+4k-1)}{48(n + k)} \ , \\
 c(k, n) &=& \frac{(n-1)(k-1)(nk+k+n)}{12(n+k)} \ .
\ee
We can covariantize the above expression by writing $h^{\vee}_{G_1} = n$, $h^{\vee}_{G_2} = k$ and $ k-1=\textrm{rank}(G_{1})$ and $n-1 = \textrm{rank}(G_2)$.
From this, we can extract the effective number of hypermultiplets and vector multiplets by using
\be
 n_h = 4(2a-c) \ , \qquad n_v = 4(5c - 4a) \ .
\ee

From this data, we can compute the effective 2d central charge via \eqref{eq:2dCCa0}, \eqref{eq:2dCCa1}. When $k$ and $n$ are coprime, we find
\be
 c_L^0 = \frac{-N(k-1)(n-1)}{n+k} \ ,   \qquad  c_R^0 = \frac{-3N(k-1)(n-1)}{2(n+k)} \ ,
\ee
and
\be
 c_L^1 = \frac{N(k-1)(n-1)(nk+n+k)}{n+k} \ , ~~  c_R^1 = \frac{N(k-1)(n-1)(2nk+2n+2k-1)}{2(n+k)} .
\ee
Here we tabulate a number of examples in table \ref{table:ADcentrals}.
\begin{table}[h]
\centering
\begin{tabular}{|c||cc|cc||cc|cc|}
\hline
theory & $a$ & $c$ & $n_h$ & $n_v$ & $c_L^0$ & $c_R^0$ & $c_L^1$ &  $c_R^1$ \\
\hline
$(A_1, A_2)$ & $43/120$ & $11/30$ & $8/5$ & $ 7/5$ & $-2N/5$ & $-3N/5$ & $22N/5$ & $21N/5$  \\
$(A_1, A_3)$ & $11/24$ & $1/2$ & $8/3$ & $5/3$ & $-2N$ & $-3N$ & $6N$ & $5N$\\
$(A_1, A_4)$ & $67/84$ & $ 17/21$ & $24/7$ & $ 22/7$ & $-4N/7$ & $-6N/7$ & $68N/7$ & $66N/7$\\
$(A_1, A_5)$ & $11/12$ & $ 23/24$ & $9/2$ & $ 7/2$ & $-2N$ & $-3N$ & $23N/2$ & $21N/2$ \\
\hline
$(A_2, A_2)$ & $7/12$ & $ 2/3$ & $4$ & $ 2$ & $-4N$ & $-3N/5$ & $8N$ & $21N/5$ \\
$(A_2, A_3)$ & $75/56$ & $ 19/14$ & $40/7$ & $ 37/7$ & $-6N/7$ & $-3N$ & $114N/7$ & $5N$ \\
\hline
\end{tabular}
\caption{4d/2d central charges for Argyres-Douglas theories. Here $(c_L^0, c_R^0) = (2(n_h - n_v), 3(n_h - n_v))$ and $(c_L^1, c_R^1) = (n_h + 2n_v, -3n_v)$. For the $N$-twist corresponding to the $\tr \CM^N$, multiply the expression by $-N$.}
\label{table:ADcentrals}
\end{table}
We find when $N=-1$, $c_L^1$ coincide with the central charge of the chiral algebra of \cite{Beem:2013sza}.

\section{Superconformal index for Lagrangian theories} \label{app:index}

The $4d$ superconformal index is evaluated by a trace formula,
\begin{equation}
\CI(\mu_i)=\Tr(-1)^Fe^{-\mu_iT_i}e^{-\beta\delta},\hspace{1cm}\delta=2\{\CQ,\CQ^\dag\},
\end{equation}
where $\CQ$ is the supercharge ``with respect to which" the index is calculated and $\{T_i\}$ a complete set of generators that commute with $\CQ$ and with each other.

For $4d$ $\CN=2$ superconformal algebra $SU(2,2|2)$, the commuting subalgebra with a single supercharge is $SU(1,1|2)$, which has rank three, so the $\CN=2$ index depends on three superconformal fugacities together with flavor fugacities. Table \ref{table:charges} summarizes the superconformal generators commuting with each $\CQ$.

\begin{table}
  \begin{centering}
  \begin{tabular}{|c|c|c|c|c|c|c|}
  \hline
${\mathcal Q}$ &$SU(2)_1$&$SU(2)_2$&$SU(2)_R$&$U(1)_r$& $\delta$ & Commuting $\delta$s \tabularnewline
  \hline
  %1
  \hline
$  {\mathcal Q}_{{\suup }-}$ &$-\half$& $0$& $\;\;\;\half$&$\half$& $\delta_{{\suup }-}=  \Delta-2j_1-2R-r$
 & $\delta_{{\sudown}{+}}$,\quad $\tilde \delta_{{\suup}\dot{+}}$,\quad $\tilde \delta_{{\suup}\dot{-}}$ \tabularnewline
  %
 %2
  \hline
$ {\mathcal Q}_{{\suup}+}$ &$\;\;\half$& $0$& $\;\;\;\half$&$\half$& $\delta_{{ \suup}+}=  \Delta+2j_1-2R-r$
 & $\delta_{{\sudown}{-}}$,\quad $\tilde \delta_{{\suup}\dot{+}}$,\quad $\tilde \delta_{{\suup}\dot{-}}$ \tabularnewline
 %
 %3
   \hline
$ {\mathcal Q}_{{\sudown}-}$ &$-\half$& $0$& $-\half$&$\half$& $ \delta_{{ \sudown}-}= \Delta-2j_1+2R-r$
 & $\delta_{{\suup}{+}}$,\quad $\tilde \delta_{{\sudown}\dot{+}}$,\quad $\tilde \delta_{{\sudown}\dot{-}}$  \tabularnewline
  %
   %4
  \hline
  ${\mathcal Q}_{{\sudown}+}$  &$\;\;\half$& $0$& $-\half$&$\half$& $\delta_{{  \sudown}+}=  \Delta+2j_1+2R-r$
 & $\delta_{{\suup}{-}}$,\quad $\tilde \delta_{{\sudown}\dot{+}}$,\quad $\tilde \delta_{{\sudown}\dot{-}}$ \tabularnewline
 \hline
 %
 %
  %1
  \hline
  $\widetilde {\mathcal Q}_{{\suup}\dot{-}}$ &$0$&$-\half$&  $\;\;\;\half$&$-\half$& $\tilde \delta_{{\suup}\dot{-}} = \Delta-2j_2-2R+r$
 & $\tilde \delta_{{\sudown}\dot{+}}$,\quad $\delta_{{\suup}{+}}$,\quad $\delta_{{\suup}{-}}$ \tabularnewline
 %2
  \hline
$\widetilde {\mathcal Q}_{{\suup}\dot{+}}$ &$0$&$\;\;\;\half$&  $\;\;\;\half$&$-\half$& $\tilde \delta_{{\suup}\dot{+}}=  \Delta+2j_2-2R+r$
 & $\tilde \delta_{{\sudown}\dot{-}}$,\quad $\delta_{{\suup}{+}}$,\quad $\delta_{{\suup}{-}}$  \tabularnewline
 %
%3
\hline
$\widetilde {\mathcal Q}_{{\sudown}\dot{-}}$ &$0$&$-\half$&  $-\half$&$-\half$& $\tilde \delta_{{ \sudown}\dot{-}}=  \Delta-2j_2+2R+r$
& $\tilde \delta_{{\suup}\dot{+}}$,\quad $\delta_{{\sudown}{+}}$,\quad $\delta_{{\sudown}{-}}$  \tabularnewline
 %4
    \hline
$\widetilde {\mathcal Q}_{{ \sudown}\dot{+}}$ &$0$&$\;\;\;\half$&  $-\half$&$-\half$& $\tilde \delta_{{  \sudown}\dot{+}}=  \Delta+2j_2+2R+r$
 & $\tilde \delta_{{\suup}\dot{-}}$,\quad $\delta_{{\sudown}{+}}$,\quad $\delta_{{\sudown}{-}}$  \tabularnewline
  \hline
  \end{tabular}
  \par  \end{centering}
  \caption{\label{table:charges} For each supercharge ${\cal Q}$, we list its quantum numbers, the associated
  $\delta \equiv 2\left\{{\mathcal Q},{\mathcal Q}^\dagger\right\}$, and the other $\delta$s commuting with it.
  Here $I = \suup,\sudown$ are $SU(2)_R$ indices and
$\alpha = \pm$, $\dot \alpha = \pm$ Lorentz indices.
 $\Delta$  is the conformal dimension,  $(j_1, j_2)$ the Cartan generators of the $SU(2)_1 \otimes SU(2)_2$ isometry group, and $(R \, ,r)$, the Cartan generators
  of  the  $SU(2)_R \otimes U(1)_r$ R-symmetry group.
  }\end{table}

We use supercharge $\CQ_{1-}$ to define the index and write it as
\begin{equation}
\CI(\rho,\sigma,\tau)=\Tr(-1)^F\rho^{\half\tilde{\delta_{1\dot{-}}}}\sigma^{\half\delta_{2+}}\tau^{\half\tilde{\delta}_{1\dot{+}}}e^{-\beta\tilde{\delta}_{1-}},
\end{equation}
or in another parametrization,
\begin{equation}
\label{eq:def:index}
\CI(p,q,t)=\Tr(-1)^Fp^{j_1-j_2+r}q^{j_1+j_2+r}t^{R-r}e^{-\beta\tilde{\delta}_{1-}} \ ,
\end{equation}
and the fugacities satisfy
\be
 |p| < 1, ~~ |q| < 1, ~~ |t| < 1, ~~ |z_i| = 1, ~~ \left| \frac{pq}{t} \right| < 1 \ .
\ee
Only states satisfying
\begin{equation}
\delta_{1-} \equiv \Delta-2j_1-2R-r=0 \ ,
\end{equation}
contribute to the index.\footnote{Our definition is slightly different from the definition given in \cite{Gadde:2011uv}, where they define the index with respect to $\tilde{\CQ}_{1\dot{-}}$, but agrees with the one given in \cite{Beem:2013sza}. }

\begin{table}
\begin{centering}
\begin{tabular}{|c|c|c|c|c|c|c|c|}
\hline
Letters & $  \Delta$ & $j_1$ & $  j_2$ & $R$ & $r$  & $\mathcal{I}(p, q, t)$ \tabularnewline
  \hline
   \hline
$  \bar{\phi}$ & $1$ & $0$ & $0$ & $0$ & $1$  & $pq/t$   \tabularnewline
  \hline
$  \lambda_{\suup+}$ & $  \frac{3}{2}$ & $  \frac{1}{2}$ & $0$ & $  \frac{1}{2}$ & $-\frac{1}{2}$ &  $-t$ \tabularnewline
  \hline
$  \bar{\lambda}_{\suup\dot{\pm}}$  & $  \frac{3}{2}$ & $0$ & $ \pm \frac{1}{2}$ & $  \frac{1}{2}$ & $  \frac{1}{2}$ &  $-q$, $-p$ \tabularnewline
  \hline
$  F_{++}$ & $2$ & $1$ & $0$ & $0$ & $0$ & $pq$ \tabularnewline
  \hline
  $  \partial_{+\dot{-}}\bar{\lambda}_{\suup\dot{+}}+  \partial_{+\dot{+}}\bar{\lambda}_{\suup\dot{-}}=0$ & $  \frac{5}{2}$ & $\half$ & $0$ & $\frac{1}{2}$ &
 $\frac{1}{2}$ & $pq$  \tabularnewline
  \hline
\hline
$q$ & $1$ & $0$ & $0$ & $  \frac{1}{2}$ & $0$ &  $\sqrt{t}$ \tabularnewline
  \hline
$  \psi_{+}$ & $  \frac{3}{2}$ & $\half$ & $0$ & $0$ & $  \frac{1}{2}$  & $-pq/\sqrt{t}$ \tabularnewline
  \hline
    \hline
$  \partial_{+\dot{\pm}}$ & $1$ & $\frac{1}{2}$ & $\pm\frac{1}{2}$ & $0$ & $0$  & $q$, $p$ \tabularnewline
\hline
\end{tabular}
\par  \end{centering}
  \caption{Contributions to the index from  ``single letters''.
  We denote by $(\phi, \bar \phi,  \lambda_{I,\alpha}, \bar\lambda_{I\,\dot \alpha},  F_{\alpha \beta}, \bar F_{\dot \alpha \dot \beta})$
 the components of the adjoint ${\cal N} = 2$ vector multiplet,  by $(q, \bar q, \psi_\alpha, \bar \psi_{\dot \alpha})$ the
 components  of the  ${\cal N} = 1$
chiral multiplet,  and by $\partial_{\alpha \dot \alpha}$ the spacetime derivatives.
}
\label{tabel:letters}
\end{table}

Contribution to the index from ``single letters" inside $\CN=2$ hyper-multiplet and vector multiplet is summarized in table \ref{tabel:letters}. The single letter index for each multiplet is
\begin{align}
 & f^{\half H} (p, q, t) = \frac{\sqrt{t} - \frac{pq}{\sqrt{t}} }{(1-p)(1-q)} \ , \\
 & f^{V} (p, q, t) = -\frac{p}{1-p} -\frac{q}{1-q} + \frac{\frac{pq}{t} - t}{(1-p)(1-q)} \ .
\end{align}
From the single letter index, we obtain partition function by taking so-called the Plethystic exponential, defined as
\be
 \textrm{PE} \left[ n x \right]_x = \frac{1}{(1-x)^n} \ ,
\ee
where $n$ is an integer and $x$ is a fugacity appear in the index.
The index after taking the plethystic exponent (with respect to all the fugacities) is
\be
 I^{H}(p, q, t; a) = \textrm{PE} \left[ f^{\half H}(p, q, t) \sum_{w \in R} a^w \right]_{p, q, t, a} =  \prod_{w \in R} \prod_{m, n=0}^{\infty} \frac{(1- a^w t^{-\half} p^{m+1} q^{n+1})}{(1- a^w t^{\half} p^m q^n)} \ ,
\ee
for the hypermultiplet and
\be
 I^{V}(p, q, t; z) &=& \prod_{\a \in \Delta} \prod_{m=0}^{\infty} (1-z^{\a} p^{m+1})(1-z^{\a} q^{m+1}) \prod_{m, n=0}^{\infty} \frac{1 - t p^m q^n z^\a }{1-t^{-1} p^{m+1} q^{n+1} z^\a } \ ,
\ee
for the vector multiplet. Here we introduced the flavor fugacity $a$ for the hypermultiplet, and gauge fugacity $z$ for the vector multiplet. Here $w \in R$ denotes weights of the representation $R$ and $\Delta$ is the set of all roots (including the Cartans) of the gauge group. $z^\a$ is a short-hand notation for $z^\a \equiv \prod_{i=1}^r z_i^{\a_i}$. For example, when the gauge group is SU(2), $z^{\a}$ can be $z^2, z^{-2}, 1$.

%%%%%%%%%%%%%%%%%%%%%%%%%%%%%%%%%%%%

\section{The tool box for monodromy traces} \label{a:toolbox}

There are various techniques to compute the
$\{\mathrm{Tr}\,\mathcal{M}(q)^N\}$.
We quickly review some of them.

\subsection{$q$--hypergeometric series}
This technique \cite{Cecotti:2010fi} works for models having a \emph{finite} BPS chamber i.e.\! a chamber whose BPS spectrum consists of just
$h<\infty$ hypermultiplets, so that the KS product $M(q)^N$ contains
finitely many factors $(q^{1/2}X_\gamma;q)_\infty$.
In this case
the most obvious technique is to expand each  BPS factor in the basis $\{X_\gamma\}_{\gamma\in\Gamma}$ of the quantum torus algebra $\mathbb{T}$ using the two Euler's sums
\begin{align}
\big(\pm q^{1/2}X_{\gamma};\,q\big)_\infty^{-1}&=\sum_{n\geq0} \frac{(\pm q^{1/2})^n}{(q;q)_n}\, X_{n\gamma},\\
\big(\pm q^{1/2}X_{\gamma};\,q\big)_\infty&=
\sum_{n\geq0}\frac{(\mp 1)^n\,q^{n^2/2}}{(q;q)_n}\,X_{n\gamma}.
\end{align}
Multiply all these factors with the rule \eqref{multx}, we get a (formal) expression of the form
\be\label{mN1}
\mathcal{M}(q)^N=\sum_{\gamma\in \Gamma} \mu(N;q)_\gamma\,X_\gamma,
\ee
for certain coefficient functions $\{\mu(N;q)_\gamma\}_{\gamma\in\Gamma}$ which are given by multiple $q$--hypergeometric sums of the general form\footnote{ Here and below, by $\mathbb{N}$ we mean the set of non--negative integers, including zero.}
\be\label{qhyper}
\mu(N;q)_\gamma=
\frac{1}{(q)_\infty^{2rN}}\sum_{n_i\in \mathbb{N}^{2h|N|}}(-1)^{\epsilon_i n_i}\;\frac{q^{n_iA_{ij}n_j/2+ B_i n_i/2}}{\prod_i(q)_{n_i}}\, \delta(\gamma, n_i\gamma_i),
\ee
where:
\begin{itemize}
\item $\gamma_i$ is the charge of the $i$--th state in the ordered KS product which (by PCT)
satisfy
\be
\gamma_{i+h}=-\gamma_i\qquad i=1,2,\dots, 2h|N|;
\ee
\item  $\delta(\gamma,\gamma^\prime)$ is the Kronecker delta in the charge lattice $\Gamma$;
\item $A$ (resp.\! $B$) is a certain integral quadratic (resp.\! linear) form
which often has a nice interpretation in terms of Cartan matrices (see \cite{Cecotti:2010fi}).
Explicitly\footnote{ The ordering of the $\gamma_i$ for $N>0$ and $N<0$ are inverse of each other, so the off--diagonal entries of $A$ differ in sign in the two cases.}
\be\label{quaforms}
\begin{aligned}
&N>0: && A_{ij}n_in_j=\sum_{i=1}^{2hN}n_i^2+
\sum_{1\leq i<j\leq 2hN} \langle \gamma_i,\gamma_j\rangle\,n_in_j, && B_i=0\\
&N<0: && A_{ij}n_in_j=
\sum_{1\leq i<j\leq 2h|N|} \langle \gamma_i,\gamma_j\rangle\,n_in_j, && B_in_i=\sum_i n_i.
\end{aligned}
\ee
\item the $\epsilon_i=0,1$ are related to the quadratic refinement; one has $\epsilon_{i+h}=\epsilon_i$.
\end{itemize}
 The Kontsevitch--Soibelman formula says that the functions $\{\mu(N;q)_\gamma\}$ are invariant under all deformations of the $\mathcal{N}=2$ central charge $Z$ provided no BPS state phase crosses the reference ray (which we fix on the positive real axis).
Since $\mathcal{M}(q)^{N+M}=\mathcal{M}(q)^N\cdot\mathcal{M}(q)^M$,
\be
\mu(N+M;q)_\gamma=\sum_{\delta\in\Gamma} q^{\langle \gamma,\delta\rangle/2}\,\mu(N;q)_{\gamma-\delta}\;\mu(M;q)_\delta,\qquad M,N\in\mathbb{Z},
\ee
so, in principle, we can compute all monodromy traces
$\{\mathrm{Tr}\,\mathcal{M}(q)^N\}_{N\in\mathbb{Z}}$ using only the coefficient functions $\mu(\pm1;q)_\gamma$.

The trace on the quantum torus algebra $\mathbb{T}$ is defined by the rule
\be\label{trtr1}
\mathrm{Tr}\,X_\gamma=\begin{cases}
y_\gamma & \gamma\ \text{is a flavor charge}\\
0 & \text{otherwise}.
\end{cases}
\ee
We recall that a charge $\gamma$ is a \emph{flavor} charge iff it belongs to the radical of the Dirac form, i.e.\!
$\langle\gamma, \nu\rangle=0$ for all $\nu\in\Gamma$. The flavor charges form a sublattice $\Gamma_f\subset\Gamma$ of the charge lattice.
Let $\{e_s\}$ be a set of generators of the lattice $\Gamma$ and let $\{\phi_a\equiv \phi_{as}e_s\}$ be a set of generators of the flavor sublattice $\Gamma_f\subset\Gamma$; as a matter of notation, given the flavor charge $\gamma\equiv \sum_an_a\phi_a$ we write
$y_{\gamma}\equiv \prod_a y_a^{n_a}$, where $y_a$ is the fugacity of the $a$--th flavor charge $\phi_a$. Then, for all elements $\gamma\equiv m_s e_s\in\Gamma$, we have (here $f=\mathrm{rank}\,\Gamma_f\equiv\mathrm{rank}\,\Gamma-2r$)
\be
\mathrm{Tr}\,X_\gamma= \sum_{k_a\in\mathbb{Z}^f} \delta\!\left(\gamma,\;\sum\nolimits_ak_a \phi_a\right)\;\prod_a y_a^{k_a}=
\sum_{k_a\in\mathbb{Z}^f}\int \prod_s \left(\frac{dx_s}{2\pi i\,x_s} x_s^{m_s-k_a\phi_{as}}\right)\prod_a y_a^{k_a}.
\ee
 From \eqref{mN1}\eqref{trtr1} the monodromy traces, as functions of $q$ and the flavor fugacities $y_a$, are
\be
\mathrm{Tr}\,\mathcal{M}(q)^N=\sum_{\phi\in\Gamma_f} \mu(N;q)_\phi\;y_\phi,
\ee
or, more explicitly,
\be
\frac{1}{(q)^{2rN}_\infty}\sum_{k_a\in\mathbb{Z}^f}\prod_ay_a^{k_a}\!\!\!\!\sum_{n_i\in \mathbb{N}^{2h|N|}} (-1)^{\epsilon_in_i}\,
\frac{q^{(n_iA_{ij}n_j+B_in_i)/2}}{\prod_i(q)_{n_i}}\int\prod_s \left(\frac{dx_s}{2\pi i\,x_s} x_s^{n_i\gamma_{is}-k_a\phi_{as}}\right),
\ee
where $\gamma_i\equiv \gamma_{is}e_s$ ($\gamma_{is}\in\mathbb{Z}$) is the charge vector of the $i$--th hypermultiplet.

Although this way of defining the monodromy traces is straightforward, and in many examples it leads to nice expressions \cite{Cecotti:2010fi},
it is far from being fully satisfactory.
 The $q$--hypergeometric sums are seldom absolutely convergent, and even when they are one would like to have expressions which are easier to handle. When the series is \emph{not} convergent, one would like  to have a controlled regularization procedure.
 This suggests replacing sums with integrals.

\subsection{The Moyal approach and Hardy norms} \label{hardynorms}
Although it is not necessary, for simplicity we shall assume our $\mathcal{N}=2$ theory has a BPS quiver description \cite{Alim:2011kw}; we fix a quiver $Q$ in the mutation class and write $B_{st}$ for its exchange matrix and $e_s$ for the dimension vector which is 1 on the $s$--th node of $Q$ and zero elsewhere. To $Q$ there is associated the quantum torus algebra $\mathbb{T}_Q$ with commutation relations
$X_{e_s}X_{e_t}=q^{B_{st}}X_{e_t}X_{e_s}$.
The quantum torus algebra $\mathbb{T}_Q$ is isomorphic to the algebra of functions on the corresponding classical torus endowed with the Moyal product $\ast$ defined by the $2$--vector $B_{ts}$.
 It is more convenient to use the equivalent description as the space of holomorphic functions on the complexified classical torus $T_Q$ (endowed with the holomorphic Moyal product). Then
 \begin{equation}\label{qqq12}\text{(quantum torus algebra) $\longleftrightarrow$ (holomorphic functions on $T_Q$ with $\ast$ product)}\end{equation}
To the $s$--th node of the quiver $Q$ we attach a $\mathbb{C}^*$ variable $u_s$, the
$s$--th simple fugacity. The $\{u_s\}$ form a global coordinate system on the torus $T_Q$. To the charge $\gamma=\sum_sn_se_s$ we associate the function $u_\gamma=\prod_s u_s^{n_s}$ called the fugacity of $\gamma$. If $\gamma$ is a flavor charge, the corresponding function is called
a flavor fugacity; in this case, to avoid confusion, we replace the symbol $u_\gamma$ with $y_\gamma$.
In the basis $\{X_\gamma\}$ of $\mathbb{T}_Q$ the
correspondence \eqref{qqq12} simply reads
\begin{equation}X_\gamma \leftrightarrow u_\gamma.\end{equation}
In particular, under \eqref{qqq12}
the operator $\mathcal{M}(q)^N$ is mapped to the holomorphic function
\be
\mu(u)^{\ast N}=\sum_{\gamma\in\Gamma} \mu(N;q)_\gamma\,u_\gamma=
\overbrace{\mu(u)\ast\mu(u)\ast\cdots\ast \mu(u)}^{N\ \text{factors}}.
\ee
The trace is then defined as the integral on
the quotient of the real torus
$|u_s|=1$ by its flavor subtorus with respect to the normalized Haar measure, i.e.\! we integrate the non--flavor fugacities along the unit circle at fixed values of the flavor fugacities $y_a$.

The function $\mu(u)$ is often easier to handle than the corresponding $q$--hypergeometric sum. $\mu(u)$
satisfies a set of functional equations,
which may be used to determine it without
bothering to give a precise sense to the poorly convergent $q$--series.
The most basic function is the one associated \emph{via} \eqref{qqq12} to the half--plane KS product:
 Given a $\mathcal{N}=2$ model, a choice of quiver $Q$, and a BPS chamber (not necessarily finite) we define its Moyal function \begin{equation}\label{moyy}\kappa(u_s\,|\,q)\end{equation} to be the holomorphic function on $T_Q$ which corresponds
  under \eqref{qqq12} to the
 inverse  KS product  associated to the upper half--plane
\be
K^{-1}\equiv(q)_\infty^r\,\prod_{\text{BPS states with}\atop
0\leq \arg Z_\gamma <\pi}^\curvearrowleft \big(q^{s+1/2}X_\gamma;q\big)_\infty^{-(-1)^{2s}},
\ee
where the normalization factor $(q)_\infty^r$ take into account the massless photons.
Equivalently, the product is taken over all stable representations of the quiver $Q$. By definition, their charges $\gamma$ belong to the positive cone $\Gamma_+\subset \Gamma$ (which is a strictly convex cone).
 Then $K^{-1}$ belongs to the positive part of the quantum torus algebra, $\mathbb{T}^+_Q\equiv \text{Span}\{X_\gamma\}_{\gamma\in \Gamma_+}$, and the function $\k(u_s\,|\,q)$
contains only non--negative powers of the simple fugacities $u_i$
\begin{equation}\label{wKap}\k(u_i\,|\,q)=\sum_{n_s\geq0} \k_{n_s}(q)\,\prod_s u_s^{n_s}.
\end{equation}
The KS wall--crossing formula states that
$\kappa(u_s\,|\,q)$ is invariant under any change of the central charge $Z$ as long as we remain in the region of parameter space covered by the given form of the quiver.
 The trace of the inverse monodromy
 is given by
 \begin{equation}\label{intint}\mathrm{Tr}\,\CM(q)^{-1}=\int_{\text{unit}\hfill\atop\text{circle}} dv_r\;\k(u_s\,|\,q)\;\k(1/u_s\,|\,q)\end{equation}
 where the integral is over the non--flavor fugacities $v_r$ at fixed value of the flavor ones.
 $\mathrm{Tr}\,\CM(q)^{-1}$ is a holomorphic function of
$q$ and the flavor fugacities $y_a$. Hence it is uniquely determined
by its restriction for $q$ real and $y_a$ on the unit circle.
So restricted,
 $\mathrm{Tr}\,\CM(q)^{-1}$ becomes the $H^2$ norm of the holomorphic
(in the non flavor fugacities $v_s$'s) Moyal function $\kappa$
\begin{equation}\label{m710}\mathrm{Tr}\,\CM(q)^{-1}=\|\kappa\|_{H^2}^2=
 \int_{\text{unit}\hfill\atop\text{circle}} dv_r\;\big|\kappa(v_r, y_a\,|\,q)\big|^2\end{equation}
Hence, if at a fixed value of $q$ and
 the flavor fugacities, the holomorphic function $\kappa$ belongs to the Hardy space $H^2(D^{2r})$ ($D$ being the unit disk) $\mathrm{Tr}\,\CM(q)^{-1}$ exists  and integration term by term in \eqref{m710} is fully justified.
 It may happen that the radius of convergence in non--flavor fugacity space is exactly 1.
 In this case the $H^2$ norm diverges but
 the integral on the slightly smaller
 circles $|v_r|=e^{-\epsilon}$ would converge, giving a way of regularizing the monodromy trace.

\paragraph{Example: pure $SU(2)$.} For $SU(2)$ SYM the associated quantum cluster mutations \cite{Cecotti:2010fi} may be seen as a functional equation  for the basic Moyal function $\kappa(u_1,u_2\,|\,q)$
\be
\kappa(u_1,u_2\,|\,q)=\kappa(u_2,u_1\,|\,q)\quad\text{and}\quad
\kappa(qu_1,u_2\,|\,q)=\frac{1-q^{1/2}u_1}{1-qu_1u_2}\,\kappa(u_1,u_2\,|\,q),
\ee
whose solution is essentially the inverse of the $q$--binomial coefficient
\be
\kappa(u_1,u_2\,|\,q)=\frac{(q)_\infty\,(qu_1u_2)_\infty}{(q^{1/2}u_1)_\infty\,(q^{1/2}u_2)_\infty}= (1-u_1u_2)^{-1} \begin{bmatrix}\alpha \\ \beta\end{bmatrix}_q^{-1},
\ee
where $q^\alpha\equiv qu_1u_2$, $q^\beta=q^{-1/2}u_1$.
Then, for pure $SU(2)$
\begin{equation}
\mathrm{Tr}\,\CM(q)^{-1}=
\int \frac{d\theta}{2\pi} \left(\left|\frac{(q)_\infty}{(q^{1/2}e^{i\theta})_\infty}\right|^2\;
\int\frac{d\phi}{2\pi}\left|\frac{(qe^{i(\theta+\phi)})_\infty}{(q^{1/2}e^{i\phi})_\infty}\right|^2\right).
\end{equation}
The internal integral is easily computed with the help of the $q$--binomial theorem
\begin{equation}
\begin{split}
\int\frac{d\phi}{2\pi}\left|\frac{(qe^{i(\theta+\phi)})_\infty}{(q^{1/2}e^{i\phi})_\infty}\right|^2&=
{}_2\phi_1(q^{1/2}e^{i\theta},q^{1/2}e^{-i\theta};q;q;q)\equiv\\
&\equiv \left|\frac{(q^{1/2}e^{i\theta})_\infty}{(q)_\infty}\right|^2
\sum_{n\geq0}\frac{q^{n/2}e^{in\theta}}{1-q^{n+1/2}e^{-i\theta}},
\end{split}
\end{equation}
so that
\be
\mathrm{Tr}\,\CM(q)^{-1}=\sum_{n\geq0}q^{n(n+1)}=\psi(1,q^2),
\ee
where $\psi(z,q)$ is Ramunajan's partial theta function (see equation \eqref{psifunction}).

\subsection{The Moyal function of
the operator $\mathcal{M}(q)^N$}

Let $A_{ij}$ be a symmetric $m\times m$ matrix.  We define the function\footnote{ The integral in the first line is  along circles of radius less than 1, $|w_i|<1$, while the one in the second line is taken along straight lines $L_i$ parallel to the imaginary axis $i\mathbb{R}$ so that $|z_i e^{t_i}|<1$.}
\be\label{fA}
\begin{split}
f_A(z_i;q)&:=\!\!\!\sum_{n_i\in\mathbb{N}^{m}}\!\!\frac{q^{A_{ij}n_in_j/2}\,z_i^{n_i}}{(q)_{n_i}}=
\oint \!\!\sum_{n_i=-\infty}^{+\infty} q^{A_{ij}n_in_j/2}(z_i/w_i)^{n_i}
\prod_i \frac{dw_i}{2\pi i\,w_i\,(w_i;q)_\infty}=\\
&=(-2\pi\tau)^{-m/2}\,(\det A)^{-1/2} \int\exp\!\left[-\frac{A^{-1}_{ij}t_i t_j}{4\pi i \tau}\right]
\prod_i \frac{dt_i}{(z_ie^{t_i};q)_\infty},
\end{split}
\ee
where we set $q=e^{2\pi i\tau}$ and used Euler's first sum and Poisson summation. In equation\,\eqref{fA} sums and integrals are convergent provided the matrix
$A_{ij}$ is positive definite. In facts, for $|z_i|<1$ the series converges absolutely under the milder condition that the quadratic form $A_{ij}n_in_j$ is \emph{weakly} semi--positive\footnote{ A quadratic form $q\colon\mathbb{Z}^n\to\mathbb{Z}$ is said to be \emph{weakly positive} (resp.\! \emph{weakly semi-positive}) if $q(v)>0$ (resp.\! $q(v)\geq0$) for all $v\in \mathbb{N}^n$
$v\neq0$.}. For general $A$'s one should define $f_A(z_i;q)$
by analytic continuation e.g.\! by modifying in the last integral  the
integration contours from the straight lines $L_i=i\mathbb{R}+c_i$ to suitable $\mathcal{C}_i$.
Series and integrals of the general form
in equation\,\eqref{fA} are familiar from the theory of integer partitions and the Thermodynamical Bethe Ansatz (TBA). We shall refer to a $q$--series of the above form as a Nahm sum for the quadratic form $A_{ij}$;   see \cite{Nahm:2004ch}
for a survey.

The Moyal functions corresponding in the sense of \eqref{qqq12} to the KS products
for the BPS phase sectors $0\leq \theta < 2\pi N$
and its inverse are of this form with $A$ as in
equation\,\eqref{quaforms} (we allow $N$ to be half--integral).
More precisely,
\be
\text{Moyal function of }\mathcal{M}(q)^N=
\frac{1}{(q)^{2rN}_\infty}\; f_A\!\Big((-1)^{\epsilon_i}q^{B_i/2}u_i;q\Big)\qquad N\in\frac{1}{2}\Z.
\ee
Unless the quadratic form $A$ is weakly semi--positive, to define this Moyal function we need a prescription such as a deformation of contours $L_i\to \mathcal{C}_i$.
The correct contour prescription should be dictated by the physical interpretation of  $\mathcal{M}(q)^N$ as a relative SCFT invariant.
We note that $A$ may be weakly semi--positive only for $N=\pm1/2$ and in this case only if the quiver is acyclic and the chamber is minimal.
\medskip

Assuming a proper contour prescription exists,
the monodromy trace is then
\be\label{traceNmoy}
\mathrm{Tr}\,\mathcal{M}(q)^N
=\frac{1}{(q)_\infty^{2rN}} \oint
f_N\!\Big((-1)^{\epsilon_i}q^{B_i/2}x_s(u,y)^{\gamma_{is}};q\Big)\prod_r\frac{du_r}{2\pi i\,u_r} ,\ee
where $x_s(u,y)$ is the $s$--th simple fugacity written as a Laurent monomial in the flavor $y_a$ and
non--flavor $u_r$ fugacities. Alternatively,
we may write the trace as the integral of the
absolute value (for $N$ odd) or the square (for $N$ even) of the Moyal function for $\mathcal{M}(q)^{N/2}$.
(This last expression has typically better convergence properties).

\subsection{Asymptotic analysis of $\mathrm{Tr}\,\mathcal{M}(q)^N$: the effective central charge $c_\text{eff}$}
We may view the monodromy traces
as powers series in $q$
\be\label{qqser}\sum_{n\geq0} a_n(y)\,q^n\ee
whose coefficients $a_n(y)$ are Laurent polynomials in the flavor fugacities $y_a$ with integral coefficients.
Given a power series $\sum_{n\geq0} a_n\,q^n$
one defines its \emph{effective central charge}
$c_\text{eff}$ as
\be\label{ceffy}
c_\text{eff}=\frac{3}{2\pi^2} \lim_{n\to\infty}\frac{(\log a_n)^2}{n},
\ee
or equivalently (setting $q=e^{2\pi i \tau}$) \emph{via} its $\tau\to0$ asymptotic behavior
\be
\sum_{n\geq0} a_n\, q^n \simeq
 \exp\!\left(\frac{2\pi i}{\tau}\,\frac{c_\text{eff}}{24}+O(1)\right)\quad\text{as }\tau\to 0.
\ee
The name `effective central charge' stems from the fact that, when the $q$--series \eqref{qqser} is a conformal block
of a (not necessarily unitary) 2d CFT, one has
\begin{equation}
c_\text{eff}\equiv c-24\min_i(h_i)
\end{equation}
 with $h_i$, $c$ the conformal weights and Virasoro central charge of the CFT; in particular, for unitary CFTs, $c_\text{eff}\equiv c$.
If there are flavor symmetries, from equation\,\eqref{ceffy} we get a function of the flavor fugacities, $c_\text{eff}(y_a)$; in this case, we define the effective
central charge $c_\text{eff}$ as the value of this function at a (suitable) critical point $\partial_{y_a} c_\text{eff}(y)=0$.

To compute $c_\text{eff}$, we study the $\tau\to0$ asymptotics of
the Moyal function $f_N(z_i;e^{2\pi i\tau})$ adapting the standard TBA methods \cite{Nahm:2004ch} to our situation.
Since \begin{equation}
\frac{1}{(w;q)_\infty}\sim \exp\!\left(-\frac{1}{2\pi i \tau}\,\mathrm{Li}_2(w)+O(1)\right)\quad\text{as }\tau\to0,\end{equation}
the integral in the second line of equation\,\eqref{fA}
is
\begin{equation}
\int \exp\!\left[-\frac{1}{2\pi i \tau}\left(A^{-1}_{ij}t_it_j/2+\sum_i \mathrm{Li}_2(z_ie^{t_i})\right)+O(1)\right] dt.
\end{equation}
As $\tau\to0$ this integral may be evaluated by saddle point. The saddle point condition is
\begin{equation}A^{-1}_{ij}t_j-\log(1-z_i\,e^{t_i})=0
\end{equation}
i.e.\! setting $U_i=e^{t_i}$,
\begin{equation}\label{onshell}U_j^{A^{-1}_{ij}}+z_i\,U_i=1\end{equation}
Let $U_i=U_i(z_j)$ be the solution. As $\tau\to0$ we have
\begin{equation}
f_N(z_i;e^{2\pi i\tau})\sim
\exp\!\left[-\frac{1}{2\pi i \tau}\sum_i
\Big(\mathrm{Li}_2(z_i\, U_i)+\frac{1}{2}\log U_i \log(1-z_i \,U_i)\Big)+O(1)\right]\end{equation}
Plugging this asymptotic expression in equation\,\eqref{traceNmoy}, we see that
the integrals in $u_r$ may also be evaluated by saddle point in the $\tau\to0$ limit, so that $c_\text{eff}(y)$
is given by the value of the exponent at a critical point with respect to the non--flavor fugacities
at fixed values of the flavor fugacities $y$.
Then $c_\text{eff}$ is the value of the exponent
extremized with respect to \emph{all} fugacities (flavor and non--flavor) $x_s=e^{v_s}$. In conclusion\footnote{ The contribution $2rN$ arises from the overall factor $(q)^{-2rN}_\infty$.}
\begin{equation}\label{sintegral}
c_\mathrm{eff}=2r N+
\frac{6}{\pi^2}\!\left(A^{-1}_{ij}t_it_j/2+\sum_i \mathrm{Li}_2(e^{\gamma_{is}v_s}\,e^{t_i}\!)\right)\qquad
\begin{array}{l}
\text{at a critical point}\\
\text{in the $t_i$'s and $v_s$'s.}\end{array}\end{equation}
At a critical point
\be
A^{-1}_{ij}t_it_j\Big|_{\text{critical}\atop
\text{point}\hfill}\equiv \sum_i(t_i+\gamma_{is}v_s)\log(1-e^{\gamma_{is}v_s+t_i}),
\ee
so $c_\text{eff}$ is written in terms of the Rogers
dilogaritm $L(z)=\mathrm{Li}_2(z)+\tfrac{1}{2}\log(z)\log(1-z)$ as
\be\label{sintegral2}
c_\text{eff}=2rN+\frac{6}{\pi^2}\sum_{i=1}^{2h|N|} L(z_i)\qquad\text{where }z_i=e^{\gamma_{is}v_s+t_i}\
\text{at the critical point}.
\ee
Eqn.\eqref{sintegral2} gives $c_\text{eff}$ as a critical value of a multi--valued function. On the appropriate Riemann surface on which its analytic continuation is uni--valued there are infinitely many critical points. We have to pick out the physically correct one.
This is a solution to the saddle point equations having all the symmetries of the physical
problem. We have various symmetries:
First we have a $\Z_{|N|}$ replica symmetry
(cfr.\! sect....) which acts on
the above variables as $t_i\to t_{i+2h}$.
Then in a $\Z_{|N|}$ symmetric critical points
the $t_i$ depend on $i$ only mod $2h$.
Next we have the CPT symmetry which acts as
$t_i\to t_{i+h}$ while $\gamma_{i+h}=-\gamma_i$;
the symmetric condition is $z_i\equiv e^{\gamma_{is}v_s+t_i}=z_{i+h}$, i.e.\!
$t_{i+h}=t_i+2\gamma_{is}v_s$.
This symmetric ansatz automatically solves the
equations $\partial_{v_s}S(t_i,v_s)=0$
($S$ being the function in the parenthesis
of \eqref{sintegral}). The bottom line is that we have
\be\label{bottom1}
c_\text{eff}= 2rN+2|N|\sum_{i=1}^h \frac{6}{\pi^2}\,L(z_i),
\ee
 where the $z_i$ are the solutions to the
 \emph{reduced} Nahm equations
 \be\label{nahm}
 z_i^2=(1-z_j)^{C^\pm_{ij}},
 \ee
where the symmetric integral matrix $C^\pm_{ij}$ is the first principal $h\times h$ minor of the $2h|N|\times 2h|N|$ matrix $2 A_{ij}$ which depends only on the sign $\pm$ of $N$. One has
\be
C^-_{ij}= 2\delta_{ij}-C^+_{ij}.
\ee

We stress that for
$|N|$ large the corresponding $q$--hypergeometric series \eqref{qhyper} would \emph{not} converge and so our monodromy operator and traces are implicitly defined through a suitable deformation of the integration contours $\mathcal{C}_i$ consistent with the CPT symmetry (i.e.\! $\mathcal{C}_{i+h}=\mathcal{C}_i$). Then eqns.\eqref{bottom1}\eqref{nahm} give the physically correct $c_\text{eff}$ \textit{independently of the details of the precise contour prescription required to define $\mathcal{M}(q)^N$, provided only such a prescription exists.} Thus $c_\text{eff}$ is a robust invariant of the monodromy traces which does not suffer ambiguities in its definition. However, from \eqref{bottom1} we see that only the values of $c_\text{eff}$ for $N=\pm 1$ yield independent information.

\paragraph{Example 1:
$(G,A_1)$ models with $N>0$.}
In this case the equations \eqref{nahm} take the form
\be\label{zzz15}
z_i^2=(1-z_j)^{C_{ij}}
\ee
where $C_{ij}$ the Cartan matrix of the simply--laced Lie algebra $G$. Writing $z_i=w_j^{-C_{ij}}$, the equations take the form
\be\label{nahm12}
1+w_j^{2\delta_{ij}-C_{ij}}=w_i^2.
\ee
The equations \eqref{zzz15} have a unique solution with $0<z_i<1$ which corresponds to the vacuum character of the
2d coset  CFT $G_2/U(1)$ \cite{Nahm:2004ch,Keegan:2011ci}.
Then $6\sum_i L(z_i)/\pi^2$ is the central charge of the  $G_2/U(1)$ coset CFT, and
\be\label{ceffNpos}
c_\text{eff}=2N\!\left(r+\frac{r_G\, h_G}{h_G+2}\right),
\ee
where $r_G$, $h_G$ are the rank and the Coxeter number of the Lie algebra $G$
(related to its dimension by the Coxeter formula $\dim G=r_G(h_G+1)$). Moreover, $2r$ is equal to $r_G$ minus the multiplicity of $h_G/2$ as an exponent of $G$.

\paragraph{Example 2: $(G,A_1)$ models
with $N<0$.} In this case the equations \eqref{nahm} take the form
\begin{equation}
z^2_i=(1-z_j)^{2\delta_{ij}-C_{ij}}.
\end{equation}
Setting $z_i=-w_j^{2\delta_{ij}-C_{ij}}$ we get back the Nahm equations \eqref{nahm12}.
For $N<0$ we are interested in a \emph{different} solution of these equation; nevertheless we may find it using the same Lie theoretic methods as in the original case  \cite{Nahm:2004ch}.
For instance, for $G=A_{2\ell}$ the
unique solution of these equations with
$0<z_j<1$ is
\begin{equation}
z_j=-\frac{\sin[(j+2)\pi(k+1)/(2k+3)]\;
\sin[j\pi(k+1)/(2k+3)]}{\big(\sin[\pi(k+1)/(2k+3)]\big)^2}.
\end{equation}
We recall the following special case of an identity by Kirillov \cite{Kirillov:1993ih}
\begin{equation}\begin{split}
\frac{48}{\pi^2}\sum_{k=1}^\ell& L\!\left(-\frac{\sin[(k+2)\pi(\ell+1)/(2\ell+3)]\,\sin[k\pi(\ell+1)/(2\ell+3)]}{\sin^2[\pi(\ell+1)/(2\ell+3)]}\right)=\\
&=4\ell+\frac{3(2\ell+1)}{2\ell+3}-1\end{split}
\end{equation}
so that, for $(A_{2\ell},A_1)$ and $N<0$
\be\label{effcc}
c_\text{eff}=\frac{2|N|\ell}{2\ell+3}.
\ee
The effective central charge of the $(p,q)$ Virasoro
minimal model is
\be\label{effeN-1}
c_\text{eff}(p,q)=c-\min_{r,s} h_{r,s}=1-\frac{6}{pq}.
\ee
Taking $p=2$ and $q=2\ell+3$, we get the effective central charge \eqref{effcc} for $N=-1$.

\paragraph{Example 3:
$(G,G^\prime)$ models with $N>0$.}
Again we reduce the saddle point conditions to systems of algebraic equations already studied in a related context by
Nahm \cite{Nahm:2004ch}. Using his results, one obtains the following formula
\be\label{eccceff}
c_\text{eff}= 2Nr+N\,\frac{r_Gr_{G^\prime}h_Gh_{G^\prime}}{h_G+h_{G^\prime}}.
\ee
which has been checked explicitly for $(G,A_2)$.

\paragraph{The general relation: $N<0$ versus $N>0$.}
The relation between $N>0$ and $N<0$ traces found in the $(G,A_1)$ example may be generalized to arbitrary $\mathcal{N}=2$ models (having a finite chamber). For simplicity,
we write $C_{ij}$ for $C^+_{ij}$ even if, in general, it is not a Cartan matrix in the Kac sense. Then, writing
\be
z_i=\begin{cases}
w_j^{-C_{ij}}  & N>0\\
-w_j^{2\delta_{ij}-C_{ij}} & N<0,
\end{cases}
\ee
we end up for both signs with equation\,\eqref{nahm12}. Using that equation, we have
\be
\frac{6}{\pi^2}\sum_{i=1}^h L(z_i)=
\left\{\begin{aligned}
\tfrac{6}{\pi^2}\sum_i L(w_j^{-C_{ij}})&=\tfrac{6}{\pi^2}\sum_i L(1-w_i^{-2})=h-
\tfrac{6}{\pi^2}\sum_i L(w_i^{-2})\\
\tfrac{6}{\pi^2}\sum_i L(-w_j^{2\delta_{ij}-C_{ij}})&=
\tfrac{6}{\pi^2}\sum_i L(1-w_i^2)=-h+\tfrac{6}{\pi^2}\sum_i L(w_i^{-2}),
\end{aligned}\right.
\ee
where we used the two functional equations
for the Rogers dilogarithm
\be
L(x)+L(1-x)=\frac{\pi^2}{6},\qquad L(x)+L(x^{-1})=\frac{\pi^2}{3}.
\ee
Hence, as a function of the $w_i$'s the central charge for $N<0$ is given by minus the expression valid for $N>0$. However, the $w_i$'s for $N<0$ correspond to a \emph{different} solution to eqns.\eqref{nahm12}
than the $w_i$'s for $N>0$. In other words, in the two cases the \emph{total} set of saddle points are the same, but switching the sign of the exponent we interchange the most dominating point with the less dominating one. Now, assuming the $N=1$ trace to correspond to a unitary CFT character, the set of all saddle point values is given in terms of this 2d CFT as
\begin{equation}
\big\{c_\text{eff}\big\}_{\text{all saddle}\atop \text{points}\hfill}= \big\{c-24\,h_i\ \text{mod }24\;\big|\; h_i\ \text{the dimension of a Virasoro primary}\big\}
\end{equation}
Then the sum of $c_\text{eff}$ for $N=+1$ and $N=-1$ is 24 times the dimension of an operator in the 2d unitary CFT.

%%%%%%%%%%%%
%%%%%%%%%%%

\section{Some $q$--hypergeometric identities}

For the benefit of the reader, in this appendix we collect various identities we used in the main body of the paper and sketch the proof of some of them.

\subsection{Expansion of $1/\Theta$ in Ramanujan's partial thetas}\label{ap:partialthetas}
In sect.\,\ref{ss:abelianmatter} we introduced a function $\Xi(z;q)$
which differs from the inverse of a Jacobi theta function only by an overall $(q)^3_\infty$ factor
\be
\Xi(z;q):= \frac{(q)^2_\infty}{(q^{1/2}z; q)_\infty\, (q^{1/2}z^{-1};q)_\infty}\equiv \frac{(q)^3_\infty}{\Theta(-z;q)}
\ee
(here $\Theta(z;q)=\sum_{n\in\mathbb{Z}}q^{n^2/2}z^n$) and used the identity
\be\label{identityinversetheta}
\Xi(z;q)=\sum_{m\in\mathbb{Z}} z^m q^{|m|/2}\,\psi(-q^{|m|},q),
\ee
where $\psi(z,q)$ is the Ramanujan \emph{partial} theta function (see \cite{lostII} sect. 6.4)
\begin{equation}\label{psifunction} \psi(z,q)=\sum_{n=0}^\infty z^n\,q^{n(n+1)/2}\end{equation}
(in a \emph{partial} theta function the sum is over the non--negative integers rather than all integers as in a complete theta function).
We present two proofs of identity \eqref{identityinversetheta} since elsewhere we use  some of the intermediate identities
of both proofs.
\medskip

\textsc{First proof.} From Euler's first sum one has
\be
\frac{1}{(q^{1/2}z; q)_\infty\, (q^{1/2}z^{-1};q)_\infty}=\sum_{m,n\geq0} \frac{q^{(m+n)/2}\,z^{m-n}}{(q)_m\,(q)_n}=
\sum_{m\in\mathbb{Z}}z^m q^{|m|/2}\sum_{n\geq0}
\frac{q^{n}}{(q)_n\,(q)_{n+|m|}}
\ee
The internal sums in the RHS may be computed using the identity in \textbf{Entry 6.3.1} of Ramunjan's
\textit{lost notebook} (see \cite{lostII} p.115)
specialized to $a=-1$ and with $b=-z$:
\medskip

\textbf{Entry 6.3.1.} \textit{For all complex $z$
\begin{equation}\label{e6.3.1}
\sum_{n=0}^\infty \frac{q^n}{(q)_n\,(zq;q)_n}=\frac{1}{(q)_\infty\,
(zq;q)_\infty}\;\sum_{m=0}^\infty (-z)^m\,q^{m(m+1)/2}\equiv \frac{\psi(-z,q)}{(q)_\infty\,(zq;q)_\infty}.
\end{equation}}
\medskip

Then for all integers $m$ one has
\begin{equation}\label{corram}
\begin{split}
\sum_{n\geq0} \frac{q^n}{(q)_n\,(q)_{n+|m|}}&=\frac{1}{(q;q)_{|m|}}\sum_{n\geq0}
\frac{q^n}{(q)_n\,(q^{|m|+1};q)_n}=\hskip 2.5cm \text{\begin{small}[by equation\,\eqref{e6.3.1}]\end{small}}\\
&=\frac{\psi(-q^{|m|},q)}{(q)_\infty\,(q;q)_{|m|}\,(q^{|m|+1};q)_\infty}\equiv
\frac{\psi(-q^{|m|},q)}{(q)_\infty^2}.
\end{split}
\end{equation}

\textsc{Second proof.} We recall
\textbf{Entry 12.2.2} of Ramanujan's \emph{lost notebook} (see \cite{lostI} p.264) which expresses
$\Xi(z,q)$ as a bilateral Lambert series
\be
\Xi(z,q)=\sum_{k\in\mathbb{Z}}
(-1)^k\,\frac{q^{k(k+1)/2}}{1-z\, q^{k+1/2}}.
\ee
The RHS may be expanded as
\begin{equation}
\begin{split}
&\sum_{k\geq0}(-1)^k q^{k(k+1)/2}
\left[1+\frac{z\,q^{k+1/2}}{1-z\,q^{k+1/2}}+
\frac{z^{-1}\, q^{k+1/2}}{1-z^{-1}\,q^{k+1/2}}\right]
=\\
&=\sum_{m\in\mathbb{Z}} z^{m}\,q^{|m|/2} \sum_{k\geq0}(-1)^k q^{k(k+1)/2+|m|k}\equiv
\sum_{m\in\mathbb{Z}}z^m\, q^{|m|/2}\,\psi(-q^{|m|},q).
\end{split}
\end{equation}

\subsection{$\mathrm{Tr}[\,\CM(q)^{-1}\,X_\gamma\,]$ for the $(A_1,A_2)$ model}\label{mminusonea2}

We wish to compute the
line operator insertions in the inverse monodromy trace, $\langle X_\gamma\rangle_{N=-1}\equiv \mathrm{Tr}[\,\CM(q)^{-1}\,X_\gamma\,]$ with respect to the quiver orientation in figure \ref{fig:A1A2quiver}.
As explained in the main body of the paper, one has
\be\label{cmm1ex}
\CM(q)^{-1}=\sum_{m,n\in
\mathbb{Z}} \langle X_{-me_1-ne_2}\rangle_{N=-1}\; X_{me_1+ne_2}.
\ee
Using the minimal chamber BPS spectrum
\begin{multline}
\CM(q)^{-1}=(q)^2_\infty\;
(q^{1/2}X_{e_1};q)^{-1}_\infty\;
(q^{1/2}X_{e_2};q)^{-1}_\infty\;
(q^{1/2}X_{e_1}^{-1};q)^{-1}_\infty\;
(q^{1/2}X_{e_2}^{-1};q)^{-1}_\infty=\\
=(q)^2_\infty
\sum_{a,b,c,d\geq0} \frac{q^{(a+b+c+d)/2}}{(q)_a(q)_b(q)_c(q)_d}
q^{[a(b-d)+bc+cd]/2}\;X_{(a-c){e_1}+(b-d)e_2}.
\end{multline}
Comparing with equation\,\eqref{cmm1ex} we have
\be\label{whichsssu}\langle X_{me_1+ne_2}\rangle_{N=-1}=
(q)^2_\infty
\sum_{a,b,c,d\geq0} \frac{q^{(a+b+c+d)/2}}{(q)_a(q)_b(q)_c(q)_d}
q^{[a(b-d)+bc+cd]/2}\,\delta_{-m,a-c}\,\delta_{-n,b-d}.
\ee
From this expression it is manifest that
\be
\langle X_{-me_1-ne_2}\rangle_{N=-1}=\langle X_{ne_1+me_2}\rangle_{N=-1}
\ee
so it suffices to consider three cases:
\textit{i)} $m\geq 0$, $n\leq0$, \textit{ii)} $m,n\geq0$ and \textit{iii)} $m\leq0$, $n\geq0$. One finds
\begin{align}\label{inser1}
&m\geq0,\ n\leq0:\quad\langle X_{me_1+ne_2}\rangle_{-1}=
q^{(m+|n|+|m|n)/2}\;G_{m+|n|+1}(q)\\
\label{inser2}&m\geq0,\ n\geq0:\quad\langle X_{me_1+ne_2}\rangle_{-1}=
q^{(m-n-mn)/2}\!\left(G_{m-n+1}(q)-\sum_{\ell=0}^{n-1}\frac{q^{\ell^2+\ell(m-n+1)}}{(q)_\ell}\right)\\
\label{inser3}&\begin{aligned}m\leq0,\ n\geq0:\quad\langle X_{me_1+ne_2}
\rangle_{-1}&=
q^{(-|m|-n+|m|n)/2}\left(G_{-|m|-n+1}(q)-\sum_{\ell=0}^{n-1}\frac{q^{\ell^2+\ell(m-n+1)}}{(q)_\ell}\right.-\\
-&\left.q^{-|m|(n+1)}\sum_{b=0}^{n-1}\;\sum_{c=0}^{n-b-1}(-1)^c
\frac{(q^{b-n+1};q)_c}{(q)_b(q)_c}q^{c(c+1)/2+c|m|+b}\right)
\end{aligned}
\end{align}
so that in all cases we have
\be
\langle X_{me_1+ne_2}\rangle_{N=-1}= q^{(m-n-mn)/2}\, G_{m-n+1}(q)+\text{\textit{finite} $q$--sum.}
\ee
Note that the RHS of eqns.\eqref{inser1},\eqref{inser2},\eqref{inser3} reduce to $G_1(q)\equiv H(q)$ for $m=n=0$.
The proofs of eqns.\eqref{inser1},\eqref{inser2},\eqref{inser3} are long and tedious.
We present only the simplest one
\eqref{inser1}; the strategy of proof for the other two is similar.
\medskip

\textbf{Lemma.} \textit{Consider the double $q$--hypergeometric sum
\begin{equation}g(w,z;q):=\sum_{k,l\geq 0} \frac{q^{kl}\,z^k\,w^l}{(q;q)_k\,(w;q)_k\,(q;q)_l\,(z;q)_l}.\end{equation}
One has
\begin{equation}
g(w,z;q)=\frac{{}_0\phi_1(-;0;q;wz)}{(w;q)_\infty\,(z;q)_\infty},
\end{equation}
where ${}_0\phi_1(-;c;q;z)$ is the basic hypergeometric series {\rm \cite{gasperrahman}}
\be
{}_0\phi_1(-;c;q;z)=\sum_{n\geq0}
\frac{q^{n(n-1)}z^n}{(q;q)_n\,(c;q)_n}.
\ee}
\medskip

Note that the Rogers--Ramanujan functions $G_\ell(q)$ may be written as
\be
G_\ell(q)= {}_0\phi_1(-;0;q;q^{\ell+1}),
\ee
so that whenever $z=q^{\ell}/w$ ($\ell\in\mathbb{Z}$) the function $g(w,z;q)$ is equal to the Rogers--Ramanujan function $G_{\ell-1}(q)$ up to the simple pre--factor
$(w;q)_\infty^{-1} (q^\ell/w;q)_\infty^{-1}$. Eqn.\eqref{inser1}
is an immediate consequence of the \textbf{Lemma}. Indeed, for $m\geq0$ and $n\leq 0$, only the terms with
$c=a+m$ and $b=d+|n|$ contribute to the sum
\eqref{whichsssu} which then reduces to
\begin{equation}
(q)^2_\infty
\sum_{a,d\geq0} \frac{q^{[2da+2a|n|+2dm+m|n|+2a+2d+m+|n|]/2}}{(q)_a(q)_d(q)_{a+m}(q)_{d+|n|}}
\equiv (q)^2_\infty\,\frac{q^{(m|n|+m+|n|)/2}}{(q;q)_m\,(q;q)_{|n|}}\; g(q^{|n|+1},q^{m+1};q).
\end{equation}
Evaluating the RHS with the help of the \textbf{Lemma} yields equation\,\eqref{inser1}.
\medskip

\textsc{Proof of the Lemma.} Repeated use of limiting forms of
Heine's transformation \cite{gasperrahman}:
\begin{equation*}
\begin{split}
&\sum_{k,l\geq 0} \frac{q^{kl}\,z^k\,w^l}{(q;q)_k\,(w;q)_k\,(q;q)_l\,(z;q)_l}=\sum_{k\geq0} \frac{z^k}{(q;q)_k\,(w;q)_k}\sum_{l\geq0}\frac{(w q^k)^l}{(q;q)_l\,(z;q)_l}=\\
&=
\sum_{k\geq0} \frac{z^k}{(q;q)_k\,(w;q)_k}\,{}_2\phi_1(0,0,z;q;wq^k)=\sum_{k\geq0} \frac{z^k}{(q;q)_k\,(w;q)_k} \frac{1}{(wq^k;q)_\infty}
\,{}_0\phi_1(-;z;q;wzq^k)=\\
&=\frac{1}{(w;q)_\infty}
\sum_{k\geq0} \frac{z^k\,(w;q)_k}{(q;q)_k\,(w;q)_k}
\,{}_0\phi_1(-;z;q;wzq^k)=\frac{1}{(w;q)_\infty}
\sum_{k\geq0}\frac{z^k}{(q)_k}
\sum_{l\geq0} \frac{q^{l(l-1)} w^l\,z^l\,q^{kl}}{(q;q)_l\,(z;q)_l}=\\
&=\frac{1}{(w;q)_\infty}\sum_{l\geq0}
\frac{q^{l(l-1)} w^l\,z^l}{(q;q)_l\,(z;q)_l}\,\frac{1}{(zq^l;q)_\infty}=\frac{1}{(w;q)_\infty\,(z;q)_\infty}\sum_{l\geq0}
\frac{q^{l(l-1)} w^l\,z^l}{(q;q)_l\,(z;q)_l}\,(z;q)_l=\\
&=\frac{1}{(w;q)_\infty\,(z;q)_\infty}\sum_{l\geq0}
\frac{q^{l(l-1)} (wz)^l}{(q;q)_l}=\frac{{}_0\phi_1(-;0;q;wz)}{(w;q)_\infty\,(z;q)_\infty}.\end{split}
\end{equation*}

\subsection{$\mathrm{Tr}\,\CM(q)^{-1}$ for $(A_1,A_{2n})$ is the character of the $(2,2n+3)$ minimal model}\label{ap:andrews}

For $n,m\in\Z_+$, we define recursively the functions $A^{(n)}(q)_m$
\be
\begin{aligned}A^{(0)}(q)_m&= 1\\
A^{(n+1)}(q)_m&=(q)_\infty\sum_{k\geq0}
\frac{q^{k(m+1)}}{(q)_k^2}\,A^{(n)}(q)_k.\end{aligned}\ee
For the model $(A_1,A_{2n})$ then we have
\be\mathrm{Tr}\,\CM(q)^{-1}=
(q)_\infty^{2n}
\sum_{\ell_1,\cdots,\ell_{2n}\geq0}
\frac{q^{\sum_i (\ell_i\ell_{i-1}+\ell_i)}}{\prod_i (q)^2_{\ell_i}}=A^{(2n)}(q)_0.
\ee

\medskip

\textbf{Lemma.} \textit{For all $n\in\Z_+$ we have
\be\label{tttlemma} A^{(2n)}(q)_m=\sum_{s_1\geq0}
\frac{(q)_{m+s_1}}{(q)^2_{s_1}}q^{s_1(s_1+1)}\!\!\sum_{s_2\geq s_1}
\frac{q^{s_2(s_2+1)}}{(q)_{s_2-s_1}}
\!\!\sum_{s_3\geq s_2}\!
\frac{q^{s_3(s_3+1)}}{(q)_{s_3-s_2}}\cdots\!\!\!\! \sum_{s_n\geq s_{n-1}}\!\!
\frac{q^{s_n(s_n+1)}}{(q)_{s_n-s_{n-1}}}.
\ee}
\medskip

\textsc{Proof.} Induction on $n$.
For $n=0$ is true.
Assume the results holds for $2n$. Then
\begin{multline}
A^{(2n+1)}(q)_m= (q)_\infty \sum_{k\geq0} \sum_{s_1\geq0}
\frac{q^{k(m+1)}}{(q)_k^2}(q)_{k+s_1}
\frac{q^{s_1(s_1+1)}}{(q)^2_{s_1}}\sum_{s_2\geq s_1}
\frac{q^{s_2(s_2+1)}}{(q)_{s_2-s_1}}\cdots\!\!\! \sum_{s_n\geq s_{n-1}}
\frac{q^{s_n(s_n+1)}}{(q)_{s_n-s_{n-1}}}=\\
= (q)_\infty \sum_{k\geq0} \sum_{s_1\geq0}
\frac{q^{k(m+1)}}{(q)_k^2}(q^{s_1+1})_k
\frac{q^{s_1(s_1+1)}}{(q)_{s_1}}\sum_{s_2\geq s_1}
\frac{q^{s_2(s_2+1)}}{(q)_{s_2-s_1}}\cdots\!\!\! \sum_{s_n\geq s_{n-1}}
\frac{q^{s_n(s_n+1)}}{(q)_{s_n-s_{n-1}}}=\\
= (q)_\infty  \sum_{s_1\geq0}
{}_2\phi_1(0,q^{s_1+1};q;q;q^{m+1})\,
\frac{q^{s_1(s_1+1)}}{(q)_{s_1}}\sum_{s_2\geq s_1}
\frac{q^{s_2(s_2+1)}}{(q)_{s_2-s_1}}\cdots\!\!\! \sum_{s_n\geq s_{n-1}}
\frac{q^{s_n(s_n+1)}}{(q)_{s_n-s_{n-1}}}=\\
= (q)_\infty  \sum_{s_1\geq0}\frac{1}{(q^{m+1})_\infty}\,
{}_1\phi_1(q^{-s_1};q;q;q^{m+s_1+2})\,
\frac{q^{s_1(s_1+1)}}{(q)_{s_1}}\sum_{s_2\geq s_1}
\frac{q^{s_2(s_2+1)}}{(q)_{s_2-s_1}}\cdots\!\!\! \sum_{s_n\geq s_{n-1}}
\frac{q^{s_n(s_n+1)}}{(q)_{s_n-s_{n-1}}}=\\
= (q)_m \sum_{s_1\geq0}\sum_{k=0}^{s_1} (-1)^k \frac{(q^{-s_1})_k}{(q)^2_k} q^{k(k-1)/2+k(m+s_1+2)}
\frac{q^{s_1(s_1+1)}}{(q)_{s_1}}\sum_{s_2\geq s_1}
\frac{q^{s_2(s_2+1)}}{(q)_{s_2-s_1}}\cdots\!\!\! \sum_{s_n\geq s_{n-1}}
\frac{q^{s_n(s_n+1)}}{(q)_{s_n-s_{n-1}}},
\end{multline}
where we used various limiting forms of Heine's ${}_2\phi_1$ transformation \cite{gasperrahman}.
Now \cite{gasperrahman}
\be
(q^{-s_1})_k=\frac{(q)_{s_1}}{(q)_{s_1-k}} (-1)^k q^{k(k-1)/2-ks_1},
\ee
so
\be
A^{(2n+1)}(q)_m=(q)_m \sum_{k\geq0} \sum_{s_1\geq k}
\frac{q^{k(k+1)+km+s_1(s_1+1)}}{(q)^2_k (q)_{s_1-k}} \sum_{s_2\geq s_1}
\frac{q^{s_2(s_2+1)}}{(q)_{s_2-s_1}}\cdots\ee
Finally,
\begin{equation}
\begin{split}
A^{(2m+2)}(q)_m&\equiv (q)_\infty \sum_{\ell\geq0} \frac{q^{\ell(m+1)}}{(q)^2_\ell}\,A^{(2n)}(q)_\ell=\\
&=(q)_\infty \sum_{\ell\geq0} \frac{q^{\ell(m+1)}}{(q)_\ell}
\sum_{k\geq0} \sum_{s_1\geq k}
\frac{q^{k(k+1)+k\ell+s_1(s_1+1)}}{(q)^2_k (q)_{s_1-k}} \sum_{s_2\geq s_1}
\frac{q^{s_2(s_2+1)}}{(q)_{s_2-s_1}}\cdots=\\
&=(q)_\infty \sum_{k\geq0} \sum_{s_1\geq k}
\frac{q^{k(k+1)+s_1(s_1+1)}}{(q)^2_k (q)_{s_1-k}} \frac{1}{(q^{m+k+1})_\infty} \sum_{s_2\geq s_1}
\frac{q^{s_2(s_2+1)}}{(q)_{s_2-s_1}}\cdots=\\
&=\sum_{k\geq0}
\frac{(q)_{m+k}}{(q)^2_k} q^{k(k+1)}\sum_{s_1\geq k}
\frac{q^{s_1(s_1+1)}}{(q)_{s_1-k}} \sum_{s_2\geq s_1}
\frac{q^{s_2(s_2+1)}}{(q)_{s_2-s_1}}\cdots
\end{split}
\end{equation}
which completes the proof of the \textbf{Lemma.}
\medskip

\textbf{Corollary.} \textit{For $(A_1,A_{2n})$ one has}
\be\label{eee108}
\mathrm{Tr}\,\mathcal{M}(q)^{-1}=
\prod_{n\neq 0,\pm1\;\mathrm{mod}(2n+3)} (1-q^n)^{-1}.
\ee
\medskip

\textsc{Proof.} Set $m=0$ in \eqref{tttlemma}; the resulting sum for $A^{(2n)}(q)_0$ is the LHS of the celebrated
Andrews--Gordon (AG) generalization of the Rogers--Ramanujan identities \cite{andrews1,andrews2}. The RHS of \eqref{eee108} is the RHS of the AG identities.

\subsection{$\mathrm{Tr}\,\CM(q)^{-1}$ for the $(A_1,A_{2n+1})$ AD model}\label{app:a2n+1}

One has
\be
\mathrm{Tr}\,\CM(q)^{-1}= \sum_{m\in\mathbb{Z}} y^m\, M(q)^{(n)}_m\qquad\text{with}\quad M(q)^{(n)}_{-m}=M(q)^{(n)}_{m}.
\ee

The zero coefficient $M(q)_0^{(n)}$ in the trace of $\CM(q)^{-1}$ in the sense of \cite{Cecotti:2010fi};
in the notation of appendix \eqref{ap:andrews} then one has
\begin{equation}\label{mzeroq}
\begin{split}
M(q)^{(n)}_0&\equiv \frac{1}{(q)_\infty}\, A^{(2n+1)}(q)_0=\\
&=\frac{1}{(q)_\infty}\sum_{s_{2n+1}\geq s_{2n}\geq \cdots\geq s_1\geq0}
\frac{q^{s_1(s_1+1)+s_2(s_2+1)+\cdots +s_{2n+1}(s_{2n+1}+1)}}{(q)_{s_{2n+1}-s_{2n}}
(q)_{s_{2n}-s_{2n-1}}\cdots (q)_{s_2-s_1}}\,
\frac{1}{(q)^2_{s_1}}.
\end{split}
\end{equation}
The expression of $M(q)^{(n)}_0$
for $m>0$ should be a generalization of this formula.
We claim the following
\medskip

\textbf{Lemma.}
\textit{For $m\geq0$ the coefficient $M(q)^{(n)}_m$ is given by}
\be\label{bbbsum}
\frac{q^m}{(q)_\infty}
\sum_{s_{2n+1}\geq \cdots\geq s_1\geq0}
\frac{q^{s_1(s_1+m+1)+s_2(s_2+m+1)+\cdots +s_{2n+1}(s_{2n+1}+m+1)}}{(q)_{s_{2n+1}-s_{2n}}
(q)_{s_{2n}-s_{2n-1}}\cdots (q)_{s_2-s_1}}\,
\frac{1}{(q)_{s_1}\,(q)_{s_1+m}}.
\ee
\medskip

The proof is similar to the previous ones in this appendix and shall be omitted for brevity.
Next we need the following fundamental result of
Andrews (\cite{andrews2} \textbf{Theorem 2})
\medskip

\noindent\textbf{Theorem}
(Andrews). \textit{Assume the two sequences $\{\alpha_n\}_{n\geq0}$, $\{\beta_n\}_{n\geq0}$ form a Bailey pair, i.e.
\be\label{baileyp}
\beta_n=\sum_{r=0}^n \frac{\alpha_r}{(q)_{n-r}(aq)_{n+r}}\qquad \text{for all }n\geq0.
\ee Then
\begin{equation}
\frac{1}{(aq)_\infty}\sum_{n\geq0} q^{kn^2} a^{kn}\alpha_n=
\sum_{s_{k}\geq s_{k-1}\geq \cdots\geq s_1\geq0}
\frac{a^{s_1+s_2+\cdots+ s_{k}}\;q^{s_1^2+s_2^2+\cdots +s_{k}^2}}{(q)_{s_{k}-s_{k-1}}
(q)_{s_{k-1}-s_{k-2}}\cdots (q)_{s_2-s_1}}\,
\beta_{s_1}.\label{andrewth}
\end{equation}}
\medskip

The sum in the RHS of \eqref{bbbsum} as the same form as the one appearing in the RHS of \eqref{andrewth} with
\be
k=2n+1,\quad a=q^{m+1},\quad \beta_{s}=\frac{1}{(q)_s (q)_{s+m}}.
\ee
To get the sequence $\{\alpha_n\}$
corresponding to the Bailey sequence $\{(q)_n^{-1}(q)_{n+m}^{-1}\}$ one uses the inversion formula of \eqref{baileyp}, see \cite{andrews2}
\be
\alpha_n=(1-aq^{2n})\sum_{j=0}^n(-1)^{n-j} \frac{(aq)_{n+j-1}}{(q)_{n-j}}\,q^{(n-j)(n-j-1)/2}\,\beta_j.
\ee
Then
\begin{multline}
M(q)_m^{(n)}=\\
=\frac{q^{|m|}}{(q)^2_\infty}\sum_{j,\ell\geq0}(-1)^\ell\,
\frac{1-q^{2\ell+2j+|m|+1}}{1-q^{|m|+1}} \,
\frac{(q)_{\ell+2j+|m|}}{(q)_\ell\, (q)_j\,(q)_{j+|m|}} \,q^{(2n+1)(\ell+j)(\ell+j+|m|+1)+\ell(\ell-1)/2}.
\end{multline}

%%%%%%%%%%%%%%%%%%%%%%%%%%%%%%%%
\bibliographystyle{jhep}
\bibliography{refs}

\providecommand{\href}[2]{#2}\begingroup\raggedright\begin{thebibliography}{10}

\bibitem{Cecotti:1992rm}
S.~Cecotti and C.~Vafa, {\it {On classification of N=2 supersymmetric
  theories}},  {\em Commun. Math. Phys.} {\bf 158} (1993) 569--644,
  [\href{http://xxx.lanl.gov/abs/hep-th/9211097}{{\tt hep-th/9211097}}].

\bibitem{Seiberg:1994rs}
N.~Seiberg and E.~Witten, {\it {Electric - magnetic duality, monopole
  condensation, and confinement in N=2 supersymmetric Yang-Mills theory}},
  {\em Nucl. Phys.} {\bf B426} (1994) 19--52,
  [\href{http://xxx.lanl.gov/abs/hep-th/9407087}{{\tt hep-th/9407087}}].
  [Erratum: Nucl. Phys.B430,485(1994)].

\bibitem{Seiberg:1994aj}
N.~Seiberg and E.~Witten, {\it {Monopoles, duality and chiral symmetry breaking
  in N=2 supersymmetric QCD}},  {\em Nucl. Phys.} {\bf B431} (1994) 484--550,
  [\href{http://xxx.lanl.gov/abs/hep-th/9408099}{{\tt hep-th/9408099}}].

\bibitem{Cecotti:2010qn}
S.~Cecotti and C.~Vafa, {\it {2D Wall-Crossing, R-Twisting, and a
  Supersymmetric Index}},  \href{http://xxx.lanl.gov/abs/1002.3638}{{\tt
  arXiv:1002.3638}}.

\bibitem{Kontsevich:2008fj}
M.~Kontsevich and Y.~Soibelman, {\it {Stability structures, motivic
  Donaldson-Thomas invariants and cluster transformations}},
  \href{http://xxx.lanl.gov/abs/0811.2435}{{\tt arXiv:0811.2435}}.

\bibitem{Cecotti:2010fi}
S.~Cecotti, A.~Neitzke, and C.~Vafa, {\it {R-Twisting and 4d/2d
  Correspondences}},  \href{http://xxx.lanl.gov/abs/1006.3435}{{\tt
  arXiv:1006.3435}}.

\bibitem{Iqbal:2012xm}
A.~Iqbal and C.~Vafa, {\it {BPS Degeneracies and Superconformal Index in
  Diverse Dimensions}},  {\em Phys. Rev.} {\bf D90} (2014), no.~10 105031,
  [\href{http://xxx.lanl.gov/abs/1210.3605}{{\tt arXiv:1210.3605}}].

\bibitem{Beem:2013sza}
C.~Beem, M.~Lemos, P.~Liendo, W.~Peelaers, L.~Rastelli, and B.~C. van Rees,
  {\it {Infinite Chiral Symmetry in Four Dimensions}},  {\em Commun. Math.
  Phys.} {\bf 336} (2015), no.~3 1359--1433,
  [\href{http://xxx.lanl.gov/abs/1312.5344}{{\tt arXiv:1312.5344}}].

\bibitem{Cordova:2015nma}
C.~Cordova and S.-H. Shao, {\it {Schur Indices, BPS Particles, and
  Argyres-Douglas Theories}},  \href{http://xxx.lanl.gov/abs/1506.0026}{{\tt
  arXiv:1506.0026}}.

\bibitem{Vafa:1989xc}
C.~Vafa, {\it {String Vacua and Orbifoldized L-G Models}},  {\em Mod. Phys.
  Lett.} {\bf A4} (1989) 1169.

\bibitem{Kinney:2005ej}
J.~Kinney, J.~M. Maldacena, S.~Minwalla, and S.~Raju, {\it {An Index for 4
  dimensional super conformal theories}},  {\em Commun. Math. Phys.} {\bf 275}
  (2007) 209--254, [\href{http://xxx.lanl.gov/abs/hep-th/0510251}{{\tt
  hep-th/0510251}}].

\bibitem{Romelsberger:2005eg}
C.~Romelsberger, {\it {Counting chiral primaries in N = 1, d=4 superconformal
  field theories}},  {\em Nucl. Phys.} {\bf B747} (2006) 329--353,
  [\href{http://xxx.lanl.gov/abs/hep-th/0510060}{{\tt hep-th/0510060}}].

\bibitem{Witten:1988ze}
E.~Witten, {\it {Topological Quantum Field Theory}},  {\em Commun. Math. Phys.}
  {\bf 117} (1988) 353.

\bibitem{Gadde:2011ik}
A.~Gadde, L.~Rastelli, S.~S. Razamat, and W.~Yan, {\it {The 4d Superconformal
  Index from q-deformed 2d Yang-Mills}},  {\em Phys. Rev. Lett.} {\bf 106}
  (2011) 241602, [\href{http://xxx.lanl.gov/abs/1104.3850}{{\tt
  arXiv:1104.3850}}].

\bibitem{Gadde:2011uv}
A.~Gadde, L.~Rastelli, S.~S. Razamat, and W.~Yan, {\it {Gauge Theories and
  Macdonald Polynomials}},  {\em Commun. Math. Phys.} {\bf 319} (2013)
  147--193, [\href{http://xxx.lanl.gov/abs/1110.3740}{{\tt arXiv:1110.3740}}].

\bibitem{Dimofte:2009bv}
T.~Dimofte and S.~Gukov, {\it {Refined, Motivic, and Quantum}},  {\em Lett.
  Math. Phys.} {\bf 91} (2010) 1,
  [\href{http://xxx.lanl.gov/abs/0904.1420}{{\tt arXiv:0904.1420}}].

\bibitem{Dimofte:2009tm}
T.~Dimofte, S.~Gukov, and Y.~Soibelman, {\it {Quantum Wall Crossing in N=2
  Gauge Theories}},  {\em Lett. Math. Phys.} {\bf 95} (2011) 1--25,
  [\href{http://xxx.lanl.gov/abs/0912.1346}{{\tt arXiv:0912.1346}}].

\bibitem{Cecotti:2014wea}
S.~Cecotti, A.~Neitzke, and C.~Vafa, {\it {Twistorial Topological Strings and a
  tt* Geometry for N=2 Theories in 4d}},
  \href{http://xxx.lanl.gov/abs/1412.4793}{{\tt arXiv:1412.4793}}.

\bibitem{Closset:2012ru}
C.~Closset, T.~T. Dumitrescu, G.~Festuccia, and Z.~Komargodski, {\it
  {Supersymmetric Field Theories on Three-Manifolds}},  {\em JHEP} {\bf 05}
  (2013) 017, [\href{http://xxx.lanl.gov/abs/1212.3388}{{\tt
  arXiv:1212.3388}}].

\bibitem{Dimofte:2011py}
T.~Dimofte, D.~Gaiotto, and S.~Gukov, {\it {3-Manifolds and 3d Indices}},  {\em
  Adv. Theor. Math. Phys.} {\bf 17} (2013) 975--1076,
  [\href{http://xxx.lanl.gov/abs/1112.5179}{{\tt arXiv:1112.5179}}].

\bibitem{Gang:2012yr}
D.~Gang, E.~Koh, and K.~Lee, {\it {Line Operator Index on $S^{1}\times
  S^{3}$}},  {\em JHEP} {\bf 05} (2012) 007,
  [\href{http://xxx.lanl.gov/abs/1201.5539}{{\tt arXiv:1201.5539}}].

\bibitem{Shapere:2008zf}
A.~D. Shapere and Y.~Tachikawa, {\it {Central charges of N=2 superconformal
  field theories in four dimensions}},  {\em JHEP} {\bf 09} (2008) 109,
  [\href{http://xxx.lanl.gov/abs/0804.1957}{{\tt arXiv:0804.1957}}].

\bibitem{ringel1984tame}
C.~M. Ringel, {\it Tame algebras and integral quadratic forms},  {\em Lecture
  Notes in Math.} {\bf 1099} (1984).

\bibitem{Bershadsky:1995vm}
M.~Bershadsky, A.~Johansen, V.~Sadov, and C.~Vafa, {\it {Topological reduction
  of 4-d SYM to 2-d sigma models}},  {\em Nucl.Phys.} {\bf B448} (1995)
  166--186, [\href{http://xxx.lanl.gov/abs/hep-th/9501096}{{\tt
  hep-th/9501096}}].

\bibitem{Kapustin:2006hi}
A.~Kapustin, {\it {Holomorphic reduction of N=2 gauge theories, Wilson-'t Hooft
  operators, and S-duality}},
  \href{http://xxx.lanl.gov/abs/hep-th/0612119}{{\tt hep-th/0612119}}.

\bibitem{Putrov:2015jpa}
P.~Putrov, J.~Song, and W.~Yan, {\it {$(0, 4)$ dualities}},
  \href{http://xxx.lanl.gov/abs/1505.0711}{{\tt arXiv:1505.0711}}.

\bibitem{Gadde:2015wta}
A.~Gadde, S.~S. Razamat, and B.~Willett, {\it {On the reduction of 4d N=1
  theories on $S^2$}},  \href{http://xxx.lanl.gov/abs/1506.0879}{{\tt
  arXiv:1506.0879}}.

\bibitem{Bhardwaj:2013qia}
L.~Bhardwaj and Y.~Tachikawa, {\it {Classification of 4d N=2 gauge theories}},
  {\em JHEP} {\bf 12} (2013) 100,
  [\href{http://xxx.lanl.gov/abs/1309.5160}{{\tt arXiv:1309.5160}}].

\bibitem{Witten:1997yu}
E.~Witten, {\it {On the conformal field theory of the Higgs branch}},  {\em
  JHEP} {\bf 9707} (1997) 003,
  [\href{http://xxx.lanl.gov/abs/hep-th/9707093}{{\tt hep-th/9707093}}].

\bibitem{Benini:2013nda}
F.~Benini, R.~Eager, K.~Hori, and Y.~Tachikawa, {\it {Elliptic genera of
  two-dimensional N=2 gauge theories with rank-one gauge groups}},  {\em
  Lett.Math.Phys.} {\bf 104} (2014) 465--493,
  [\href{http://xxx.lanl.gov/abs/1305.0533}{{\tt arXiv:1305.0533}}].

\bibitem{Benini:2013xpa}
F.~Benini, R.~Eager, K.~Hori, and Y.~Tachikawa, {\it {Elliptic Genera of 2d
  ${\mathcal{N}}$ = 2 Gauge Theories}},  {\em Commun.Math.Phys.} {\bf 333}
  (2015), no.~3 1241--1286, [\href{http://xxx.lanl.gov/abs/1308.4896}{{\tt
  arXiv:1308.4896}}].

\bibitem{Gadde:2013dda}
A.~Gadde and S.~Gukov, {\it {2d Index and Surface operators}},  {\em JHEP} {\bf
  1403} (2014) 080, [\href{http://xxx.lanl.gov/abs/1305.0266}{{\tt
  arXiv:1305.0266}}].

\bibitem{Benini:2015noa}
F.~Benini and A.~Zaffaroni, {\it {A topologically twisted index for
  three-dimensional supersymmetric theories}},  {\em JHEP} {\bf 07} (2015) 127,
  [\href{http://xxx.lanl.gov/abs/1504.0369}{{\tt arXiv:1504.0369}}].

\bibitem{Gaiotto:2008cd}
D.~Gaiotto, G.~W. Moore, and A.~Neitzke, {\it {Four-Dimensional Wall-Crossing
  via Three-Dimensional Field Theory}},  {\em Commun. Math. Phys.} {\bf 299}
  (2010) 163--224, [\href{http://xxx.lanl.gov/abs/0807.4723}{{\tt
  arXiv:0807.4723}}].

\bibitem{tHooft:1977hy}
G.~'t~Hooft, {\it {On the Phase Transition Towards Permanent Quark
  Confinement}},  {\em Nucl. Phys.} {\bf B138} (1978) 1.

\bibitem{tHooft:1979uj}
G.~'t~Hooft, {\it {A Property of Electric and Magnetic Flux in Nonabelian Gauge
  Theories}},  {\em Nucl. Phys.} {\bf B153} (1979) 141.

\bibitem{tHooft:1981ht}
G.~'t~Hooft, {\it {Topology of the Gauge Condition and New Confinement Phases
  in Nonabelian Gauge Theories}},  {\em Nucl. Phys.} {\bf B190} (1981) 455.

\bibitem{Witten:1979ey}
E.~Witten, {\it {Dyons of Charge E Theta/2 Pi}},  {\em Phys. Lett.} {\bf B86}
  (1979) 283--287.

\bibitem{Alim:2011kw}
M.~Alim, S.~Cecotti, C.~Cordova, S.~Espahbodi, A.~Rastogi, and C.~Vafa, {\it
  {$\mathcal{N} = 2$ Quantum Field Theories and Their BPS Quivers}},  {\em Adv.
  Theor. Math. Phys.} {\bf 18} (2014) 27--127,
  [\href{http://xxx.lanl.gov/abs/1112.3984}{{\tt arXiv:1112.3984}}].

\bibitem{Cecotti:2012va}
S.~Cecotti, {\it {Categorical Tinkertoys for ${\mathcal{N}}\!=2$ Gauge
  Theories}},  {\em Int. J. Mod. Phys.} {\bf A28} (2013) 1330006,
  [\href{http://xxx.lanl.gov/abs/1203.6734}{{\tt arXiv:1203.6734}}].

\bibitem{fad}
L.~D. Faddeev and A.~{\relax Yu}. Volkov, {\it {Abelian current algebra and the
  Virasoro algebra on the lattice}},  {\em Phys. Lett.} {\bf B315} (1993)
  311--318, [\href{http://xxx.lanl.gov/abs/hep-th/9307048}{{\tt
  hep-th/9307048}}].

\bibitem{Cecotti:2011gu}
S.~Cecotti and M.~Del~Zotto, {\it {On Arnold's 14 `exceptional' N=2
  superconformal gauge theories}},  {\em JHEP} {\bf 10} (2011) 099,
  [\href{http://xxx.lanl.gov/abs/1107.5747}{{\tt arXiv:1107.5747}}].

\bibitem{Alim:2011ae}
M.~Alim, S.~Cecotti, C.~Cordova, S.~Espahbodi, A.~Rastogi, and C.~Vafa, {\it
  {BPS Quivers and Spectra of Complete N=2 Quantum Field Theories}},  {\em
  Commun. Math. Phys.} {\bf 323} (2013) 1185--1227,
  [\href{http://xxx.lanl.gov/abs/1109.4941}{{\tt arXiv:1109.4941}}].

\bibitem{Cecotti:2013sza}
S.~Cecotti and M.~Del~Zotto, {\it {The BPS spectrum of the 4d N=2 SCFT's $H_1,
  H_2, D_4, E_6, E_7, E_8$}},  {\em JHEP} {\bf 06} (2013) 075,
  [\href{http://xxx.lanl.gov/abs/1304.0614}{{\tt arXiv:1304.0614}}].

\bibitem{Cecotti:2015qha}
S.~Cecotti and M.~Del~Zotto, {\it {Galois covers of N=2 BPS spectra and quantum
  monodromy}},  \href{http://xxx.lanl.gov/abs/1503.0748}{{\tt
  arXiv:1503.0748}}.

\bibitem{meltzer}
H.~Meltzer, {\em Exceptional vector bundles, tilting sheaves and tilting
  complexes for weighted projective lines}, vol.~808.
\newblock American Mathematical Soc., 2004.

\bibitem{Cecotti:2015hca}
S.~Cecotti and M.~Del~Zotto, {\it {Higher S-dualities and Shephard-Todd
  groups}},  {\em JHEP} {\bf 09} (2015) 035,
  [\href{http://xxx.lanl.gov/abs/1507.0179}{{\tt arXiv:1507.0179}}].

\bibitem{Cecotti:2014zga}
S.~Cecotti and M.~Del~Zotto, {\it {$Y$ systems, $Q$ systems, and 4D
  $\mathcal{N}=2$ supersymmetric QFT}},  {\em J. Phys.} {\bf A47} (2014),
  no.~47 474001, [\href{http://xxx.lanl.gov/abs/1403.7613}{{\tt
  arXiv:1403.7613}}].

\bibitem{Nahm:2004ch}
W.~Nahm, {\it Conformal field theory and torsion elements of the bloch group},
  2004.
\newblock \href{http://xxx.lanl.gov/abs/hep-th/0404120}{{\tt hep-th/0404120}}.

\bibitem{Hori:2000ck}
K.~Hori, A.~Iqbal, and C.~Vafa, {\it {D-branes and mirror symmetry}},
  \href{http://xxx.lanl.gov/abs/hep-th/0005247}{{\tt hep-th/0005247}}.

\bibitem{Keegan:2011ci}
S.~Keegan and W.~Nahm, {\it {Nahm's Conjecture and Coset Models}},
  \href{http://xxx.lanl.gov/abs/1103.4986}{{\tt arXiv:1103.4986}}.

\bibitem{Cecotti:2011rv}
S.~Cecotti and C.~Vafa, {\it {Classification of complete N=2 supersymmetric
  theories in 4 dimensions}},  {\em Surveys in differential geometry} {\bf 18}
  (2013) [\href{http://xxx.lanl.gov/abs/1103.5832}{{\tt arXiv:1103.5832}}].

\bibitem{Beem:2014zpa}
C.~Beem, M.~Lemos, P.~Liendo, L.~Rastelli, and B.~C. van Rees, {\it {The
  ${\mathcal N}=2$ Superconformal Bootstrap}},
  \href{http://xxx.lanl.gov/abs/1412.7541}{{\tt arXiv:1412.7541}}.

\bibitem{Feigin:1993qr}
B.~L. Feigin and A.~V. Stoyanovsky, {\it {Quasi-Particles Models for the
  Representation of Lie Algebras and Geometry of Flag Manifold}},  {\em Funct.
  Anal. Appl.} {\bf 28} (1994) 68--90,
  [\href{http://xxx.lanl.gov/abs/hep-th/9308079}{{\tt hep-th/9308079}}].

\bibitem{GIS}
K.~Garrett, M.~E. Ismail, and D.~Stanton, {\it Variants of the
  rogers--ramanujan identities},  {\em Advances in Applied Mathematics} {\bf
  23} (1999), no.~3 274--299.

\bibitem{Feigin:1991wv}
B.~L. Feigin, T.~Nakanishi, and H.~Ooguri, {\it {The Annihilating ideals of
  minimal models}},  {\em Int. J. Mod. Phys.} {\bf A7S1A} (1992) 217--238.
  [Int. J. Mod. Phys.A7,217(1992)].

\bibitem{Nahm:1992sx}
W.~Nahm, A.~Recknagel, and M.~Terhoeven, {\it {Dilogarithm identities in
  conformal field theory}},  {\em Mod. Phys. Lett.} {\bf A8} (1993) 1835--1848,
  [\href{http://xxx.lanl.gov/abs/hep-th/9211034}{{\tt hep-th/9211034}}].

\bibitem{Song:2015wta}
J.~Song, {\it {Superconformal indices of generalized Argyres-Douglas theories
  from 2d TQFT}},  \href{http://xxx.lanl.gov/abs/1509.0673}{{\tt
  arXiv:1509.0673}}.

\bibitem{etafunctions}
S.~O. Warnaar and W.~Zudilin, {\it Dedekind's $\eta$-function and
  rogers--ramanujan identities},  {\em Bulletin of the London Mathematical
  Society} {\bf 44} (2012), no.~1 1--11.

\bibitem{Fortin:2006dn}
J.-F. Fortin, P.~Mathieu, and S.~O. Warnaar, {\it {Characters of Graded
  Parafermion Conformal Field Theory}},  {\em Adv. Theor. Math. Phys.} {\bf 11}
  (2007) [\href{http://xxx.lanl.gov/abs/hep-th/0602248}{{\tt hep-th/0602248}}].

\bibitem{conwarnaar}
O.~Warnaar, ``unpublished notes.''

\bibitem{Buican:2015ina}
M.~Buican and T.~Nishinaka, {\it {On the Superconformal Index of
  Argyres-Douglas Theories}},  \href{http://xxx.lanl.gov/abs/1505.0588}{{\tt
  arXiv:1505.0588}}.

\bibitem{Buican:2015tda}
M.~Buican and T.~Nishinaka, {\it {Argyres-Douglas Theories, the Macdonald
  Index, and an RG Inequality}},  \href{http://xxx.lanl.gov/abs/1509.0540}{{\tt
  arXiv:1509.0540}}.

\bibitem{Benini:2013cda}
F.~Benini and N.~Bobev, {\it {Two-dimensional SCFTs from wrapped branes and
  c-extremization}},  {\em JHEP} {\bf 06} (2013) 005,
  [\href{http://xxx.lanl.gov/abs/1302.4451}{{\tt arXiv:1302.4451}}].

\bibitem{Gadde:2013sca}
A.~Gadde, S.~Gukov, and P.~Putrov, {\it {Fivebranes and 4-manifolds}},
  \href{http://xxx.lanl.gov/abs/1306.4320}{{\tt arXiv:1306.4320}}.

\bibitem{Bah:2015nva}
I.~Bah and V.~Stylianou, {\it {Gravity duals of N=(0,2) SCFTs from M5-branes}},
   \href{http://xxx.lanl.gov/abs/1508.0413}{{\tt arXiv:1508.0413}}.

\bibitem{Xie:2012hs}
D.~Xie, {\it {General Argyres-Douglas Theory}},  {\em JHEP} {\bf 01} (2013)
  100, [\href{http://xxx.lanl.gov/abs/1204.2270}{{\tt arXiv:1204.2270}}].

\bibitem{Xie:2013jc}
D.~Xie and P.~Zhao, {\it {Central charges and RG flow of strongly-coupled N=2
  theory}},  {\em JHEP} {\bf 1303} (2013) 006,
  [\href{http://xxx.lanl.gov/abs/1301.0210}{{\tt arXiv:1301.0210}}].

\bibitem{Kirillov:1993ih}
A.~N. Kirillov, {\it {Dilogarithm Identities, Partitions and Spectra in
  Conformal Field Theory. 1.}},
  \href{http://xxx.lanl.gov/abs/hep-th/9212150}{{\tt hep-th/9212150}}.

\bibitem{lostII}
G.~E. Andrews and B.~C. Berndt, {\em Ramanujan's Lost Notebook. Part II.}
\newblock Springer, 2009.

\bibitem{lostI}
G.~E. Andrews and B.~C. Berndt, {\em Ramanujan's Lost Notebook. Part I.}
\newblock Springer, 2005.

\bibitem{gasperrahman}
G.~Gasper and M.~Rahman, {\em Basic hypergeometric series}, vol.~96.
\newblock Cambridge university press, 2004.

\bibitem{andrews1}
G.~E. Andrews, {\it An analytic generalization of the rogers-ramanujan
  identities for odd moduli},  {\em Proceedings of the National Academy of
  Sciences} {\bf 71} (1974), no.~10 4082--4085.

\bibitem{andrews2}
G.~Andrews, {\it Multiple series rogers-ramanujan type identities},  {\em
  Pacific journal of mathematics} {\bf 114} (1984), no.~2 267--283.

\end{thebibliography}\endgroup

\end{document}